\documentclass[a4paper,11pt]{article}
\usepackage{jheppub}
\pdfoutput=1 
\usepackage[most]{tcolorbox}
\usepackage{bm}
\usepackage{mathtools}
\usepackage{shuffle}
\usepackage{comment}
\usepackage[dvipsnames]{xcolor}
\usepackage{appendix}
\usepackage{tikz}
\usepackage{tikz-3dplot}
\usepackage{xcolor}
\usepackage{float}
\usetikzlibrary{shapes.geometric}
\usetikzlibrary{patterns, decorations.markings, arrows, decorations.pathmorphing}
\usepackage{hhline}
\usetikzlibrary{decorations.markings}
\usetikzlibrary{calc}
\tikzset{
	chalkdust/.style={
		decorate,
		decoration={
        markings,
        mark=between positions 0 and 1 step 1pt
			with {
				\foreach \i in {1,5,9,13,17} {
					\pgfmathsetmacro{\dx}{(rand-0.1)*0.65}
					\pgfmathsetmacro{\dy}{(rand-0.1)*0.65}
					\fill[Maroon, opacity=1]
					($(0,0)+(\dx pt,\dy pt)$) circle (0.5pt);
	}}}}}
\usetikzlibrary{arrows.meta}
\usepackage{mathabx}
\usepackage[export]{adjustbox}
\allowdisplaybreaks
\usepackage[backgroundcolor=green!5, linecolor=blue!60]{todonotes}

\usepackage{booktabs,tabularx}

\newcommand{\la}{\langle}
\newcommand{\ra}{\rangle}
\newcommand{\ang}[1]{\langle #1\rangle}

\definecolor{almond}{rgb}{0.94,0.87,0.8}

\title{Multi-Loop Negative Geometries}
\preprint{
  \begin{tabular}{l}
   MPP-2026-100,
   USTC-ICTS/PCFT-26-10
  \end{tabular}
}

\author[a]{Lance J. Dixon,}\emailAdd{lance@slac.stanford.edu}
\author[b]{Umut Oktem,}\emailAdd{ucoktem@ucdavis.edu}
\author[c]{Shruti Paranjape,}\emailAdd{shruti$\_$paranjape@brown.edu}
\author[b]{Jaroslav Trnka,}\emailAdd{trnka@ucdavis.edu}
\author[d]{Yongqun Xu,}\emailAdd{yongqunxu@mail.ustc.edu.cn}
\author[e]{Shun-Qing Zhang}\emailAdd{sqzhang@mpp.mpg.de}

\affiliation[a]{SLAC National Accelerator Laboratory, Stanford University Stanford, CA 94309, USA}

\affiliation[b]{Center for Quantum Mathematics and Physics (QMAP),\\Department of Physics, University of California, Davis, CA 95616, USA}

\affiliation[c]{Department of Physics, Brown University, Providence, RI 02912, USA}

\affiliation[d]{Interdisciplinary Center for Theoretical Study, University of Science and Technology of China, Hefei, Anhui 230026, China}

\affiliation[e]{Max-Planck-Institut für Physik, Werner-Heisenberg-Institut, 85748 Garching, Germany}

\abstract{Scattering amplitudes in planar ${\cal N}=4$ supersymmetric Yang-Mills theory are dual to expectation values of null polygonal Wilson loops. The Amplituhedron provides a geometric construction for the all-loop integrand as the canonical form on the geometric region in the Grassmannian defined by a certain set of inequalities. For a closely related object, the logarithm of the scattering amplitude, the integrand is reproduced in a similar way using negative geometries. When integrated over all loop momenta except one, the result is infrared (IR) finite and equal to the expectation value of a certain Wilson loop with a Lagrangian insertion. At four points, this quantity, ${\cal F}(g,z)$ only depends on a single cross ratio $z$ and the 't Hooft coupling $g$. At weak coupling, it is known up to three loops from perturbative Wilson loop computations and at strong coupling through the AdS/CFT correspondence at leading order. In this paper, we explore this object further through the lens of the Amplituhedron and negative geometries, which provide very natural IR finite building blocks. We perform an explicit three-loop computation of all negative geometries and show that the number of internal cycles in the diagram is closely linked to the depth of polylogarithms. We calculate the cusp anomalous dimension $\Gamma_{\rm cusp}$ by integrating ${\cal F}(g,z)$ over $z$. We show that the higher-cycle diagrams are suppressed if we consider separate odd and even zeta contributions. Furthermore, we focus on certain convergent infinite series of one-cycle diagrams, perform all-loop order resummations of such contributions, and discuss various features of the result.}

\begin{document} 

\maketitle
\newpage
\section{Introduction}
Scattering amplitudes in the planar limit of ${\cal N}=4$ super Yang-Mills (SYM) theory exhibit many fascinating features and provide an extremely fruitful playground for testing new theoretical ideas. This includes powerful unitarity methods \cite{Bern:1994zx,Bern:1994cg,Bern:2005iz,Bern:2009kd,Bern:2008ap,Bern:2012uc,Carrasco:2021otn,Bourjaily:2017wjl}, Wilson-loop/amplitudes duality and dual conformal symmetry \cite{Alday:2007hr, Drummond:2007aua, Brandhuber:2007yx, Drummond:2007cf}, loop recursion relations \cite{Arkani-Hamed:2010pyv,ArkaniHamed:2010kv}, symbols \cite{Goncharov:2010jf, Golden:2013xva, Golden:2014xqa}, cluster algebras and the symbol bootstrap \cite{Dixon:2014xca,Dixon:2014iba,Caron-Huot:2019vjl, Caron-Huot:2019bsq, Caron-Huot:2020bkp,Basso:2020xts}, on-shell diagrams, and the connection to the positive Grassmannian \cite{Arkani-Hamed:2009ljj, Arkani-Hamed:2009nll, Arkani-Hamed:2009kmp, Arkani-Hamed:2009pfk, Mason:2009qx, Arkani-Hamed:2016byb,Herrmann:2016qea,Bourjaily:2016mnp,Paranjape:2022ymg,Lisitsyn:2025prd}. Many of these recent discoveries involve an intimate connection between amplitudes and intriguing mathematical structures. The prime example of that is the amplituhedron \cite{Arkani-Hamed:2013jha,Arkani-Hamed:2013kca,Arkani-Hamed:2017vfh,Damgaard:2019ztj,Ferro:2022abq,He:2023rou,Arkani-Hamed:2023epq} which computes all tree-level amplitudes and loop integrands in the planar ${\cal N}=4$ SYM theory as properly defined volume forms. This mathematical reformulation of the physical problem of particle scattering provides a novel tool to attack some of the most important problems in this field; see refs.~\cite{Franco:2014csa,Arkani-Hamed:2014dca,Ferro:2015grk,Dennen:2016mdk,Ferro:2016zmx,Arkani-Hamed:2018rsk,Herrmann:2020qlt,Dian:2022tpf} for some applications. The amplituhedron is also interesting from a mathematical perspective as a generalization of the positive Grassmannian. It has many combinatorial properties and connections to cluster algebras \cite{Karp:2016uax,Karp:2017ouj,Galashin:2018fri,Lukowski:2020dpn,Parisi:2021oql,Parisi:2024psm,Lam:2024gyg,Akhmedova:2023wcf,Even-Zohar:2025ydi,Ranestad:2024svp,Mazzucchelli:2025kzy}. Recently, the conjecture that the BCFW recursion relations triangulate the amplituhedron space has been proven~\cite{Even-Zohar:2021sec,Even-Zohar:2023del,Even-Zohar:2024nvw,Galashin:2024ttp}.

The amplituhedron provides a new formulation for loop integrands for planar ${\cal N}=4$ SYM amplitudes. However, these objects are divergent after integration, and need to be regulated. The use of standard dimensional regularization with $D=4-2\epsilon$ is a bit problematic in this context: it breaks symmetries of the result, including dual conformal symmetry which is hard-wired in the geometric formulation, and it also requires the knowledge of the $D$-dimensional integrand. Alternative regularization schemes, such as mass regularization~\cite{Alday:2009zm,Henn:2010bk,Henn:2010ir,Bourjaily:2019jrk,Arkani-Hamed:2023epq}, provide a promising path but still need to be further explored. Ideally, we would like to calculate an infrared (IR) finite object which requires no regularization and can be computed directly in $D=4$. Examples of such objects are the remainder function, which strips the IR divergences from amplitudes by dividing by the BDS ansatz~\cite{Bern:2005iz}, and the ratio function of two amplitudes with different helicities (the IR divergence also cancels in this ratio). These objects vanish at $n=4,5$ and so the first non-trivial results are for $n=6$. Both objects have been in the center of symbology, namely the hexagon and heptagon bootstraps \cite{Dixon:2011pw, Dixon:2011nj,Drummond:2014ffa,Dixon:2015iva,Caron-Huot:2016owq,Dixon:2016nkn,Drummond:2018caf,Caron-Huot:2019vjl,Caron-Huot:2019bsq,Caron-Huot:2020bkp,He:2025tyv}. While the $L$-loop integrand for the ratio function is a combination of products of integrands of various loop orders, the existence of the integrand and a geometric construction for the remainder function is still an open question. 

Another very interesting object is the logarithm of the amplitude. Because of the exponentiation of IR divergences, it only diverges as $1/\epsilon^2$ to all orders in perturbation theory. Furthermore, if the $(L{+}1)$-loop integrand is only integrated over $L$ loop momenta, it produces an IR finite expression. At four points, the result is a function with transcendental weight $2L$ that depends only on a single cross-ratio. In the amplitudes/Wilson-loop duality, this quantity is equal to the expectation value of the null polygonal Wilson loop with a Lagrangian insertion (WLI), at an arbitrary but fixed insertion point, divided by the same Wilson loop without an insertion. Because of dual conformal symmetry, the position of the insertion point, together with the position of the four cusps, gives rise only to a single cross ratio, $z$. In both definitions, we get a function of the 't Hooft coupling $g$ and one parameter, ${\cal F}(g,z)$. If we integrate over the position of the insertion point, or over the fixed (frozen) loop momentum variable in the momentum space, we hit the IR divergence and reproduce the logarithm of the four-point amplitude.

The leading IR divergence of the amplitude logarithm is regulator-independent and controlled by the cusp anomalous dimension, schematically
\begin{equation}
    {\rm log}\,A_4\ \propto\ \frac{\Gamma_{\rm cusp}}{\epsilon^2} + \dots \label{Gamma}
\end{equation}
where $\Gamma_{\rm cusp}$ is known from integrability via the BES equation \cite{Beisert:2006ez}. We can expand it at both weak and strong coupling to an arbitrary order. The cusp anomalous dimension provides an important link between the worlds of integrability and amplitudes, but currently we do not have any first-principle derivation of this object from the amplitudes perspective. As a result, the connection between the BES equation and the leading IR divergence is still conjecture, although a lot of evidence supports it. Obviously, $\Gamma_{\rm cusp}$ can be extracted from ${\cal F}(g,z)$ after integrating over the insertion point, or frozen loop momentum, and reading off the leading IR divergence from eq.~(\ref{Gamma}). There is even a better way to do it, which bypasses all divergences, by integrating over a different contour \cite{Alday:2013ip,Henn:2019swt}.

It is worth emphasizing that while integrability provides $\Gamma_{\rm cusp}$ from the BES equation, it is not clear how it can be used to calculate more complicated objects. One successful program is the flux tube S-matrix bootstrap \cite{Basso:2015uxa, Basso:2014hfa, Basso:2013vsa}, which aims to calculate amplitudes and Wilson loops at finite coupling by doing an expansion around special kinematical regions, the near-collinear limits. While this program provides excellent numerical control over the amplitude, it does not incorporate many mathematical structures discovered in perturbation theory and believed to be applicable to the full non-perturbative object.

This brings us to the main motivation of this program, the Wilson loop ratio ${\cal F}(g,z)$ as the \emph{simplest} non-trivial object that we eventually aim to calculate to all loop orders (not in this paper), and explore the role of integrability and exact solutions in amplitudes. The entry point to this ambitious goal could be provided by $\Gamma_{\rm cusp}$. The study of this particular Wilson loop with a Lagrangian insertion has been a very active and fruitful area of research. At weak coupling, explicit calculations using the Wilson loop technology have been done up to three loops~\cite{Alday:2012hy,Alday:2013ip,Henn:2019swt}, while at strong coupling the leading order contribution was calculated via the AdS/CFT correspondence as a minimal surface problem~\cite{Alday:2007hr}. We will stick to the perturbation theory and use the method of \emph{negative geometries} discovered in ref.~\cite{Arkani-Hamed:2021iya}, and later used in refs.~\cite{Brown:2023mqi,He:2023rou,Henn:2023pkc,He:2023exb,Lagares:2024epo,Li:2024lbw,Chicherin:2024hes,Glew:2024zoh,Brown:2025plq,Chicherin:2025cua,Paranjape:2026kix} to obtain interesting results both in the planar maximally supersymmetric Yang-Mills and ABJM theories. The main idea is to use the refinement of the Amplituhedron picture for the integrand of the logarithm of the amplitude. As a result, we get an expansion in terms of canonical dlog forms on Amplituhedron-like geometries. This provides an alternative expansion to Feynman diagrams with very different term-wise properties. Most importantly, these objects are IR finite term by term and have a pure maximal transcendental weight when integrated over all loop momenta (except the frozen one). Note that in the standard approaches it is often highly non-trivial to expand the IR finite object in terms of finite building blocks; ${\cal F}(g,z)$ is ``barely finite'' and any standard diagrammatic expansion will contain divergent terms that only cancel in the sum.  IR finiteness is essential in our approach, because the whole geometric problem is formulated in $D=4$ and there is no space for any $D=4-2\epsilon$ terms. 

The negative geometry expansion can be recast using graphs with vertices (loop momenta) and edges (quadratic geometric constraints). The set of all such graphs is naturally organized in terms of the number of internal cycles (in this constrained space).  The more cycles there are, the more complicated functions one gets. While this organization has no physical analogue, it provides a very powerful mathematical tool. In particular, certain subsets of graphs can be calculated to all loop orders~\cite{Arkani-Hamed:2021iya,Brown:2023mqi}, and in this paper we will push this frontier even further. As with any amplitude computation, the problem of calculating ${\cal F}(g,z)$ boils down to two parts: calculating the integrand for all negative geometries that contribute at a fixed loop order, and integrating them. The mathematical definition of negative geometries provides us with a set of inequalities that we need to solve. In certain cases this problem is solvable for arbitrary loop order; in other cases we have to work diligently to determine the integrand form. How to do it in general is an open problem and a big challenge, and in this paper we push the technology to the next level. The second part, performing the integration, is an even less explored subject. In cases where the contribution to ${\cal F}_{\rm geom}(z)$ from a given negative geometry satisfies a differential equation, we can solve for the contribution directly, without integrating over the frozen loop momentum. In other cases, we have to rely on the integration technology developed in the context of Feynman integrals. The differential equation method is very natural in this context, although finding an equation for each individual geometry is an open problem. 

In this paper, we focus on two different aspects of the negative geometry expansion of the Wilson loop ${\cal F}(g,z)$. We will construct the integrands for all negative geometries contributing at three loops.  We will then evaluate these integrals explicitly, both at the level of the symbol (in the paper) and the actual functions (in an ancillary file), and confirm that the sum agrees with the result in the literature~\cite{Alday:2012hy,Alday:2013ip,Henn:2019swt}. This agreement not only provides strong evidence that the negative geometry expansion conjecture is correct, but it will also give us very important insights into the hierarchy of transcendental functions that contribute to geometries with no cycles, one cycle, etc. The second topic of this paper is the resummation of certain classes of negative geometries. In ref.~\cite{Arkani-Hamed:2021iya}, all ``tree'' geometries, i.e. ~with no cycles, were resummed to all orders. Here we will discuss the resummation procedure in more generality, and provide a number of examples involving geometries with more cycles. Both of these threads, explicit fixed loop order calculations and all-loop resummations, are important components of the problem and move us a bit closer to the final goal of determining ${\cal F}(g,z)$ exactly. 

The paper is organized as follows: in section \ref{sec:preliminaries}, we will give an overview of the Wilson loop with a Lagrangian insertion. In section \ref{sec:neg_geom}, we discuss the negative geometry expansion, and provide new formulas for all four-loop negative geometries. In section \ref{sec:integrated_neg_geom}, we discuss the integrated negative geometries for the three-loop WLI and the correspondence between geometries and the depths of polylogarithms. In section \ref{sec:resum}, we resum certain classes of one-cycle geometries. In section \ref{sec:cusp}, we discuss the contributions of individual negative geometries to $\Gamma_{\rm cusp}$. We conclude in section \ref{sec:conclusion}.  More technical parts are moved to appendices that cover the 3-cycle integrand~(\ref{app:3-cycle}) and details of the triangle-ladder computation~(\ref{sec:tri-ladder}).

\newpage

\section{Preliminaries}
\label{sec:preliminaries}

We begin by reviewing basic concepts, notation and all necessary ingredients used in the rest of the paper.

\subsection{Momentum twistors} 

Scattering amplitudes in planar $\mathcal{N}=4$ SYM theory and the underlying Amplituhedron geometry \cite{Arkani-Hamed:2013jha} are naturally described using momentum twistor variables~\cite{Hodges:2009hk}. We start with the dual momentum variables $x_i$, which are associated with a particular cyclic ordering of the $n$ external momenta $p_i$ in $\mathbb{R}^{3,1}$, according to
\begin{equation}
    p_i = x_i - x_{i-1} \,,
\end{equation}
where $i = 1,2, \dots,n$ labels the external particles.

Momentum conservation is automatically satisfied if we identify $x_{n+1} \equiv x_1$ and hence the dual points $x_i$ form a closed polygon with light-like edges. The complexified momenta for $n$ particles can also be encoded in $n$ complex vectors $Z_i\in \mathbb{C}^4$, called \textit{momentum twistors}. The dual conformal symmetry \cite{Drummond:2007aua} then acts as ${\rm SL}(4)$ and the invariant multi-linear form is given by the twistor bracket
\begin{equation}\label{four_bracket}
    \langle ijkl \rangle := \det(Z_i Z_j Z_k Z_l ) = \epsilon_{IJKL} Z^{I}_i Z^{J}_j Z^{K}_k Z^{L}_l \ ,
\end{equation}
where $\epsilon_{IJKL}$ is the four-dimensional Levi-Civita tensor. More concretely, we can think about these \emph{four-brackets} as $4\times 4$ determinants of four momentum twistors. Any $n$ generic momentum twistors correspond to cyclically ordered lightlike momenta that satisfy momentum conservation. Hence, the $Z_i$ are unconstrained and all four-brackets are generically non-zero. The four-brackets do satisfy certain identities as derived from the properties of Levi-Civita tensors, for example, the Schouten identity
\begin{equation}
    \la abcd\ra \la abef\ra + \la abcf\ra\la abde\ra = \la abce\ra\la abdf\ra \,.
\end{equation}
Due to dual conformal invariance, we can treat the $Z_j$'s as points in the projective space $\mathbb{P}^{3}$. In order to describe the loop integrands, we introduce (dual) loop momenta $y_i \in \mathbb{R}^{3,1}$, $i=1,2,\ldots,L$, which in momentum-twistor space are associated with lines $AB_\ell$ in $\mathbb{P}^3$. The choice of points $A_i$, $B_i$ on this line is arbitrary, and the collection of all lines in $\mathbb{P}^3$ is called the {\it Grassmannian} $G(2,4)$.

\subsection{Loop integrand and the four-point Amplituhedron}

The Amplituhedron is a geometric picture for scattering amplitudes in planar ${\cal N}=4$ SYM theory, namely the tree-level amplitudes and the loop integrands which are reconstructed as canonical differential forms. The focus of this paper is the four-point scattering process, so we will restrict the discussion to this case. 

At tree-level, the four-point amplitude is given by a Parke-Taylor factor, 
\begin{equation}
    {\cal A}_4^{\rm tree} = \frac{\delta^4(P)\delta^8(Q)}{\la12\ra\la23\ra\la34\ra\la41\ra}   \,,
\end{equation}
where the delta function $\delta^4(P)$ stands for momentum conservation, and $\delta^8(Q)$ for super-momentum conservation. Because of dual conformal symmetry, the four-point amplitude at $L$ loops can be written as 
\begin{equation}
    {\cal A}_4^{(L)} =   {\cal A}_4^{\rm tree} \times \frac{1}{L!}\int {\cal I}^{(L)}_4 \,,
\end{equation}
where ${\cal I}^{(L)}_4$ is the four-point $L$-loop integrand which depends on momentum twistors and $L$ copies of lines $AB_i$, $i=1,2,{\dots},L$. The integrand ${\cal I}^{(L)}_4$ can be obtained as a canonical differential form on the Amplituhedron geometry $\Omega^{(L)}_4$. The symmetry factor $1/L!$ comes from the fact that the integrand ${\cal I}_4^{(L)}$ is fully symmetrized in all loop momenta. Using the topological definition of the Amplituhedron \cite{Arkani-Hamed:2017vfh} we can first define a configuration space of $L$ lines $AB_i$ in projective space $\mathbb{P}^3$. Each of the lines lives in a special subspace, a positive Grassmannian $G_+(2,4)$, and is subject to the following constraints,
\begin{equation}
\la AB_i 12\ra, \la AB_i 23\ra, \la AB_i 34\ra, \la AB_i 14\ra > 0; \quad \la AB_i 13\ra, \la AB_i 24\ra < 0. \label{oneloop}
\end{equation}
Each line has four degrees of freedom and lives in the (four-dimensional) top cell of $G_+(2,4)$. This space has four codimension-1 boundaries where one of the brackets $\la AB_i k\,k{+}1\ra=0$ for $k=1,2,3,4$. Each line can be conveniently parametrized by four variables $x_i$, $y_i$, $z_i$, $w_i$ as
\begin{equation}
    AB_i = \left(\begin{array}{cccc} 1 & x_i & 0 & -y_i\\ 0 & z_i & 1 & w_i \end{array}\right)  \,,\label{param}
\end{equation}
where the row denotes an expansion of points $A_i$, $B_i$ in terms of four momentum twistors $Z_j$, namely,
\begin{equation}
    A_i = Z_1 + x_i Z_2 - y_i Z_4, \quad B_i = z_i Z_2 + Z_3 + w_i Z_4 \,.
\end{equation}
The inequalities (\ref{oneloop}) force all the parameters to be positive, $x_i,y_i,z_i,w_i >0$. For $L=1$ there are no other conditions and the resulting geometry is a simple `first quadrant' of the four-dimensional space. The canonical dlog form on this space \cite{Arkani-Hamed:2017tmz} is then 
\begin{equation}
    \Omega_4^{(1)} = \frac{dx}{x} \frac{dy}{y} \frac{dz}{z} \frac{dw}{w} \,.
\end{equation}
After changing the variables back to the momentum twistors, we find that this expression is equal to 
\begin{equation}
    \Omega_4^{(1)} = \frac{\la AB\,d^2 A\ra\la AB\,d^2B\ra\la 1234\ra^2}{\la AB12\ra\la AB23\ra\la AB34\ra\la AB41\ra} = \la AB\,d^2 A\ra\la AB\,d^2B\ra \ {\cal I}_4^{(1)} \,,
\end{equation}
where ${\cal I}_4^{(1)}$ is a properly normalized box integrand. For $L>1$ the definition of the Amplituhedron contains additional \emph{mutual inequalities} between two loops $AB_i$ and $AB_j$,
\begin{equation}
\la AB_i AB_j\ra > 0\quad \mbox{for any pair $i,j=1,2,\ldots,L$.}
\end{equation}
In the parametrization (\ref{param}) this imposes additional quadratic conditions on the positive parameters $x_i,y_i,z_i,w_i$, 
\begin{equation}
    -(x_i-x_j)(w_i-w_j) - (y_i-y_j)(z_i-z_j) > 0. 
\end{equation}
Although the space of $x_i,y_i,z_i,w_i$ can be defined in a very compact way for any $L$, the geometry is extremely non-trivial, and finding the canonical form $\Omega^{(L)}_4$ is a very difficult task. From the physics perspective, this is not surprising, because the loop integrand contains information about all scattering amplitudes to all loops. (Higher-point amplitudes are encoded here on the cuts of four-point amplitudes.) 

The canonical form for general $L$ can be written as 
\begin{equation}
    \Omega_4^{(L)} =  \prod_{i=1}^{L} \, \langle AB_i \, d^{2}A_i \rangle \, \langle AB_i \, d^{2}B_i \rangle \, \, {\cal I}_4^{(L)} \,,
\end{equation}
where ${\cal I}_n^{(L)}$ is a rational function, the \emph{loop integrand}, written using four-brackets of the twistor coordinates $AB_i$ and the $Z_i$. 

The loop integrand $\Omega_4^{(L)}$ can be expanded in terms of planar diagrams and determined via methods of generalized unitarity \cite{Bern:2006ew,Bern:2007ct,Bourjaily:2011hi}, $f$-graphs, or other very efficient methods which pushed the computational frontier to a remarkable 12 loops \cite{Eden:2011we,Eden:2012tu, Bourjaily:2015bpz, Bourjaily:2016evz,He:2024cej,Bourjaily:2025iad}. At the same time, the BCFW recursion relation for the integrand is believed to yield tilings of the Amplituhedron~\cite{Arkani-Hamed:2012zlh}. The latter statement has recently been proven rigorously at tree level~\cite{Even-Zohar:2021sec,Even-Zohar:2023del,Galashin:2024ttp}, and the loop version of the proof is in progress. It is also known that the one-loop MHV Amplituhedron is a positive geometry \cite{Ranestad:2024svp}, while the two-loop Amplituhedron is a \emph{weighted positive geometry} \cite{Dian:2022tpf} with non-trivial internal boundaries. The precise geometric description of the Amplituhedron for higher $L$ is still an open question.

\subsection{BDS ansatz and the IR divergence}

IR divergences in gauge theories exhibit interesting exponentiation properties. In planar ${\cal N}=4$ SYM theory, the situation is even more special: The IR finite parts of the four- and five-point amplitudes exponentiate as well, up to constants.  For $n=4$ and 5, the amplitude divided by the tree, $A_n$, is given by the Bern-Dixon-Smirnov (BDS) ansatz~\cite{Bern:2005iz},
\begin{equation}
A_n = \exp\left[\sum_{L=1}^\infty g^{2L}\left(f^{(L)}(\epsilon) A_n^{(1)}(L\epsilon) + C_n^{(L)} + E_n^{(L)}(\epsilon)\right)\right] \,,\label{BDS}
\end{equation}
where $A_n^{(1)}(\epsilon)$ is the one-loop amplitude (divided by the tree) calculated in dimensional regularization, and $g^2 = N_c g_{\rm YM}^2/(4\pi)^2 = \lambda/(4\pi)^2$ is the 't Hooft coupling. (At $L$ loops, the one-loop amplitude enters the expression with the rescaling $\epsilon\rightarrow L\epsilon$.) The non-vanishing terms for $n=4$ in the $\epsilon\rightarrow0$ limit are given by
\begin{equation}
     A_4^{(1)}(\epsilon) = e^{-\gamma \epsilon} \left[ -\frac{4}{\epsilon^2} + \frac{2}{\epsilon}\log(st) - 2\log(s)\log(t) +\frac{4}{3} \pi^2 \right] + {\cal O}(\epsilon),
\end{equation}
where $\gamma$ is the Euler-Mascheroni constant.
The $L$-loop coefficient $f^{(L)}(\epsilon)$ is given by
\begin{equation}
    f^{(L)}(\epsilon) = f^{(L)}_0 + \epsilon f^{(L)}_1 + \epsilon^2 f^{(L)}_2  \,.
\end{equation}
The coefficients $C_n^{(L)}={\cal O}(1)$ and $E_n^{(L)}(\epsilon)={\cal O}(\epsilon)$ in dimensional regularization. We can calculate the logarithm of (\ref{BDS}) and expand it as $\epsilon\rightarrow0$, which gives
\begin{equation}
\label{eq:logA4L}
    (\log A_4)^{(L)} = - \frac{\Gamma_{\rm cusp}^{(L)}}{L^2 \epsilon^2} - \frac{\Gamma_{\rm col}^{(L)}}{L \epsilon} + \frac{\Gamma_{\rm cusp}^{(L)}}{2 L \epsilon} \log(st)   + {\cal O}(1),
\end{equation}
where the leading $1/\epsilon^2$ divergence of the amplitude's logarithm at $L$ loops is given by the $L$-loop coefficient of the \emph{cusp anomalous dimension} $\Gamma_{\rm cusp}(g)$, and it is regulator independent. The sub-leading IR divergence is controlled by the \emph{collinear anomalous dimension} $\Gamma_{\rm col}(g)$, which is regulator-dependent.  The first two coefficients in $f^{(L)}(\epsilon)$ are given by these quantities, 
$f_0^{(L)} = \Gamma_{\rm cusp}^{(L)}/4$
and $f_1^{(L)} = L/4\times \Gamma_{\rm col}^{(L)}$.

Very importantly, $\Gamma_{\rm cusp}(g)$ is predicted for arbitrary values of coupling from integrability via Beisert-Eden-Staudacher (BES) equation \cite{Beisert:2006ez}. The result can be expanded at both weak and strong coupling 
\begin{align}
    \Gamma_{\rm cusp}(g)|_{g\ll1} &= 4g^2 -\frac{4\pi^2}{3}g^4 + \frac{44\pi^4}{45}g^6 - 8 \left(\frac{73\pi^6}{630}+4\zeta_3^2\right)g^8 + {\cal O}(g^{10})\\
    \Gamma_{\rm cusp}(g)|_{g\gg1}& = 2g - \frac{3\log2}{2\pi} + {\cal O}\left(\frac{1}{g}\right) \,.
\end{align}
The $L$-loop logarithm can be expanded in $g^2$ at weak coupling in terms of products of amplitudes,
\begin{equation}
    \log A = g^2 A^{(1)} + g^4\left(A^{(2)} - \frac12(A^{(1)})^2\right) + g^6\left(A^{(3)} - A^{(2)}A^{(1)} +\frac{1}{3}(A^{(1)})^3\right) + {\cal O}(g^8), \label{logA}
\end{equation}
where each term in the sum at the given order $g^{2L}$ is $1/\epsilon^{2L}$ divergent, while the sum is only $1/\epsilon^2$ divergent. For example, for $L=2$ both the two-loop amplitude $A^{(2)}$ and the square of the one-loop amplitude $(A^{(1)})^2$ are individually $1/\epsilon^4$ divergent but their combination in the two-loop logarithm is $1/\epsilon^2$ divergent.  
%
\begin{equation}
		\log{A}\,\Big|_{g^4} \, \propto
		\underbrace{\begin{tikzpicture}[baseline={(0,0.5)cm}]
				\draw(0,0)--(-0.5,-0.5);
				\draw(0,0)--(2,0);
				\draw(0,1)--(2,1);
				\draw(0,0)--(0,1);
				\draw(0,1)--(-0.5,1.5);
				\draw(1,0)--(1,1);
				\draw(2,0)--(2,1);
				\draw(2,0)--(2.5,-0.5);
				\draw(2,1)--(2.5,1.5);
				\draw[color=white](0,0)--(0,-0.7);
		\end{tikzpicture}}_{\displaystyle \frac{1}{\epsilon^4}}\ - \frac{1}{2}\; \underbrace{
		\left(\ \begin{tikzpicture}[baseline={(0,0.45)cm}]
				\draw(0,0)--(1,0);
				\draw(0,0)--(0,1);
				\draw(0,1)--(1,1);
				\draw(1,0)--(1,1);
				\draw(1,0)--(1.5,-0.5);
				\draw(0,0)--(-0.5,-0.5);
				\draw(0,1)--(-0.5,1.5);
				\draw(1,1)--(1.5,1.5);
			\end{tikzpicture}\ \right)^2
			\begin{tikzpicture}[baseline={(0,0.49)cm}]
				\draw[color=white](0,0)--(0,-0.7);
			\end{tikzpicture}
		}_{\displaystyle \frac{1}{ \epsilon^4}} \sim \frac{1}{\epsilon^2}\,.
	\end{equation}

\subsection{Integrand of the logarithm}

The mild divergence of the $L$-loop logarithm $(\log A)^{(L)}$ in eq.~(\ref{eq:logA4L}) motivates us to study this object at the integrand level. We can combine the (loop-symmetrized) combinations of amplitudes integrands from (\ref{logA}) and define the $L$-loop integrand of the logarithm $\widetilde{\cal I}_n^{(L)}$,
\begin{equation}
    (\log A_n)^{(L)} = \frac{1}{L!}\int \widetilde{\cal I}_n^{(L)} = {\cal O}\left(\frac{1}{\epsilon^2}\right) \,.
\end{equation}
where $1/L!$ again comes from the symmetrization in all loop momenta. The integrand $\widetilde{\cal I}_n^{(L)}$ can be written using momentum-twistor variables, similar to the $L$-loop integrand ${\cal I}_n^{(L)}$, although it cannot be expanded in terms of planar graphs (only sums of products of planar graphs). The very mild IR divergence post-integration is encoded in the cut structure of the integrand $\widetilde{\cal I}_n^{(L)}$, namely in the collinear region \cite{Arkani-Hamed:2013kca}. 

The collinear region of the integrand corresponds to the configuration where the loop line $AB$ cuts two adjacent propagators $\la AB\,i{-}1\,i\ra{=}\la AB\,i\,i{+}1\ra{=}0$ including the Jacobian. This localizes the line in the plane $(Z_{i{-}1}Z_iZ_{i{+}1})$ and passes through the point $Z_i$, 
%
\begin{equation}
\begin{tikzpicture}[scale=0.78]
    \draw[ultra thick](0.191,0.160)--(200:3.2) node [pos=0.9] {\Huge$\bullet$} node [pos=0.9, below right] {$Z_3$};
    \draw[ultra thick, Maroon](0.235,-0.085)--(170:3.2) node [pos=0.9, above] {$(AB)$};
    \draw[ultra thick](-0.043,-0.246)--(130:3.2) node [pos=0.9, black] {\Huge$\bullet$}  node [pos=0.9, above right, black] {$Z_1$};
    \node at (-0.2,-0.05) {\Huge$\bullet$};
    \node[below] at (-0.2,-0.1) {$Z_2$};
\end{tikzpicture}
\end{equation}
which leaves one unfixed degree of freedom,
\begin{equation}
    Z_A = Z_i,\quad Z_B = Z_{i{-}1} + \alpha Z_{i{+}1}\,.
\end{equation}
In momentum space this region corresponds to the loop momentum being proportional to an external momentum, $\ell\sim p_i$. If a term in the one-loop integrand has a vanishing residue in this region, then the associated integral is IR finite. The finiteness can be due to either the absence of a massless corner in the term in the integrand, or a special integrand numerator such as a chiral pentagon \cite{Arkani-Hamed:2010pyv,ArkaniHamed:2010kv}. 

For higher loops, the situation is a bit more subtle. At two loops, the integrand for the logarithm of the amplitudes (after combining all contributing terms) is given by,
\begin{equation}
\label{eq:2loopint}
\widetilde{\cal I}_4^{(2)} = \frac{-\la1234\ra^3(\la AB13\ra\la CD24\ra+\la AB24\ra\la CD13\ra)}{\la AB12\ra\la AB23\ra\la AB34\ra\la AB14\ra\la CD12\ra\la CD23\ra\la CD34\ra\la CD14\ra\la ABCD\ra} \,,
\end{equation}
where lines $AB$, $CD$ represent two loop momenta. We can identify dangerous collinear regions in the denominator from massless corners; for example, a pair of propagators $\la AB12\ra,\la AB23\ra$ which allow the line $AB$ to be localized in the plane $(123)$ passing through $Z_2$, i.e. $Z_A = Z_2$ and $Z_B = Z_1+\alpha Z_3$. The numerator vanishes in this configuration and we cannot directly localize $AB$ to this configuration. However, we can consider combined cuts of $AB$ and $CD$ and localize both the lines in the same collinear region by cutting $\la AB12\ra=\la AB23\ra=\la CD12\ra=\la CD23\ra=\la ABCD\ra=0$ and associated Jacobians, 
\begin{equation}
\begin{tikzpicture}[scale=0.78]
    \draw[ultra thick](0.191,0.160)--(200:3.2) node [pos=0.9] {\Huge$\bullet$} node [pos=0.9, below right] {$Z_3$};
    \draw[ultra thick, Maroon](0.235,-0.085)--(170:3.2) node [pos=0.9, above] {$(CD)$};
    \draw[ultra thick, Maroon](0.125,-0.216)--(150:3.2) node [pos=0.9, above] {$(AB)$};
    \draw[ultra thick](-0.043,-0.246)--(130:3.2) node [pos=0.9, black] {\Huge$\bullet$}  node [pos=0.9, above right, black] {$Z_1$};
    \node at (-0.2,-0.05) {\Huge$\bullet$};
    \node[below] at (-0.2,-0.1) {$Z_2$};
\end{tikzpicture}
\end{equation}

As a consequence of this non-vanishing cut, the integral generates the expected $1/\epsilon^2$ divergence. If the integrand did not allow any localization in the collinear region of either of the lines $AB$, $CD$, the integral would be IR finite (which is not the case).

At $L$ loops we have a configuration of $L$ loop lines $AB_i$, $i=1,2,{\dots},L$. The cuts of the $L$-loop integrand of the logarithm $\widetilde{\cal I}_4^{(L)}$ satisfy a very special property when the lines are localized in the collinear region: If we consider any cut which involves fewer than $L$ lines $AB_i$, then we can never localize any of the lines in the collinear region. The only way to localize any line $AB_i$ in the collinear region is to cut all of the loops $AB_i$ and move them collectively to the collinear region. A consequence of this very restricted behavior in the collinear regions of the integrand is the mild $1/\epsilon^2$ IR divergence post-integration. Note that the integrand of the $L$-loop amplitude ${\cal I}_4^{(L)}$ has exactly the opposite property: we can localize each of the loop lines $AB_i$ to the collinear region without touching the others. This property leads to the very strong $1/\epsilon^{2L}$ IR divergence. 
\begin{equation}
\begin{tikzpicture}[scale=0.78]
    \draw[ultra thick](0.191,0.160)--(200:4) node [pos=0.9] {\Huge$\bullet$} node [pos=0.9, below right] {$Z_3$};
    \draw[ultra thick, Maroon](0.25,0)--(180:4) node [pos=0.9, above] {$(AB)_1$};
    \draw[ultra thick, Maroon](0.216,-0.125)--(165:4) node [pos=0.9, above] {$(AB)_2$} node[pos=0.75,sloped,above] {$\vdots$};
    \draw[ultra thick, Maroon](0.125,-0.216)--(150:4) node [pos=0.9, above] {$(AB)_L$};
    \draw[ultra thick](-0.043,-0.246)--(130:4) node [pos=0.9, black] {\Huge$\bullet$}  node [pos=0.9, above right, black] {$Z_1$};
    \node at (-0.2,-0.05) {\Huge$\bullet$};
    \node[below] at (-0.2,-0.1) {$Z_2$};
\end{tikzpicture}
\end{equation}

This very special property of the integrand $\widetilde{\cal I}_4^{(L)}$ has a dramatic consequence on the integration: we need to integrate over all of the loops $AB_i$ to generate an IR divergence. If we freeze any of the loops and integrate over the others, the result is still IR finite. In particular, if we `freeze' the loop $AB_1$ and integrate over $AB_2,AB_3,{\dots},AB_L$ we get a $(L{-}1)$ loop function ${\cal F}^{(L{-}1)}$ which depends on four external twistors $Z_i$ and the line $AB_1$. Due to dual conformal symmetry, ${\cal F}^{(L{-}1)}$ is a function of a single variable $z$:
\begin{equation} \label{eq:F_from_geometry}
    {\cal F}^{(L{-}1)}(z) = \frac{1}{(L{-}1)!}\int {\rm d}AB_2\,{\rm d}AB_3{\dots} {\rm d}AB_L \,\widetilde{\cal I}_4^{(L)} \,,
\end{equation}
where ${\rm d}AB \equiv \la AB\,d^2A\ra\la AB\,d^2B\ra$ is the integration measure of one of the loops, and 
\begin{equation}
\label{eq:zdef}
    z = \frac{\la AB12\ra\la AB34\ra}{\la AB23\ra\la AB14\ra} \,,
\end{equation}
where we identified $AB\equiv AB_1$. The function ${\cal F}^{(L{-}1)}$ has a uniform transcendental weight $2L-2$. For $L=1$ and $L=2$ it is given by
\begin{equation}
    {\cal F}^{(0)}(z) = -1,\quad {\cal F}^{(1)}(z) = \log^2z + \pi^2 \,.
\end{equation}
If we resum the all-loop contributions, we get an exact function ${\cal F}(g,z)$,
\begin{equation}
    {\cal F}(g,z) = \sum_{L=0}^\infty g^{2L} {\cal F}^{(L)}(z).
\label{eq:fexact}
\end{equation}
The logarithm of the four-point amplitude $\log A_4$ is then obtained by integrating the last loop line $AB$,
\begin{equation}
    \log A_4 = \int {\rm d}AB\,{\cal F}(g,z).
\end{equation}
However, this last integration produces an IR divergence and needs a regularization. In particular, if we want to stay in momentum-twistor space, we cannot use a standard dimensional regularization, and we have to regulate directly in four dimensions. One option is to calculate amplitudes on the Coulomb branch \cite{Flieger:2025ekn} -- that is, consider massive amplitudes -- which also correspond (for a special mass configuration) to deformations of the Amplituhedron \cite{Arkani-Hamed:2023epq}. Other alternatives that preserve dual conformal symmetry have also been proposed \cite{Alday:2009zm,Henn:2010bk,Henn:2010ir,Caron-Huot:2014lda,Bourjaily:2019jrk}. 

In a sense, the function ${\cal F}(g,z)$ is an ideal integrated object to study, as it is IR finite and only depends on one parameter. At the same time, it is only one integration away from the amplitude. 

\subsection{Wilson loop and IR finite object}

One of the most surprising properties of planar ${\cal N}=4$ SYM amplitudes is the duality to expectation values of null polygonal Wilson loops \cite{Drummond:2007aua,Brandhuber:2007yx, Alday:2007hr}. Namely, the $n$-point scattering amplitude $A_n(p_1,\dots,p_n)$ is dual to $W_n(x_1,\dots,x_n)$, the $n$-sided light-like polygonal Wilson loop, with cusps located at points $x_i$ in the dual momentum space.

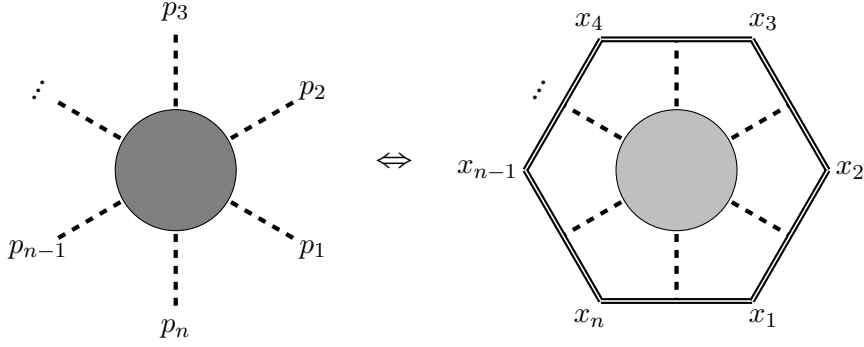
\begin{center}
	\begin{tikzpicture}[baseline={(0,0)cm}]
		\draw[dashed, ultra thick] (0,0) --(30:1.8);
		\draw[dashed, ultra thick] (0,0) --(90:1.8);
		\draw[dashed, ultra thick] (0,0) --(150:1.8);
		\draw[dashed, ultra thick] (0,0) --(210:1.8);
		\draw[dashed, ultra thick] (0,0) --(270:1.8);
		\draw[dashed, ultra thick] (0,0) --(330:1.8);
		\node at (330:2.1){$p_1$};
		\node at (30:2.1){$p_2$};
		\node at (90:2.1){$p_3$};
		\node at (147:2.1){\Large$\cdot$};
		\node at (150:2.095){\Large$\cdot$};
		\node at (153:2.1){\Large$\cdot$};
		\node at (210:2.1){$p_{n-1}$};
		\node at (270:2.1){$p_{n}$};
		\filldraw[fill= gray, draw=black] (0,0) circle (0.8cm);
	\end{tikzpicture} 
	~~{~\Large$ \Leftrightarrow$ ~}
	\begin{tikzpicture}[baseline={(0,0)cm}]
		\draw[dashed, ultra thick] (0,0) --(30:1.7);
		\draw[dashed, ultra thick] (0,0) --(90:1.7);
		\draw[dashed, ultra thick] (0,0) --(150:1.7);
		\draw[dashed, ultra thick] (0,0) --(210:1.7);
		\draw[dashed, ultra thick] (0,0) --(270:1.7);
		\draw[dashed, ultra thick] (0,0) --(330:1.7);
		\draw[thick, double] (1,1.73) -- (2,0);
		\draw[thick, double] (1,1.73) -- (-1,1.73);
		\draw[thick, double] (-2,0) -- (-1, 1.73);
		\draw[thick, double] (-2,0) -- (-1,-1.73);
		\draw[thick, double] (1,-1.73) -- (-1,-1.73);
		\node at (147:2.1){\Large$\cdot$};
		\node at (150:2.095){\Large$\cdot$};
		\node at (153:2.1){\Large$\cdot$};
		\node at (1.15,-1.96)	{$x_1$};
		\node at (2.3,0)		{$x_2$};
		\node at (1.15,1.96)	{$x_3$};
        \node at (-1.15,1.96)	{$x_4$};
		\node at (-1.15,-1.96)	{$x_n$};
		\node at (-2.5,0)		{$x_{n-1}$};
		\draw[thick, double] (1,-1.73) -- (2,0);
		\filldraw[fill= lightgray, draw=black] (0,0) circle (0.8cm);
	\end{tikzpicture}
\end{center}


The conformal symmetry of the Wilson loop then translates to the dual conformal symmetry of the amplitude. Also, the UV divergence of the Wilson loop (from virtual gluons emitted near each corner of the polygon) corresponds to the IR divergence of the amplitude (from virtual gluons exchanged at long distances between color-adjacent external particles). The same duality also holds for the logarithm of the amplitude, which is related to a ratio of two Wilson loops,
\begin{equation}
F_n(x_1,\ldots,x_n;z_0) := \pi^2\frac{\langle{W_n(x_1,\cdots,x_n) {\cal L}(z_0)\rangle}}{\langle W_n(x_1,\ldots,x_n)\rangle} \,,
\label{eq:WLIn}
\end{equation}
where $\mathcal{L}(z_0)$ denotes the Lagrangian inserted at a point $z_0$.
This observable has been considered extensively in previous works \cite{Alday:2011ga,Alday:2012hy,Alday:2013ip,Engelund:2011fg,Engelund:2012re,Henn:2019swt,Chicherin:2022bov,Chicherin:2022zxo,Chicherin:2024hes,Carrolo:2025pue,Chicherin:2025jej,Abreu:2024yit} and it enjoys a number of special properties. The leading term at strong coupling was calculated as a minimal surface problem using the AdS/CFT correspondence \cite{Alday:2011ga}. More specifically, it corresponds to classical solutions in supergravity. However, the computation of the sub-leading term in the strong-coupling expansion is very difficult as it would require quantizing the Nambu-Goto action in the curved AdS background.  

Most importantly, eq.~(\ref{eq:WLIn}) is IR finite at each order in perturbation theory, and because of the duality between Wilson loops and scattering amplitudes \cite{Alday:2007hr,Drummond:2007aua,Brandhuber:2007yx}, the integrand of ${\cal F}_n$ at $L$ loops is equal to the integrand of the logarithm of the $n$-point MHV amplitude at $(L{+}1)$ loops, after treating $z_0$ as an integration variable of the ``frozen'' loop momentum. In the four-point case, ${\cal F}_n$ reduces to the object ${\cal F}(g,z)$ we constructed earlier. Furthermore, the function $\mathcal{F}(g,z)$ is related to the object on the left-hand side above, namely $F(x_1,...,x_n;z_0)$, through 
\begin{align}
\label{eq:TwoFnormalizations}
    {F}(g,z)=-g^2\mathcal{F}(g,z)\,. 
\end{align}
As a result, ${\cal F}(g,z)$ is an ideal object for testing new ideas and exploring features of integrated expressions that are related to planar ${\cal N}=4$ SYM amplitudes. It is a function of a single cross ratio, it has a direct definition in terms of Wilson loops, and it is only one integration away from being the MHV amplitude's logarithm. When we talk about strong coupling results however, we will consider $F(g,z)$ instead of $\mathcal{F}(g,z)$ as it is more directly related to Wilson loops.

\section{Negative geometry expansion}
\label{sec:neg_geom}

As remarked earlier, the four-point $L$-loop Amplituhedron space is defined as a configuration of $L$ lines $AB_i$ in $\mathbb{P}^3$ which satisfy certain conditions with respect to four external twistors $Z_1$, $Z_2$, $Z_3$ and $Z_4$,
\begin{equation}
\la AB_i 12\ra, \la AB_i 23\ra, \la AB_i 34\ra, \la AB_i 14\ra > 0; \quad \la AB_i 13\ra, \la AB_i 24\ra < 0 
\end{equation}
and mutual conditions $\la AB_i AB_j\ra >0$ for any pair of lines. The details were summarized in section 2.2. Here we employ a graphical notation for this space and associated canonical form,
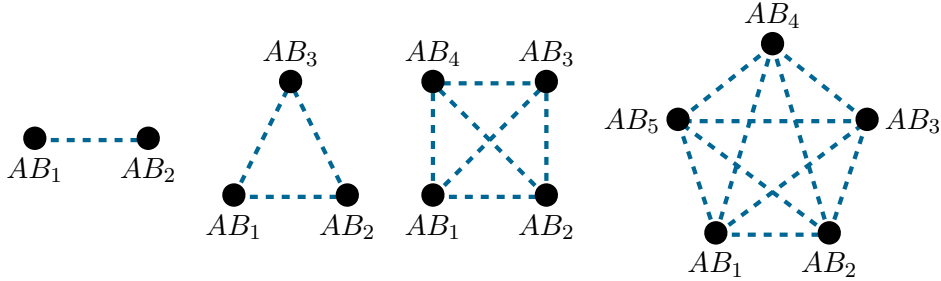
\begin{center}
    \begin{tikzpicture}[baseline={(0,0)cm}]
        \draw[dashed, ultra thick, MidnightBlue] (0,0) -- (1.5,0) node[at start, black] {\Huge$\bullet$} node[at end, black] {\Huge$\bullet$};
        \node[below, black] at (0,-0.1)  {$AB_1$};
        \node[below, black] at (1.5,-0.1)  {$AB_2$};
    \end{tikzpicture}\ \begin{tikzpicture}[baseline={(0,0.75)cm}]
        \draw[dashed, ultra thick, MidnightBlue] (1.5,0) -- (0.75,1.5);
        \draw[dashed, ultra thick, MidnightBlue] (0,0) -- (0.75,1.5) node[at end, black] {\Huge$\bullet$};
        \draw[dashed, ultra thick, MidnightBlue] (0,0) -- (1.5,0) node[at start, black] {\Huge$\bullet$} node[at end, black] {\Huge$\bullet$};
        \node[below, black] at (0,-0.1)  {$AB_1$};
        \node[below, black] at (1.5,-0.1)  {$AB_2$};
        \node[above, black] at (0.75,1.6)  {$AB_3$};
    \end{tikzpicture}\ \begin{tikzpicture}[baseline={(0,0.75)cm}]
        \draw[dashed, ultra thick, MidnightBlue] (1.5,0) -- (0,1.5);
        \draw[dashed, ultra thick, MidnightBlue] (0,0) -- (1.5,1.5);
        \draw[dashed, ultra thick, MidnightBlue] (1.5,1.5) -- (0,1.5);
        \draw[dashed, ultra thick, MidnightBlue] (1.5,0) -- (1.5,1.5) node[at end, black] {\Huge$\bullet$};
        \draw[dashed, ultra thick, MidnightBlue] (0,0) -- (0,1.5) node[at end, black] {\Huge$\bullet$};
        \draw[dashed, ultra thick, MidnightBlue] (0,0) -- (1.5,0) node[at start, black] {\Huge$\bullet$} node[at end, black] {\Huge$\bullet$};
        \node[below, black] at (0,-0.1)  {$AB_1$};
        \node[below, black] at (1.5,-0.1)  {$AB_2$};
        \node[above, black] at (1.5,1.6)  {$AB_3$};
        \node[above, black] at (0,1.6)  {$AB_4$};
    \end{tikzpicture}\ \begin{tikzpicture}[baseline={(0,1.25)cm}]
        \draw[dashed, ultra thick, MidnightBlue] (0,0) -- (2,1.5);
        \draw[dashed, ultra thick, MidnightBlue] (1.5,0) -- (-0.5,1.5);
        \draw[dashed, ultra thick, MidnightBlue] (0,0) -- (-0.5,1.5);
        \draw[dashed, ultra thick, MidnightBlue] (1.5,0) -- (2,1.5);
        \draw[dashed, ultra thick, MidnightBlue] (0.75,2.5) -- (-0.5,1.5);
        \draw[dashed, ultra thick, MidnightBlue] (0.75,2.5) -- (2,1.5);
        \draw[dashed, ultra thick, MidnightBlue] (2,1.5) -- (-0.5,1.5) node[at end, black] {\Huge$\bullet$} node[at start, black] {\Huge$\bullet$};
        \draw[dashed, ultra thick, MidnightBlue] (1.5,0) -- (0.75,2.5);
        \draw[dashed, ultra thick, MidnightBlue] (0,0) -- (0.75,2.5) node[at end, black] {\Huge$\bullet$};
        \draw[dashed, ultra thick, MidnightBlue] (0,0) -- (1.5,0) node[at start, black] {\Huge$\bullet$} node[at end, black] {\Huge$\bullet$};
        \node[below, black] at (0,-0.1)  {$AB_1$};
        \node[below, black] at (1.5,-0.1)  {$AB_2$};
        \node[above, black] at (0.75,2.6)  {$AB_4$};
        \node[right, black] at (2.1,1.5)  {$AB_3$};
        \node[left, black] at (-0.6,1.5)  {$AB_5$};
    \end{tikzpicture}
\end{center}
where each vertex corresponds to one of the lines $AB_i$, which satisfies all one-loop Amplituhedron conditions (\ref{oneloop}), and the dashed line stands for the mutual condition $\la AB_i AB_j\ra >0$ between respected lines. Note that this graph is complete (all vertices are connected to all other vertices). The next step is to express the canonical form for this space in terms of canonical forms of other geometries. To do that we express each \emph{positive} (dashed blue) link as a difference between no link and \emph{negative} (solid red) link. 
\begin{equation}
\label{eq:linkrel}
    \begin{tikzpicture}[baseline={(0,0)cm},scale=1.]
        \draw[dashed, ultra thick, MidnightBlue] (0,0) -- (1.5,0) node[at start, black] {\Huge$\bullet$} node[at end, black] {\Huge$\bullet$};
        \node[below, black] at (0,-0.1)  {$AB_1$};
        \node[below, black] at (1.5,-0.1)  {$AB_2$};
        \node[below] at (0.75,-0.75) {$\la AB_1 AB_2\ra>0$};
    \end{tikzpicture}\ +\ \begin{tikzpicture}[baseline={(0,0)cm},scale=1.]
        \draw[ultra thick, Maroon] (0,0) -- (1.5,0) node[at start, black] {\Huge$\bullet$} node[at end, black] {\Huge$\bullet$};
        \node[below, black] at (0,-0.1)  {$AB_1$};
        \node[below, black] at (1.5,-0.1)  {$AB_2$};
        \node[below] at (0.75,-0.75) {$\la AB_1 AB_2\ra<0$};
    \end{tikzpicture}\ =\ \begin{tikzpicture}[baseline={(0,0)cm},scale=1.]
        \draw[dashed, ultra thick, white] (0,0) -- (1.5,0) node[at start, black] {\Huge$\bullet$} node[at end, black] {\Huge$\bullet$};
        \node[below, black] at (0,-0.1)  {$AB_1$};
        \node[below, black] at (1.5,-0.1)  {$AB_2$};
        \node[below] at (0.75,-0.75) {no sign restriction};
    \end{tikzpicture}
\end{equation}

At the level of geometries, eq.~(\ref{eq:linkrel}) describes the union of the two spaces with the conditions $\la AB_i AB_j\ra>0$ and $\la AB_iAB_j\ra<0$ as the space with no restriction on the sign of $\la AB_iAB_j\ra$. At the level of canonical forms,  eq.~(\ref{eq:linkrel}) describes the relation between the forms on these three spaces. We can now make this replacement for all dashed links and get the expansion for the canonical form of the amplitude as 
\begin{align}
    \label{fig4}
    \Omega_L\ =\ \begin{tikzpicture}[baseline={(0,1.25)cm}]
        \draw[dashed, ultra thick, MidnightBlue] (0,0) -- (2,1.5);
        \draw[dashed, ultra thick, MidnightBlue] (1.5,0) -- (-0.5,1.5);
        \draw[dashed, ultra thick, MidnightBlue] (0,0) -- (-0.5,1.5);
        \draw[dashed, ultra thick, MidnightBlue] (1.5,0) -- (2,1.5);
        \draw[dashed, ultra thick, MidnightBlue] (0.75,2.5) -- (-0.5,1.5);
        \draw[dashed, ultra thick, MidnightBlue] (0.75,2.5) -- (2,1.5);
        \draw[dashed, ultra thick, MidnightBlue] (2,1.5) -- (-0.5,1.5) node[at end, black] {\Huge$\bullet$} node[at start, black] {\Huge$\bullet$};
        \draw[dashed, ultra thick, MidnightBlue] (1.5,0) -- (0.75,2.5);
        \draw[dashed, ultra thick, MidnightBlue] (0,0) -- (0.75,2.5) node[at end, black] {\Huge$\bullet$};
        \draw[dashed, ultra thick, MidnightBlue] (0,0) -- (1.5,0) node[at start, black] {\Huge$\bullet$} node[at end, black] {\Huge$\bullet$};
    \end{tikzpicture}\ =\ \sum_{\text{all }G} (-1)^{E(G)}\ \begin{tikzpicture}[baseline={(0,1.25)cm}]
        \draw[ultra thick, Maroon] (0,0)--(-0.5,1.5)--(0.75,2.5)--(0,0);
        \draw[ultra thick, Maroon] (1.5,0)--(2,1.5);
        \node[black] at (0,0)  {\Huge$\bullet$};
        \node[black] at (1.5,0)  {\Huge$\bullet$};
        \node[black] at (0.75,2.5)  {\Huge$\bullet$};
        \node[black] at (2,1.5)  {\Huge$\bullet$};
        \node[black] at (-0.5,1.5)  {\Huge$\bullet$};
    \end{tikzpicture}
\end{align}

The sum of all graphs with $L$ vertices can be written as the exponential of the sum of only connected graphs also with $L$ vertices. We use this fact from graph theory and write the canonical form for the logarithm of the amplitude $\widetilde{\Omega}_4^{(L)}$,
\begin{align}
\label{eq:negconn}
    \widetilde{\Omega}_L\ =\ \sum_{G} (-1)^{E(G)}\ \begin{tikzpicture}[baseline={(0,1.25)cm}]
        \draw[ultra thick, Maroon] (0,0)--(-0.5,1.5)--(2,1.5)--(0,0)--(1.5,0);
        \draw[ultra thick, Maroon] (0.75,2.5)--(2,1.5);
        \node[black] at (0,0)  {\Huge$\bullet$};
        \node[black] at (1.5,0)  {\Huge$\bullet$};
        \node[black] at (0.75,2.5)  {\Huge$\bullet$};
        \node[black] at (2,1.5)  {\Huge$\bullet$};
        \node[black] at (-0.5,1.5)  {\Huge$\bullet$};
    \end{tikzpicture}
\end{align}
where the differential form $\widetilde{\Omega}_4^{(L)}$ differs from the loop integrand of the amplitude's logarithm $\widetilde{\cal I}_4^{(L)}$ only by the measure,
\begin{equation}
    \widetilde{\Omega}_4^{(L)} =  \prod_{i=1}^{L} \langle AB_i \, d^{2}A_i \rangle \, \langle AB_i \, d^{2}B_i \rangle \, \, \widetilde{\cal I}_4^{(L)} \,.
\end{equation}
Each term in the sum is a canonical form on a \emph{negative geometry} (this name just refers to the fact that the mutual conditions are negative). We denote by $\Omega_G$ a canonical form on the negative geometry for the graph $G$. Note that each negative geometry has an independent definition as the space of $L$ lines in $\mathbb{P}^3$ where each line $AB_i$ satisfies the one-loop Amplituhedron conditions (\ref{oneloop}), and the mutual conditions
\begin{equation}
    \la AB_iAB_j\ra < 0\quad \mbox{for each edge $(i,j)$ in the graph.}
\end{equation}
It is obvious that more edges the graph has, more mutual conditions are imposed and more complicated the geometry we get. Note that we can also define a mixed geometry where some of the links are positive and some of the links are negative. These objects might be interesting to explore too but we will not need them for our discussion here. 

There is a natural organization of the sum over all connected negative geometries (\ref{eq:negconn}) into groups based on the number of edges. The minimal number of edges for a connected graph with $L$ vertices is $L{-}1$ and the maximal number is $L(L{-}1)/2$. The graphs with $L{-}1$ edges are \emph{tree graphs}, {\it i.e.}~they do not have any closed cycles. Adding an edge to a tree graph we get a graph with one closed cycle, adding two edges leads to two closed cycles, etc. As a result, we get a natural expansion in terms of the number ${\cal C}$ of closed cycles (or `loops' in this space of loops, hence we also refer to this expansion as being in loops of loops): 
\begin{equation}
    \widetilde{\Omega}_4^{(L)} = \sum_{{\cal C}=0}^{{\cal C}_{max}} \sum_{G_{\cal C}} (-1)^{E(G)}\Omega_{G_{\cal C}} \,,
\end{equation}
where ${\cal C}_{max} = (L{-}1)(L{-}2)/2$ and $G_{\cal C}$ is a graph with ${\cal C}$ closed cycles. Note that bubble subgraphs, {\it i.e.}~graphs where two vertices are connected by more than one edge, are not allowed.

Let us write down explicitly the expansions for $L=1,2,3,4$ in terms of negative geometries, 
\begin{equation} 
\widetilde{\Omega}_4^{(1)} =\quad \begin{tikzpicture}[baseline={(0,-0.1)cm}]
    \node at (0,0) {\Huge$\bullet$};
\end{tikzpicture} \,,\qquad\qquad
  \widetilde{\Omega}_4^{(2)} =-\begin{tikzpicture}[baseline={(0,-0.1)cm}]
      \draw[ultra thick, Maroon] (0,0)--(1,0) node[at end, black] {\Huge$\bullet$} node[at start, black] {\Huge$\bullet$};
  \end{tikzpicture} \,,
\end{equation}
\begin{equation} \label{eq:omega3}
  \widetilde{\Omega}_4^{(3)} = \underbrace{\begin{tikzpicture}[baseline={(0,0.5)cm}]
      \draw[ultra thick, Maroon] (0,0)--(1.5,0) node[at end, black] {\Huge$\bullet$};
      \draw[ultra thick, Maroon] (0,0)--(0.75,1) node[at end, black] {\Huge$\bullet$} node[at start, black] {\Huge$\bullet$};
  \end{tikzpicture}}_{\displaystyle \widetilde{\Omega}_3^{\rm tree}}\ +\ \underbrace{\begin{tikzpicture}[baseline={(0,0.5)cm}]
      \draw[ultra thick, Maroon] (0.75,1)--(1.5,0);
      \draw[ultra thick, Maroon] (0,0)--(1.5,0) node[at end, black] {\Huge$\bullet$};
      \draw[ultra thick, Maroon] (0,0)--(0.75,1) node[at end, black] {\Huge$\bullet$} node[at start, black] {\Huge$\bullet$};
  \end{tikzpicture}}_{\displaystyle \widetilde{\Omega}_3^\text{1-cycle}}
  \,,
\end{equation}
\begin{align} \label{eq:omega4}
\widetilde{\Omega}_4^{(4)}=&\ \underbrace{-\begin{tikzpicture}[baseline={(0,0.5)cm}]
    \draw[ultra thick, Maroon] (0,1.25)--(1.375,1.25) node[at end, black] {\Huge$\bullet$};
    \draw[ultra thick, Maroon] (0,0)--(1.375,0) node[at end, black] {\Huge$\bullet$};
    \draw[ultra thick, Maroon] (0,0)--(0,1.25) node[at end, black] {\Huge$\bullet$} node[at start, black] {\Huge$\bullet$};
\end{tikzpicture}\ -\ \begin{tikzpicture}[baseline={(0,0.5)cm}]
    \draw[ultra thick, Maroon] (0,0)--(1.375,1.25) node[at end, black] {\Huge$\bullet$};
    \draw[ultra thick, Maroon] (0,0)--(1.375,0) node[at end, black] {\Huge$\bullet$};
    \draw[ultra thick, Maroon] (0,0)--(0,1.25) node[at end, black] {\Huge$\bullet$} node[at start, black] {\Huge$\bullet$};
\end{tikzpicture}}_{ \displaystyle \widetilde{\Omega}_4^{\rm tree}}\ +\ \underbrace{\begin{tikzpicture}[baseline={(0,0.5)cm}]
    \draw[ultra thick, Maroon] (1.375,0)--(1.375,1.25);
    \draw[ultra thick, Maroon] (0,1.25)--(1.375,1.25) node[at end, black] {\Huge$\bullet$};
    \draw[ultra thick, Maroon] (0,0)--(1.375,0) node[at end, black] {\Huge$\bullet$};
    \draw[ultra thick, Maroon] (0,0)--(0,1.25) node[at end, black] {\Huge$\bullet$} node[at start, black] {\Huge$\bullet$};
\end{tikzpicture}\ +\ \begin{tikzpicture}[baseline={(0,0.5)cm}]
    \draw[ultra thick, Maroon] (0,1.25)--(1.375,1.25);
    \draw[ultra thick, Maroon] (0,0)--(1.375,1.25) node[at end, black] {\Huge$\bullet$};
    \draw[ultra thick, Maroon] (0,0)--(1.375,0) node[at end, black] {\Huge$\bullet$};
    \draw[ultra thick, Maroon] (0,0)--(0,1.25) node[at end, black] {\Huge$\bullet$} node[at start, black] {\Huge$\bullet$};
\end{tikzpicture}}_{\displaystyle \widetilde{\Omega}_4^\text{1-cycle}}\nonumber\\
&\ \underbrace{-\begin{tikzpicture}[baseline={(0,0.5)cm}]
    \draw[ultra thick, Maroon] (0,1.25)--(1.375,1.25);
    \draw[ultra thick, Maroon] (1.375,0)--(1.375,1.25);
    \draw[ultra thick, Maroon] (0,0)--(1.375,1.25) node[at end, black] {\Huge$\bullet$};
    \draw[ultra thick, Maroon] (0,0)--(1.375,0) node[at end, black] {\Huge$\bullet$};
    \draw[ultra thick, Maroon] (0,0)--(0,1.25) node[at end, black] {\Huge$\bullet$} node[at start, black] {\Huge$\bullet$};
\end{tikzpicture}}_{\displaystyle \widetilde{\Omega}_4^\text{2-cycle}}\ + \ \underbrace{\begin{tikzpicture}[baseline={(0,0.5)cm}]
    \draw[ultra thick, Maroon] (0,1.25)--(1.375,1.25);
    \draw[ultra thick, Maroon] (1.375,0)--(1.375,1.25);
    \draw[ultra thick, Maroon] (0,1.25)--(1.375,0);
    \draw[ultra thick, Maroon] (0,0)--(1.375,1.25) node[at end, black] {\Huge$\bullet$};
    \draw[ultra thick, Maroon] (0,0)--(1.375,0) node[at end, black] {\Huge$\bullet$};
    \draw[ultra thick, Maroon] (0,0)--(0,1.25) node[at end, black] {\Huge$\bullet$} node[at start, black] {\Huge$\bullet$};
\end{tikzpicture}}_{\displaystyle \widetilde{\Omega}_4^\text{3-cycle}}\,
\end{align}

This organization of the sum over all graphs $G$ turns out to be very useful. In particular, we can solve for the canonical forms for all tree graphs (${\cal C}=0$) and one-cycle graphs (${\cal C}=1$) to any loop order $L$. As it turns out, the number of vertices $L$ is not the main problem when calculating a form for a given graph, the complexity is tied more to the number of cycles ${\cal C}$. 

In the quest to find the canonical form for a given negative geometry represented by a $L$-loop graph $G$, we can always start with a general structure
\begin{equation}
    \Omega_G = \frac{d\mu\,{\cal N}_G}{D_1D_2{\dots}D_L \prod_{i,j}\la AB_i AB_j\ra}
\end{equation}
where $d\mu$ is the integration measure over $L$ lines $AB_i$ in momentum-twistor space,
\begin{equation}
d\mu = \prod_{i=1}^L \la AB_id^2A_i\ra\la AB_id^2B_i\ra
\end{equation}
The denominator can always be read off directly from the graph: $D_k$ is the product of all external propagators for a loop $AB_k$, 
\begin{equation}
    D_k = \la AB_k12\ra\la AB_k23\ra\la AB_k34\ra\la AB_k14\ra \,.
\end{equation}
The denominator also contains a product of all mutual propagators $\la AB_iAB_j\ra$ for all edges $(i,j)$ in the graph. Hence the only unknown part of the canonical form is the numerator ${\cal N}_G$ which can be very complex for a general graph. In fact, for the complete graph it is as complicated as the numerator for the whole amplitude.

Our method to determine the canonical forms is a hybrid of a generalized unitarity approach (cuts), together with imposing constraints from geometric inequalities. In particular, we start with an ansatz for the numerator in terms of a linear combination of building blocks that individually satisfy certain constraints. Then we apply other constraints to determine the coefficients, very much like in generalized unitarity. However, because we are not constructing the integrand for an amplitude, but the integrands for negative geometries, the physical constraints on the amplitude are replaced here by constraints from negative geometry inequalities. There are three types of constraints on the numerator:
\begin{enumerate}
    \item No double poles are allowed for the form when taking subsequent residues.
    \item The form must vanish at special kinematical points when the negative geometry inequalities are violated.
    \item The form must reduce to a simpler (known) negative-geometry form on certain residues.
\end{enumerate}
It turns out that these conditions suffice to fix uniquely the canonical form for cases that are understood: tree graphs, one-cycle graphs, and the new results obtained in this paper. We conjecture that the uniqueness holds for any negative geometry, and that we never need to consider any other data to fix the numerator of the form uniquely. The main bottleneck is to enumerate enough of the constraints (of the three types we listed) to fix the numerator -- the list that suffices for a generic negative geometry is not clear {\it a priori}. 

It is worth noting that the cleanest method to calculate the canonical forms would be to triangulate the associated geometry. However, this is a difficult task (at least using current technology) and the method outlined above is the most efficient way to get the result.

\subsection{Canonical forms for all tree and one-cycle graphs}

Here we will review the results of \cite{Arkani-Hamed:2021iya,Brown:2023mqi} and show the closed formulas for the  canonical forms of all tree graphs and one-cycles. The simplest $L=1$ geometry just gives a one-loop four-point integrand form,
\begin{equation}
    \begin{tikzpicture}[baseline=-0.6ex, scale=1.]
      \node () at (0,0) {\Huge$\bullet$};
  \end{tikzpicture}  \ = \frac{\la 1234\ra^2}{D_1} \,.
\end{equation}

For the only $L=2$ negative geometry we reproduce the integrand of the two-loop logarithm $\widetilde{\cal I}_4^{(2)}$ written in a compact notation,
\begin{equation}
    \begin{tikzpicture}[baseline={(0,0)cm}, scale=1.]
      \draw[ultra thick, Maroon] (0,0)--(1,0) node[at end, black] {\Huge$\bullet$} node[at start, black] {\Huge$\bullet$};
  \end{tikzpicture}  \ = \frac{\la 1234\ra^3\,n_{12}}{D_1D_2\la AB_1AB_2\ra}\,,
\end{equation}
where we denote
\begin{equation}
    n_{ij} = \la AB_i13\ra\la AB_j24\ra + \la AB_j24\ra\la AB_i13\ra \,.
\end{equation}
At $L=3$ we have a tree graph and a one-cycle graph. The tree graph has a surprisingly simple numerator,
\begin{equation}
   \begin{tikzpicture}[baseline={(0,0)cm}, scale=1.]
      \draw[ultra thick, Maroon] (0,0)--(2,0) node[at end, black] {\Huge$\bullet$} node[at start, black] {\Huge$\bullet$} node[pos=0.5, black] {\Huge$\bullet$};
  \end{tikzpicture}   \ = \frac{\la 1234\ra^4\,n_{12}n_{23}}  {D_1D_2D_3\la AB_1AB_2\ra\la AB_2AB_3\ra}  \ .
\end{equation}
Here we see the factorization property of the numerator as the product of two $L=2$ numerators $n_{ij}$, one for each link. This rule generalizes to an arbitrary complicated tree graph, 
\begin{equation}
    \begin{tikzpicture}[baseline={(0,0.5)cm}, scale=1.]
        \draw[ultra thick, Maroon] (1,0)--(1.75,-0.5) node[at end, black] {\Huge$\bullet$};
        \draw[ultra thick, Maroon] (1,0)--(1.5,0.75);
        \draw[ultra thick, Maroon] (1.5,0.75)--(2.25,1) node[at end, black] {\Huge$\bullet$};
        \draw[ultra thick, Maroon] (1.5,0.75)--(0.75,1) node[at end, black] {\Huge$\bullet$} node[at start, black] {\Huge$\bullet$};
        \draw[ultra thick, Maroon] (0,0)--(3,0) node[at end, black] {\Huge$\bullet$} node[at start, black] {\Huge$\bullet$} node[pos=0.33, black] {\Huge$\bullet$} node[pos=0.66, black] {\Huge$\bullet$};
    \end{tikzpicture}\ =\ \frac{\ang{1234}^{L+1}\prod_{i,j}n_{ij}}{D_1D_2\cdots D_L\prod_{i,j}\ang{AB_iAB_j}}\,,
\end{equation}
where the products are over all links $(i,j)$ in the graph. 

To tackle the problem of the canonical forms for one-cycle geometries we first consider a core $L$-gon diagram,
\begin{equation}
    \begin{tikzpicture}[scale=1.]
        \draw[ultra thick, Maroon] (0,0)--(-0.75,1);
        \draw[ultra thick, Maroon] (0,2)--(-0.75,1) node [at end, black] {\Huge$\bullet$};
        \draw[ultra thick, Maroon] (0,2)--(1.25,2) node [at start, black] {\Huge$\bullet$};
        \draw[ultra thick, Maroon] (1.25,2)--(2,1) node [at start, black] {\Huge$\bullet$} node[pos=0.5, sloped, above, black] {$\cdots$};
        \draw[ultra thick, Maroon] (1.25,0)--(2,1) node [at end, black] {\Huge$\bullet$};
        \draw[ultra thick, Maroon] (0,0)--(1.25,0) node[at end, black] {\Huge$\bullet$} node[at start, black] {\Huge$\bullet$};
        \node[below] at (0,-0.1) {$AB_2$};
        \node[below] at (1.25,-0.1) {$AB_1$}; 
        \node[right] at (2.1,1) {$AB_L$}; 
        \node[above] at (1.25,2.1) {$AB_5$};
        \node[above] at (0,2.1) {$AB_4$}; 
        \node[left] at (-0.76,1) {$AB_3$};  
    \end{tikzpicture}
\end{equation}
Based on the result for tree graphs, we can write a ``naive'' numerator,
\begin{equation}
    {\cal N}_{\rm naive} = \prod_{i,j\in {\rm core}}n_{ij} \,.
\end{equation}
While this numerator satisfies many of the constraints from the negative geometry list, it fails to preserve the logarithmicity of the form -- in other words, there are double poles in the cut structure of the corresponding form. This is not an issue for the tree graphs but it appears for one-cycle graphs (and beyond). Hence the correct numerator has an additional compensating term,
\begin{equation}
    {\cal N}_{\rm core} = {\cal N}_{\rm naive} + {\cal R}_G \,,
\end{equation}
where ${\cal R}_G$ is a `remainder' which is very constrained (for example, it vanishes on many of the cuts identically), schematically
\begin{equation}
    \begin{tikzpicture}
        \draw[ultra thick, Maroon, dashed] (0,0)--(-0.75,1);
        \draw[ultra thick, Maroon, dashed] (0,2)--(-0.75,1) node [at end, black] {\Huge$\bullet$};
        \draw[ultra thick, Maroon, dashed] (0,2)--(1.25,2) node [at start, black] {\Huge$\bullet$};
        \draw[ultra thick, Maroon, dashed] (1.25,2)--(2,1) node [at start, black] {\Huge$\bullet$} node[pos=0.5, sloped, above, black] {$\cdots$};
        \draw[ultra thick, Maroon, dashed] (1.25,0)--(2,1) node [at end, black] {\Huge$\bullet$};
        \draw[ultra thick, Maroon, dashed] (0,0)--(1.25,0) node[at end, black] {\Huge$\bullet$} node[at start, black] {\Huge$\bullet$};
        \node[below] at (0,-0.1) {$AB_2$};
        \node[below] at (1.25,-0.1) {$AB_1$}; 
        \node[right] at (2.1,1) {$AB_L$}; 
        \node at (3.5,1) {$+$};
        \node[above] at (1.25,2.1) {$AB_5$};
        \node[above] at (0,2.1) {$AB_4$}; 
        \node[left] at (-0.86,1) {$AB_3$};  
    \end{tikzpicture}
    \begin{tikzpicture}
        \draw[chalkdust] (0,0)--(-0.75,1);
        \draw[chalkdust] (0,2)--(-0.75,1) node [at end, black] {\Huge$\bullet$};
        \draw[chalkdust] (0,2)--(1.25,2) node [at start, black] {\Huge$\bullet$};
        \draw[chalkdust] (1.25,2)--(2,1) node [at start, black] {\Huge$\bullet$} node[pos=0.5, sloped, above, black] {$\cdots$};
        \draw[chalkdust] (1.25,0)--(2,1) node [at end, black] {\Huge$\bullet$};
        \draw[chalkdust] (0,0)--(1.25,0) node[at end, black] {\Huge$\bullet$} node[at start, black] {\Huge$\bullet$};
        \node[below] at (0,-0.1) {$AB_2$};
        \node[below] at (1.25,-0.1) {$AB_1$}; 
        \node[right] at (2.1,1) {$AB_L$}; 
        \node[above] at (1.25,2.1) {$AB_5$};
        \node[above] at (0,2.1) {$AB_4$}; 
        \node[left] at (-0.86,1) {$AB_3$};  
    \end{tikzpicture}
\end{equation}
The construction of this remainder was provided in ref.~\cite{Brown:2023mqi}. Once we calculate this core diagram, it is simple to determine the numerator for any one-cycle graph: it is given by a simple multiplication by $n_{ij}$ for each edge in the branches that extend from the core,
\begin{equation}
    {\cal N}= {\cal N}_{\rm core} \times \prod_{i,j\in{\rm branches}} n_{ij} \,.\label{general}
\end{equation}
where the product is over all edges in the branches. As an example, let us consider $L=12$,
%
%
\begin{equation}
    \begin{tikzpicture}[scale=1.]
        \draw[ultra thick, Maroon] (0,0)--(1,1);
        \draw[ultra thick, Maroon] (0,0)--(-1,1) node[at start, black] {\Huge$\bullet$};
        \draw[ultra thick, Maroon] (1,1)--(1,2) node[at start, black] {\Huge$\bullet$};
        \draw[ultra thick, Maroon] (-1,1)--(-1,2);
        \draw[ultra thick, Maroon] (-1,2)--(0,3);
        \draw[ultra thick, Maroon] (1,2)--(0,3) node[at end, black] {\Huge$\bullet$};
        \draw[ultra thick, Maroon] (-1,2)--(-3,2) node[at end, black] {\Huge$\bullet$} node[at start, black] {\Huge$\bullet$} node[pos=0.5, black] {\Huge$\bullet$};
        \draw[ultra thick, Maroon] (-1,1)--(-2,1) node[at end, black] {\Huge$\bullet$} node[at start, black] {\Huge$\bullet$};
        \draw[ultra thick, Maroon] (1,2)--(2,2) node[at start, black] {\Huge$\bullet$};
        \draw[ultra thick, Maroon] (2,2)--(2.75,2.75) node[at end, black] {\Huge$\bullet$};
        \draw[ultra thick, Maroon] (2,2)--(2.75,1.25) node[at end, black] {\Huge$\bullet$} node[at start, black] {\Huge$\bullet$};
        \node[below] at (0,-0.2) {$AB_{1}$};
        \node[below] at (1,0.8) {$AB_{2}$};
        \node[below] at (2.85,1.10) {$AB_{12}$};
        \node[below] at (-1,0.8) {$AB_{6}$};
        \node[below] at (-2,0.8) {$AB_{7}$};
        \node[above] at (0,3.2) {$AB_{4}$};
        \node[above] at (1,2.25) {$AB_{3}$};
        \node[above] at (1.9,2.25) {$AB_{10}$};
        \node[above] at (-1,2.25) {$AB_{5}$};
        \node[above] at (-2,2.25) {$AB_{8}$};
        \node[above] at (-3,2.25) {$AB_{9}$};
        \node[above] at (2.85,2.90) {$AB_{11}$};
    \end{tikzpicture}
\end{equation}
The numerator is given by
\begin{align}
{\cal N}_G &= n_{67}\,n_{58}\,n_{89}\,n_{3,10}\,n_{10,11}\,n_{10,12} \times \bigg\{n_{12}\,n_{23}\,n_{34}\,n_{45}\,n_{56}\,n_{61} + R^\text{1-cycle}_{123456}\bigg\}\,,
\end{align}
where $R^\text{1-cycle}_{123456}$ is given in eq.~(5.39) of ref.~\cite{Brown:2023mqi}. This strategy works also for higher cycles: For an arbitrary graph, we can write the numerator of the canonical form as the product between the core numerator and the product of $n_{ij}$ for all edges on the tree branches (\ref{general}). Hence the problem reduces to the calculation of the numerator of the core part of the diagram.

We do not yet have a formula for the integrand of a general 2-cycle or higher-cycle diagram. However, in the next section we will focus on the specific case of $L=4$ and determine all contributions, including 2-cycle and 3-cycle terms.

\subsection{All four-loop negative geometries}

The list of all contributing $L=4$ negative geometries is 

\begin{align}
\widetilde{\Omega}_4^{(4)}=&\ \underbrace{-\begin{tikzpicture}[baseline={(0,0.5)cm}]
    \draw[ultra thick, Maroon] (0,1.25)--(1.375,1.25) node[at end, black] {\Huge$\bullet$};
    \draw[ultra thick, Maroon] (0,0)--(1.375,0) node[at end, black] {\Huge$\bullet$};
    \draw[ultra thick, Maroon] (0,0)--(0,1.25) node[at end, black] {\Huge$\bullet$} node[at start, black] {\Huge$\bullet$};
\end{tikzpicture}\ -\ \begin{tikzpicture}[baseline={(0,0.5)cm}]
    \draw[ultra thick, Maroon] (0,0)--(1.375,1.25) node[at end, black] {\Huge$\bullet$};
    \draw[ultra thick, Maroon] (0,0)--(1.375,0) node[at end, black] {\Huge$\bullet$};
    \draw[ultra thick, Maroon] (0,0)--(0,1.25) node[at end, black] {\Huge$\bullet$} node[at start, black] {\Huge$\bullet$};
\end{tikzpicture}}_{ \displaystyle \widetilde{\Omega}_4^{\rm tree}}\ +\ \underbrace{\begin{tikzpicture}[baseline={(0,0.5)cm}]
    \draw[ultra thick, Maroon] (1.375,0)--(1.375,1.25);
    \draw[ultra thick, Maroon] (0,1.25)--(1.375,1.25) node[at end, black] {\Huge$\bullet$};
    \draw[ultra thick, Maroon] (0,0)--(1.375,0) node[at end, black] {\Huge$\bullet$};
    \draw[ultra thick, Maroon] (0,0)--(0,1.25) node[at end, black] {\Huge$\bullet$} node[at start, black] {\Huge$\bullet$};
\end{tikzpicture}\ +\ \begin{tikzpicture}[baseline={(0,0.5)cm}]
    \draw[ultra thick, Maroon] (0,1.25)--(1.375,1.25);
    \draw[ultra thick, Maroon] (0,0)--(1.375,1.25) node[at end, black] {\Huge$\bullet$};
    \draw[ultra thick, Maroon] (0,0)--(1.375,0) node[at end, black] {\Huge$\bullet$};
    \draw[ultra thick, Maroon] (0,0)--(0,1.25) node[at end, black] {\Huge$\bullet$} node[at start, black] {\Huge$\bullet$};
\end{tikzpicture}}_{\displaystyle \widetilde{\Omega}_4^\text{1-cycle}}\nonumber\\
&\ \underbrace{-\begin{tikzpicture}[baseline={(0,0.5)cm}]
    \draw[ultra thick, Maroon] (0,1.25)--(1.375,1.25);
    \draw[ultra thick, Maroon] (1.375,0)--(1.375,1.25);
    \draw[ultra thick, Maroon] (0,0)--(1.375,1.25) node[at end, black] {\Huge$\bullet$};
    \draw[ultra thick, Maroon] (0,0)--(1.375,0) node[at end, black] {\Huge$\bullet$};
    \draw[ultra thick, Maroon] (0,0)--(0,1.25) node[at end, black] {\Huge$\bullet$} node[at start, black] {\Huge$\bullet$};
\end{tikzpicture}}_{\displaystyle \widetilde{\Omega}_4^\text{2-cycle}}\ + \ \underbrace{\begin{tikzpicture}[baseline={(0,0.5)cm}]
    \draw[ultra thick, Maroon] (0,1.25)--(1.375,1.25);
    \draw[ultra thick, Maroon] (1.375,0)--(1.375,1.25);
    \draw[ultra thick, Maroon] (0,1.25)--(1.375,0);
    \draw[ultra thick, Maroon] (0,0)--(1.375,1.25) node[at end, black] {\Huge$\bullet$};
    \draw[ultra thick, Maroon] (0,0)--(1.375,0) node[at end, black] {\Huge$\bullet$};
    \draw[ultra thick, Maroon] (0,0)--(0,1.25) node[at end, black] {\Huge$\bullet$} node[at start, black] {\Huge$\bullet$};
\end{tikzpicture}}_{\displaystyle \widetilde{\Omega}_4^\text{3-cycle}}\,
\end{align}

The zero-cycle and one-cycle contributions can be found using the methods from previous section and we will review the results in appendix \ref{app:3-cycle}. A strategy for the computation of canonical forms for higher cycle geometries was proposed in section 6 of \cite{Brown:2023mqi}. The 2-cycle contribution $\widetilde{\Omega}_4^{\rm 2-cycle}$ only involves one negative geometry. The numerator decomposition takes the form
\begin{align}
{\cal N}_G = &\ n_{12}n_{23}n_{34}n_{14}n_{13}  \nonumber\\
&\hspace{1.5cm}- R_{123}^\text{1-cycle} n_{34}n_{14} - R_{134}^\text{1-cycle} n_{12} n_{23}+ R_{1234}^\text{1-cycle} n_{13} - R_{1234,13}^\text{2-cycle} \,,\label{twoloopdec}
\end{align}
or graphically, 
%
\begin{equation}
\label{eq:cycledecomp4}
    \mathcal{N}_G =\ \begin{tikzpicture}[baseline={(0,0.5cm)}]
        \draw[Maroon,ultra thick, dashed] (0,1.25)--(1.375,0);
        \draw[Maroon,ultra thick, dashed] (0,0)--(1.375,0);
        \draw[Maroon,ultra thick, dashed] (0,0)--(0,1.25) node[at start, black] {\Huge$\bullet$};
        \draw[Maroon,ultra thick, dashed] (0,1.25)--(1.375,1.25) node[at start, black] {\Huge$\bullet$};
        \draw[Maroon,ultra thick, dashed] (1.375,1.25)--(1.375,0) node[at end, black] {\Huge$\bullet$} node[at start, black] {\Huge$\bullet$};
        \node[below] at (0,-0.1) {$AB_2$};
        \node[below] at (1.375,-0.1) {$AB_1$};
        \node[above] at (0,1.35) {$AB_3$};
        \node[above] at (1.375,1.35) {$AB_4$};
    \end{tikzpicture}\ -\ \begin{tikzpicture}[baseline={(0,0.5cm)}]
        \draw[chalkdust] (0,1.25)--(1.375,0);
        \draw[chalkdust] (0,0)--(1.375,0);
        \draw[chalkdust] (0,0)--(0,1.25) node[at start, black] {\Huge$\bullet$};
        \draw[Maroon,ultra thick, dashed] (0,1.25)--(1.375,1.25) node[at start, black] {\Huge$\bullet$};
        \draw[Maroon,ultra thick, dashed] (1.375,1.25)--(1.375,0) node[at end, black] {\Huge$\bullet$} node[at start, black] {\Huge$\bullet$};
        \node[below] at (0,-0.1) {$AB_2$};
        \node[below] at (1.25,-0.1) {$AB_1$};
        \node[above] at (0,1.35) {$AB_3$};
        \node[above] at (1.25,1.35) {$AB_4$};
    \end{tikzpicture}\ -\ \begin{tikzpicture}[baseline={(0,0.5cm)}]
        \draw[chalkdust] (0,1.25)--(1.375,0);
        \draw[Maroon,ultra thick, dashed] (0,0)--(1.375,0);
        \draw[Maroon,ultra thick, dashed] (0,0)--(0,1.25) node[at start, black] {\Huge$\bullet$};
        \draw[chalkdust] (0,1.25)--(1.375,1.25) node[at start, black] {\Huge$\bullet$};
        \draw[chalkdust] (1.375,1.25)--(1.375,0) node[at end, black] {\Huge$\bullet$} node[at start, black] {\Huge$\bullet$};
        \node[below] at (0,-0.1) {$AB_2$};
        \node[below] at (1.25,-0.1) {$AB_1$};
        \node[above] at (0,1.35) {$AB_3$};
        \node[above] at (1.25,1.35) {$AB_4$};
    \end{tikzpicture}\ +\ \begin{tikzpicture}[baseline={(0,0.5cm)}]
        \draw[Maroon,ultra thick, dashed] (0,1.25)--(1.375,0);
        \draw[chalkdust] (0,0)--(1.375,0);
        \draw[chalkdust] (0,0)--(0,1.25) node[at start, black] {\Huge$\bullet$};
        \draw[chalkdust] (0,1.25)--(1.375,1.25) node[at start, black] {\Huge$\bullet$};
        \draw[chalkdust] (1.375,1.25)--(1.375,0) node[at end, black] {\Huge$\bullet$} node[at start, black] {\Huge$\bullet$};
        \node[below] at (0,-0.1) {$AB_2$};
        \node[below] at (1.25,-0.1) {$AB_1$};
        \node[above] at (0,1.35) {$AB_3$};
        \node[above] at (1.25,1.35) {$AB_4$};
    \end{tikzpicture}\ -\ \begin{tikzpicture}[baseline={(0,0.5cm)}]
        \draw[chalkdust] (0,1.25)--(1.375,0);
        \draw[chalkdust] (0,0)--(1.375,0);
        \draw[chalkdust] (0,0)--(0,1.25) node[at start, black] {\Huge$\bullet$};
        \draw[chalkdust] (0,1.25)--(1.375,1.25) node[at start, black] {\Huge$\bullet$};
        \draw[chalkdust] (1.375,1.25)--(1.375,0) node[at end, black] {\Huge$\bullet$} node[at start, black] {\Huge$\bullet$};
        \node[below] at (0,-0.1) {$AB_2$};
        \node[below] at (1.25,-0.1) {$AB_1$};
        \node[above] at (0,1.35) {$AB_3$};
        \node[above] at (1.25,1.35) {$AB_4$};
    \end{tikzpicture}\,.
\end{equation}
The general rule is to consider all possible core subgraphs, shown with solid lines, with the remaining (dashed) edges denoting a trivial $n_{ij}$ numerator. The sign of each term in eq.~(\ref{twoloopdec}) is determined by the number of solid lines in the graph. In addition to one-cycle irreducible numerators, there is a new two-cycle term $R^\text{2-cycle}_{1234,13}$, which satisfies the same condition that it vanishes on any cut that fixes one or more signs of $\la AB_iAB_j\ra$, $\{i,j\}\in\{\{1,2\},\{2,3\},\{3,4\},\{1,4\},\{2,4\}\}$. The irreducible numerator $R^\text{2-cycle}_{1234,13}$ was explicitly constructed in ref.~\cite{Brown:2023mqi}; see appendix B of that paper. We will review the result in appendix \ref{app:3-cycle} of this paper too.

Finally, we construct the numerator for the $L=4$ three-cycle geometry.  We use the same decomposition outlined in ref.~\cite{Brown:2023mqi} (where the graphs were drawn a little differently), 
\begin{align}
    \mathcal{N}_G=&\,\begin{tikzpicture}[baseline={(0,0.75)cm}]
        \draw[Maroon, ultra thick, dashed] (0,0)--(1.75,0);
        \draw[Maroon, ultra thick, dashed] (0,0)--(0.87,1.5);
        \draw[Maroon, ultra thick, dashed] (0.87,1.5)--(1.75,0);
        \draw[Maroon, ultra thick, dashed] (0,0)--(0.87,0.65) node[at start, black] {\Huge$\bullet$};
        \draw[Maroon, ultra thick, dashed] (1.75,0)--(0.87,0.65) node[at start, black] {\Huge$\bullet$};
        \draw[Maroon, ultra thick, dashed] (0.87,1.5)--(0.87,0.65) node[at start, black] {\Huge$\bullet$} node[at end, black] {\Huge$\bullet$};
    \node[below left] at (0,0) {$AB_1$};
    \node[below right] at (1.75,0) {$AB_4$};
    \node[above] at (0.87,1.6) {$AB_2$};
    \node[below] at (0.87,0.55) {$AB_3$};
    \end{tikzpicture}\,-\sum_{\pi_1}\,\begin{tikzpicture}[baseline={(0,0.75)cm}]
        \draw[chalkdust] (0,0)--(1.75,0);
        \draw[chalkdust] (0,0)--(0.87,1.5);
        \draw[chalkdust] (0.87,1.5)--(1.75,0);
        \draw[Maroon, ultra thick, dashed] (0,0)--(0.87,0.65) node[at start, black] {\Huge$\bullet$};
        \draw[Maroon, ultra thick, dashed] (1.75,0)--(0.87,0.65) node[at start, black] {\Huge$\bullet$};
        \draw[Maroon, ultra thick, dashed] (0.87,1.5)--(0.87,0.65) node[at start, black] {\Huge$\bullet$} node[at end, black] {\Huge$\bullet$};
    \node[below left] at (0,0) {$AB_1$};
    \node[below right] at (1.75,0) {$AB_4$};
    \node[above] at (0.87,1.6) {$AB_2$};
    \node[below] at (0.87,0.55) {$AB_3$};
    \end{tikzpicture}\,+\sum_{\pi_2}\,\begin{tikzpicture}[baseline={(0,0.75)cm}]
        \draw[chalkdust] (0,0)--(1.75,0);
        \draw[chalkdust] (0,0)--(0.87,1.5);
        \draw[Maroon, ultra thick, dashed] (0.87,1.5)--(1.75,0);
        \draw[Maroon, ultra thick, dashed] (0,0)--(0.87,0.65) node[at start, black] {\Huge$\bullet$};
        \draw[chalkdust] (1.75,0)--(0.87,0.65) node[at start, black] {\Huge$\bullet$};
        \draw[chalkdust] (0.87,1.5)--(0.87,0.65) node[at start, black] {\Huge$\bullet$} node[at end, black] {\Huge$\bullet$};
    \node[below left] at (0,0) {$AB_1$};
    \node[below right] at (1.75,0) {$AB_4$};
    \node[above] at (0.87,1.6) {$AB_2$};
    \node[below] at (0.87,0.55) {$AB_3$};
    \end{tikzpicture}\nonumber\\
    &-\sum_{\pi_3}\,\begin{tikzpicture}[baseline={(0,0.75)cm}]
        \draw[chalkdust] (0,0)--(1.75,0);
        \draw[chalkdust] (0,0)--(0.87,1.5);
        \draw[chalkdust] (0.87,1.5)--(1.75,0);
        \draw[Maroon, ultra thick, dashed] (0,0)--(0.87,0.65) node[at start, black] {\Huge$\bullet$};
        \draw[chalkdust] (1.75,0)--(0.87,0.65) node[at start, black] {\Huge$\bullet$};
        \draw[chalkdust] (0.87,1.5)--(0.87,0.65) node[at start, black] {\Huge$\bullet$} node[at end, black] {\Huge$\bullet$};
    \node[below left] at (0,0) {$AB_1$};
    \node[below right] at (1.75,0) {$AB_4$};
    \node[above] at (0.87,1.6) {$AB_2$};
    \node[below] at (0.87,0.55) {$AB_3$};
    \end{tikzpicture}\,+\,\begin{tikzpicture}[baseline={(0,0.75)cm}]
        \draw[chalkdust] (0,0)--(1.75,0);
        \draw[chalkdust] (0,0)--(0.87,1.5);
        \draw[chalkdust] (0.87,1.5)--(1.75,0);
        \draw[chalkdust] (0,0)--(0.87,0.65) node[at start, black] {\Huge$\bullet$};
        \draw[chalkdust] (1.75,0)--(0.87,0.65) node[at start, black] {\Huge$\bullet$};
        \draw[chalkdust] (0.87,1.5)--(0.87,0.65) node[at start, black] {\Huge$\bullet$} node[at end, black] {\Huge$\bullet$};
    \node[below left] at (0,0) {$AB_1$};
    \node[below right] at (1.75,0) {$AB_4$};
    \node[above] at (0.87,1.6) {$AB_2$};
    \node[below] at (0.87,0.55) {$AB_3$};
    \end{tikzpicture}\,.
\end{align}
The $\pi_i$ label permutations that describe different embeddings of a given topology.  There are 4 elements of $\pi_1$, corresponding to four different $AB_i$ being at the center of that graph; 3 elements of $\pi_2$; and 6 elements of $\pi_3$. The corresponding decomposition of the graph numerator is 
\begin{align}
\mathcal{N}_G &= n_{12}n_{23}n_{34}n_{14}n_{24}n_{13}  - \sum_{\pi_1} R_{124}^\text{1-cycle} n_{13} n_{23} n_{34} + \sum_{\pi_2} R_{1234}^\text{1-cycle} n_{13}n_{24}\nonumber\\ 
&\hspace{0.3cm} - \sum_{\pi_3} R_{1234,24}^\text{2-cycle}n_{13} + R_{1234}^\text{3-cycle}\,,
\end{align}
with only one new object, the 3-cycle term $R_{1234}^\text{3-cycle}$. Using the generalized unitarity approach together with the geometric conditions (the same strategy as used in ref.~\cite{Brown:2023mqi}) we can construct this object explicitly. See appendix \ref{app:3-cycle} for an explicit construction. As a result, we have the canonical forms for all $L=4$ negative geometries.

\section{Integrated negative geometries}
\label{sec:integrated_neg_geom}

Starting from the $L{+}1$ loop integrand for the amplitude logarithm, $\widetilde{\cal I}_4^{(L+1)}$, we can now integrate over $L$ of the loop momenta $AB_i$, $i=2,3,\ldots,L+1$ and get ${\cal F}^{(L)}(z)$, which also can be interpreted as the expectation value of the Wilson loop with a Lagrangian insertion, as discussed in the previous section. This function is IR finite and has a uniform (and maximal) transcendentality $2L$. The same integration operation can be performed for each negative geometry $G$ separately. We get an IR finite function ${\cal F}_G(z)$ of the same uniform maximal transcendentality. If we sum over all functions ${\cal F}_G(z)$ for all graphs in the negative geometry expansion of the integrand at $L+1$ loops, we get ${\cal F}^{(L)}(z)$ as expected.

To distinguish between the canonical form for a graph $\Omega_G$ and the transcendental function ${\cal F}_G(z)$, we use a modified graphical notation for ${\cal F}_G(z)$ where the frozen loop $AB_1\equiv AB$ is denoted by the cross circle (marked point),
while all other vertices correspond to loops over which we integrate. For most of the graphs, there are multiple ways to freeze the loop, and hence one negative geometry generates several integrated negative geometries.
At $L=1$, we get only one term,
\vspace{0.5cm}
\begin{equation}
    \begin{tikzpicture}[baseline=-0.6ex,scale=1.,transform shape]
        \draw[ultra thick, Maroon] (0,0)--(1,0) node[at end, black] {\Huge$\bullet$} node[at start, black] {\Huge$\bullet$};
    \end{tikzpicture}\ \rightarrow\ \begin{tikzpicture}[baseline=-0.6ex,scale=1.,transform shape]
        \draw[ultra thick, Maroon] (0,0)--(1,0) node[at end, black] {\Huge$\bullet$};
        \filldraw[white] (0,0) circle (5pt) node[]{};
        \node[black] at (0,0) {\Large$\otimes$};
    \end{tikzpicture}
\end{equation}

At $L=2$, we have two terms: the tree graph produces two different integrated geometries and the loop graph produces one integrated geometry,
\begin{align}
    \begin{tikzpicture}[baseline=-0.6ex,scale=1.,transform shape]
  \draw[Maroon, ultra thick] (1,0)--(3,0) node[pos=0.5,black]{\Huge$\bullet$} node[at end,black]{\Huge$\bullet$};
  \node[black] at (1,0) {\Huge$
  \bullet$};
\end{tikzpicture}&\ \to\ \begin{tikzpicture}[baseline=-0.6ex,scale=1.,transform shape]
  \draw[Maroon, ultra thick] (1,0)--(3,0) node[pos=0.5,black]{\Huge$\bullet$} node[at end,black]{\Huge$\bullet$};
   \filldraw[white] (1,0) circle (5pt) node[]{};
  \node[black] at (1,0) {\Large$
  \otimes$};
\end{tikzpicture}+ \frac{1}{2!}\begin{tikzpicture}[baseline=-0.6ex,scale=1.,transform shape]
  \draw[Maroon, ultra thick] (1,0)--(2,0.6) node[at end,black]{\Huge$\bullet$};
  \draw[Maroon, ultra thick] (1,0)--(2,-0.6) node[at end,black]{\Huge$\bullet$};
   \filldraw[white] (1,0) circle (5pt) node[]{};
  \node[black] at (1,0) {\Large$\otimes$};
\end{tikzpicture}\nonumber\\
\begin{tikzpicture}
[baseline=-0.6ex,scale=1.,transform shape]
    \draw[Maroon, ultra thick] (0,-0.5)--(0,0.5);
    \draw[Maroon, ultra thick](0,-0.5)--(1,0) node[at start,black] {\Huge$\bullet$}; 
    \draw[Maroon, ultra thick](0,0.5)--(1,0) node[at start,black] {\Huge$\bullet$};
    \node[black] at (1,0) {\Huge$\bullet$};
\end{tikzpicture}&\ \to\ \frac{1}{2!}\begin{tikzpicture}
[baseline=-0.6ex,scale=1.,transform shape]
    \draw[Maroon, ultra thick] (0,-0.5)--(0,0.5);
    \draw[Maroon, ultra thick](0,-0.5)--(1,0) node[at start,black] {\Huge$\bullet$}; 
    \draw[Maroon, ultra thick](0,0.5)--(1,0) node[at start,black] {\Huge$\bullet$};
     \filldraw[white] (1,0) circle (5pt) node[]{};
    \node[black] at (1,0) {\Large$\otimes$};
\end{tikzpicture}
\end{align}
There are also symmetry factors involved which are given by the number of repetitions we get when marking the point, divided by $L!$ from the symmetrization of integrated loops. 
Such an analysis is easy to do for any negative geometry. The integration over all unfrozen loop momenta generally gives a rational function (leading singularity) multiplied by a transcendental function,
%
%
%
\begin{equation}
    \begin{tikzpicture}[baseline=8ex,scale=0.9]
        \draw[ultra thick, Maroon] (0,0)--(1,1);
        \draw[ultra thick, Maroon] (0,0)--(-1,1) node[at start, black] {\Huge$\bullet$};
        \draw[ultra thick, Maroon] (1,1)--(1,2) node[at start, black] {\Huge$\bullet$};
        \draw[ultra thick, Maroon] (-1,1)--(-1,2);
        \draw[ultra thick, Maroon] (-1,2)--(0,3);
        \draw[ultra thick, Maroon] (1,2)--(0,3) node[at end, black] {\Huge$\bullet$};
        \draw[ultra thick, Maroon] (-1,2)--(-3,2) node[at end, black] {\Huge$\bullet$} node[at start, black] {\Huge$\bullet$} node[pos=0.5, black] {\Huge$\bullet$};
        \draw[ultra thick, Maroon] (-1,1)--(-2,1) node[at end, black] {\Huge$\bullet$} node[at start, black] {\Huge$\bullet$};
        \draw[ultra thick, Maroon] (1,2)--(2,2) node[at start, white] {\Huge$\bullet$};
        \draw[ultra thick, Maroon] (2,2)--(2.75,2.75) node[at end, black] {\Huge$\bullet$};
        \draw[ultra thick, Maroon] (2,2)--(2.75,1.25) node[at end, black] {\Huge$\bullet$} node[at start, black] {\Huge$\bullet$};
       \filldraw [white] (1,2) circle (5.5pt);
       \node[black] at (1,2) {\Large$\otimes$};
    \end{tikzpicture}=\quad{\cal R}\times {\cal T}(z)\,.
\end{equation}
At four points, the little group weights allow only one leading singularity, which is precisely the four-point one-loop amplitude,
\begin{equation}
    {\cal R} = \frac{\la1234\ra^2}{\la AB12\ra\la AB23\ra\la AB34\ra\la AB14\ra} \label{LS}\,.
\end{equation}
Higher point leading singularities were explored and for the full ${\cal F}(z)$ they were fully classified in ref.~\cite{Brown:2025plq}. Because all negative geometries produce the same (\ref{LS}), we just factor it out, and do not consider it in our discussion here. Hence, the $0$-loop integrated geometry (just the marked point itself) is normalized to $1$, 
\begin{equation}
\begin{tikzpicture}[baseline=-0.6ex,scale=1.,transform shape]
   \filldraw[white] (0,0) circle (5pt) node[]{};
  \node[black] at (0,0) {\Large$
  \otimes$};
\end{tikzpicture}= 1\,,
\end{equation}

while the one-loop negative geometry gives 
\begin{equation}
\begin{tikzpicture}[baseline=-0.6ex,scale=1.,transform shape]
  \draw[Maroon, ultra thick] (1,0)--(2,0) node[at end,black]{\Huge$\bullet$};
   \filldraw[white] (1,0) circle (5pt) node[]{};
  \node[black] at (1,0) {\Large$
  \otimes$};
\end{tikzpicture}= \int_{CD} \frac{-(\la AB13\ra\la CD24\ra{+}\la AB24\ra\la CD13\ra)\la 1234\ra}{\la CD12\ra\la CD23\ra\la CD34\ra\la CD14\ra\la ABCD\ra} = \log^2z + \pi^2 \,.\label{L2}
\end{equation}
where $z$ is defined in eq.~(\ref{eq:zdef}), and the actual integral was an IR finite pentagon considered already in refs.~\cite{Arkani-Hamed:2010pyv,ArkaniHamed:2010kv}.

Generically, we denote the contribution of an individual graph $G_C$  with $C$ cycles by $\mathcal{G}_{G_C}(z)$, not including the overall factor of $(-1)^{E(G)}$. Including that factor gives a contribution to $\mathcal{F}(z)$. Then, contribution to the $L-$loop $\mathcal{F}$ is given by
\begin{equation}
    \mathcal{F}^{(L)}(z) =\sum_{{\cal C}=0}^{{\cal C}_{max}} \sum_{G_{\cal C}} (-1)^{E(G)}\mathcal{G}_{G_{\cal C}}(z)\,
\end{equation}
and this is the $\mathcal{F}^{(L)}(z)$ that enters into \eqref{eq:fexact}.
\subsection{Differential equation method}
\label{sec:boxing}

The special form of the tree numerator allows us to find a powerful differential equation \cite{Arkani-Hamed:2021iya}. Suppose the frozen loop $AB$ is connected to the rest of the graph through a link to another loop $CD$. Then we can write for the whole integral 
\begin{equation}
    \mathcal{G}(z) = -\int_{CD} \frac{(\ang{AB13}\ang{CD24}+\ang{AB24}\ang{CD13})\ang{1234}}{\ang{CD12}\ang{CD23}\ang{CD34}\ang{CD41}\ang{ABCD}} H(z_{CD})\,,
\end{equation}
where $H(z_{CD})$ is the rest of the diagram integrated with the loop $CD$ frozen, with $z_{CD} = \la CD12\ra\la CD34\ra/ (\la CD14\ra\la CD23\ra)$. Let's consider the first piece of this integral, namely
\begin{equation}
    I(z)=-\int_{CD} \frac{(\ang{AB13}\ang{CD24})\ang{1234}}{\ang{CD12}\ang{CD23}\ang{CD34}\ang{CD41}\ang{ABCD}} H(z_{CD})\,,
\end{equation}
We can now act with the Laplacian 
\begin{equation}
    \Box=\frac12(z\partial_z)^2\,,
\end{equation}
such that the action on the integral $I(z)$ is
\begin{equation}
    (z \partial_z)^2 I(z) = H(z)\,. \label{diff2}
\end{equation}
Note that this equation also holds separately for the second piece with $\ang{AB24}\ang{CD13}$ in the numerator, hence we normalize the operator $\Box$ with a factor $\frac{1}{2}$ in front and get
\begin{equation}
\label{eq:BoxFeqmH}
    \Box\,\mathcal{G}(z)=H(z)\,.
\end{equation}
Effectively, the action of the operator produces a delta function which reduces the graph to a simpler one with the link $(ABCD)$ removed. In the dual variables, this is 
%
	\begin{equation} \label{diff}
		\Box_{x_0}\ \begin{tikzpicture}[baseline={-0.7ex}]
			\draw[Maroon, ultra thick] (0,0)--(1,0);
			\draw[white, fill=white] (0,0) circle [radius=0.18];
			\node[black] at (0,0) {\Large$\otimes$};
			\draw[black, fill=gray] (1.2,0) circle [radius=0.6];
		\end{tikzpicture}\ =\ \begin{tikzpicture}[baseline={-0.7ex}]
			\node[black,left] at (0.63,0) {\Large$\otimes$};
			\draw[black, fill=gray] (1.09,0) circle [radius=0.6];
		\end{tikzpicture} 
	\end{equation}
We can demonstrate it on eq.~(\ref{L2}). In this case the right hand side of the equation is just given by the $L=0$ diagram, which evaluates to 1. Since the $L=1$ diagram enters $\mathcal{F}$ with an extra minus sign, the equation at hand is
\begin{equation}
\Box\,\mathcal{F}(z) = -1\,,
\end{equation}
which has a general solution 
\begin{equation}
    \mathcal{F}(z) = -(\log^2z + A) \label{eq1}\,.
\end{equation}
The boundary condition for every diagram that has a branch, such as the LHS diagram in (\ref{diff}), is 
\begin{equation}
    \mathcal{F}(z=-1) = 0 \label{eq2}\,.
\end{equation}
This follows from the factorized form of the numerator, and the fact that $\la AB13\ra=\la AB24\ra=0$ sets $z=-1$ (see eq.~(\ref{eq:zdef})), for which the integrand vanishes identically. Using (\ref{eq2}) the solution (\ref{eq1}) then reads 
\begin{equation}
    \mathcal{F}(z) = -(\log^2z + \pi^2)\,,
\end{equation}
in agreement with a direct integration of eq.~(\ref{L2}) -- note that the sign difference comes from the extra minus sign in the numerator of (\ref{L2}) compared to \eqref{eq:2loopint}.  In fact, all tree graphs can be solved using the same differential equation method. Consider a general diagram where the marked point is located somewhere in the middle of the diagram,
%
%
\begin{align}
\begin{tikzpicture}[baseline=-0.6ex,scale=1.,transform shape]
  \draw[Maroon, ultra thick] (1,0)--(1.6,0.6) node[at end,black]{\Huge$\bullet$};
  \draw[Maroon, ultra thick] (1,0)--(1.6,-0.6) node[at end,black]{\Huge$\bullet$};
  \draw[Maroon, ultra thick] (0,0)--(1.,0) node[at end,black]{\Huge$\bullet$};  
    \draw[Maroon, ultra thick] (-0.4,0.8)--(0.4,0.8) node[at end,black]{\Huge$\bullet$};
  \draw[Maroon, ultra thick] (-1.2,0)--(-0.4,0.8) node[at end,black]{\Huge$\bullet$};
  \draw[Maroon, ultra thick] (-1.2,0)--(-0.4,-0.8) node[at end,black]{\Huge$\bullet$};
  \draw[Maroon, ultra thick] (0,0)--(-1.2,0) node[at end,black]{\Huge$\bullet$};
   \filldraw[white] (0,0) circle (5pt) node[]{};
  \node[black] at (0,0) {\Large$\otimes$};
\end{tikzpicture}  
\;\;=\;\;
\begin{tikzpicture}[baseline=-0.6ex,scale=1.,transform shape]
    \draw[Maroon, ultra thick] (-0.4,0.8)--(0.4,0.8) node[at end,black]{\Huge$\bullet$};
  \draw[Maroon, ultra thick] (-1.2,0)--(-0.4,0.8) node[at end,black]{\Huge$\bullet$};
  \draw[Maroon, ultra thick] (-1.2,0)--(-0.4,-0.8) node[at end,black]{\Huge$\bullet$};
  \draw[Maroon, ultra thick] (0,0)--(-1.2,0) node[at end,black]{\Huge$\bullet$};
   \filldraw[white] (0,0) circle (5pt) node[]{};
  \node[black] at (0,0) {\Large$\otimes$};
\end{tikzpicture} 
\times
\begin{tikzpicture}[baseline=-0.6ex,scale=1.,transform shape]
  \draw[Maroon, ultra thick] (1,0)--(1.6,0.6) node[at end,black]{\Huge$\bullet$};
  \draw[Maroon, ultra thick] (1,0)--(1.6,-0.6) node[at end,black]{\Huge$\bullet$};
  \draw[Maroon, ultra thick] (0,0)--(1.,0) node[at end,black]{\Huge$\bullet$};  
   \filldraw[white] (0,0) circle (5pt) node[]{};
  \node[black] at (0,0) {\Large$\otimes$};
\end{tikzpicture}  
\label{tree_fac}
\end{align}
where the integrand factorizes as
\begin{equation}
    {\cal I}(AB_1,\dots, AB_L) = {\cal R}^{-1}\times  
    {\cal I}_1(AB_1,\dots,AB_{L'},AB_{0}) \times {\cal I}_2 (AB_{0},AB_{L'+1},\dots,AB_L)\,,
\end{equation}
where ${\cal R}^{-1}$ accounts for double counting of the denominator factor of the frozen loop $AB_0$. After integration we get 
\begin{equation}
    \mathcal{F}(z) = \mathcal{F}_1(z) \times \mathcal{F}_2(z) \,,
\end{equation}
where we stripped off the overall ${\cal R}$ factor as usual. Because both factorized parts of the diagrams now have the marked point in the correct position to use eq.~(\ref{diff}), both $\mathcal{F}_1(z)$ and $\mathcal{F}_2(z)$ now satisfy the differential equation (\ref{diff2}). Of course, if there are more legs connected to the frozen loop in (\ref{tree_fac}), the integrand initially factorizes into more components, each satisfying the differential equation. As a result, the function $\mathcal{F}(z)$ for any tree graph factorizes into a product of functions $\mathcal{F}_k(z)$, each satisfying the differential equation (\ref{diff2}).

\subsection{Two-loop results and the depth of the polylogarithm}

The full two-loop Wilson-loop result was computed in ref.~\cite{Alday:2013ip} and the negative geometry expansion was evaluated term-by-term in ref.~\cite{Arkani-Hamed:2021iya}. 
This expansion involves three terms, and the explicit evaluation reveals the major difference between two tree graphs and the one-cycle graph. The two tree graphs can be evaluated using the differential equation method. For the `ladder' geometry, the equation reads 
%
%
%
\begin{equation}
   \Box \;
\Big[\begin{tikzpicture}[baseline=-0.6ex,scale=1.,transform shape]
  \draw[Maroon, ultra thick] (1,0)--(3,0) node[pos=0.5,black]{\Huge$\bullet$} node[at end,black]{\Huge$\bullet$};
   \filldraw[white] (1,0) circle (5pt) node[]{};
  \node[black] at (1,0) {\Large$
  \otimes$};
\end{tikzpicture}\Big]=  
\begin{tikzpicture}[baseline=-0.6ex,scale=1.,transform shape]
  \draw[Maroon, ultra thick] (1,0)--(2,0) node[at end,black]{\Huge$\bullet$};
   \filldraw[white] (1,0) circle (5pt) node[]{};
  \node[black] at (1,0) {\Large$
  \otimes$};
\end{tikzpicture}
\end{equation}
which translates into 
\begin{equation}
    \Box\,\mathcal{F}(z) = (\log^2z + \pi^2) \,,
\end{equation}
with the same boundary condition (\ref{eq2}). As a result, we get 
\begin{equation}
    \begin{tikzpicture}[baseline=-0.6ex,scale=1.,transform shape]
  \draw[Maroon, ultra thick] (1,0)--(3,0) node[pos=0.5,black]{\Huge$\bullet$} node[at end,black]{\Huge$\bullet$};
   \filldraw[white] (1,0) circle (5pt) node[]{};
  \node[black] at (1,0) {\Large$
  \otimes$};
\end{tikzpicture}=  \frac{1}{6} \left[ \pi^2 + \log^2 z \right] \times \left[ 5 \pi^2 +   \log^2 z \right]\,.
\end{equation}
The second diagram is even simpler, as it factorizes as a product of two (\ref{L2}) diagrams,
\begin{equation}
\begin{tikzpicture}[baseline=-0.6ex,scale=1.,transform shape]
  \draw[Maroon, ultra thick] (1,0)--(2,0.6) node[at end,black]{\Huge$\bullet$};
  \draw[Maroon, ultra thick] (1,0)--(2,-0.6) node[at end,black]{\Huge$\bullet$};
   \filldraw[white] (1,0) circle (5pt) node[]{};
  \node[black] at (1,0) {\Large$\otimes$};
\end{tikzpicture}=  \left[ \pi^2 + \log^2 z \right]^2 .
\end{equation}

The one-cycle diagram does not satisfy the same differential equation, and we have to perform a direct integration. As a result, we get
\begin{align}
\begin{tikzpicture}
[baseline=-0.6ex,scale=1.,transform shape]
    \draw[Maroon, ultra thick] (0,-0.5)--(0,0.5);
    \draw[Maroon, ultra thick](0,-0.5)--(1,0) node[at start,black] {\Huge$\bullet$}; 
    \draw[Maroon, ultra thick](0,0.5)--(1,0) node[at start,black] {\Huge$\bullet$};
     \filldraw[white] (1,0) circle (5pt) node[]{};
    \node[black] at (1,0) {\Large$\otimes$};
\end{tikzpicture}
&= 8\displaystyle H_{\text{0,0,0,0}}+8 H_{\text{-1,0,0,0}}-16 H_{\text{-1,-1,0,0}}+8 H_{\text{-2,0,0}} - 8 \zeta_3 \left(2 H_{-1}-H_0\right)\nonumber \\
&+ 4\pi ^2 \left(H_{\text{-1,0}}- 2\,H_{\text{-1,-1}}+H_{-2}\right) +\frac{13 \pi ^4}{45}\,,\label{L3c}
\end{align}
where we use the notation for the harmonic polylogarithms (HPLs) \cite{Remiddi:1999ew} (with the argument $z$ omitted for brevity).
If an HPL does not have a trailing $0$ index, then it admits a convergent power series expansion for real $z$ with $-1<z<1$,
\begin{align}
H_{-a_d,\dots,-a_1} (z)  
\equiv H_{\underbrace{\scriptstyle 0,\dots,0}_{a_d-1},-1,\dots,\underbrace{\scriptstyle 0,\dots,0}_{a_1-1},-1}(z)
=(-1)^d\sum\limits_{j_d>\dots>j_1\ge 1}  \frac{\left(-z\right)^{j_d}}{j_d^{a_d}} \frac{1}{j_{d-1}^{a_{d-1}}}\dots  \frac{1}{j_1^{a_1}}\,,
\end{align}
where the number of nonzero indices is the \emph{depth} $d$.
Trailing $0$'s can be removed using the shuffle algebra and the fact that
\begin{equation}
H_{\underbrace{\scriptstyle 0,\dots,0}_{n}}(z) = 
\frac{1}{n!} \, \log^n z \,.
\end{equation}

In the case of (\ref{L3c}), all HPLs reduce to classical polylogarithms~\cite{Arkani-Hamed:2021iya},
\begin{align}
&\begin{tikzpicture}
[baseline=-0.6ex,scale=1.,transform shape]
    \draw[Maroon, ultra thick] (0,-0.5)--(0,0.5);
    \draw[Maroon, ultra thick](0,-0.5)--(1,0) node[at start,black] {\Huge$\bullet$}; 
    \draw[Maroon, ultra thick](0,0.5)--(1,0) node[at start,black] {\Huge$\bullet$};
     \filldraw[white] (1,0) circle (5pt) node[]{};
    \node[black] at (1,0) {\Large$\otimes$};
\end{tikzpicture}
=\frac{1}{3} \log^4 z  - 2\log^2 z \left[ -\frac{2}{3} \text{Li}_2\left(\frac{1}{1+z}\right)-\frac{2}{3} \text{Li}_2\left(\frac{z}{1+z}\right)+\frac{\pi ^2}{9} \right]  \\
&- 2\log z \left[ 4 \text{Li}_3\left(\frac{z}{1+z}\right)-4 \text{Li}_3\left(\frac{1}{1+z}\right) \right] 
+\frac{4}{3} \left[ \text{Li}_2\left(\frac{1}{1+z}\right)+\text{Li}_2\left(\frac{z}{1+z}\right)-\frac{\pi ^2}{6}\right]^2 \nonumber\\
& + \frac{16}{3} \pi ^2 \left[\text{Li}_2\left(\frac{1}{1+z}\right)+\text{Li}_2\left(\frac{z}{1+z}\right)-\frac{\pi ^2}{6}\right]
+16 \text{Li}_4\left(\frac{1}{1+z}\right)+16 \text{Li}_4\left(\frac{z}{1+z}\right)+\frac{\pi ^4}{9}\,.\nonumber 
\end{align}
The reduction to classical polylogarithms does not generalize to higher loops. Note that all three diagrams produce transcendentality four functions, but the two tree graphs are much simpler than the triangle one-cycle diagram. 

The main difference between the tree result (which are just powers of $\log z$) and the one-cycle formula is the \emph{depth} of the polylogarithms. This is best evident from the \emph{symbol}~\cite{Goncharov:2010jf}. The symbol can be read off from the uncompressed list of HPL indices belonging to $\{0,-1\}$, by reversing the order and letting $0\to z$ and $-1 \to 1+z$.  We also write ${\rm SB}(a,b,c,d)$ for $a\otimes b\otimes c\otimes d$.   The symbols of the contributions to $\mathcal{F}(z)$ at two loops are,
\begin{align}
    {\rm Symbol}\left[\begin{tikzpicture}[baseline=-0.6ex,scale=1.,transform shape]
  \draw[Maroon, ultra thick] (1,0)--(3,0) node[pos=0.5,black]{\Huge$\bullet$} node[at end,black]{\Huge$\bullet$};
   \filldraw[white] (1,0) circle (5pt) node[]{};
  \node[black] at (1,0) {\Large$
  \otimes$};
\end{tikzpicture}\right]
&=  4\,\text{SB}(z,z,z,z)\,,\\
    {\rm Symbol}\left[\begin{tikzpicture}[baseline=-0.6ex,scale=1.,transform shape]
  \draw[Maroon, ultra thick] (1,0)--(2,0.6) node[at end,black]{\Huge$\bullet$};
  \draw[Maroon, ultra thick] (1,0)--(2,-0.6) node[at end,black]{\Huge$\bullet$};
  \filldraw[white] (1,0) circle (5pt) node[]{};
  \node[black] at (1,0) {\Large$\otimes$};
\end{tikzpicture}\right]
&=  24\,\text{SB}(z,z,z,z)\,,\\ {\rm Symbol}\left[\begin{tikzpicture}
[baseline=-0.6ex,scale=1.,transform shape]
    \draw[Maroon, ultra thick] (0,-0.5)--(0,0.5);
    \draw[Maroon, ultra thick](0,-0.5)--(1,0) node[at start,black] {\Huge$\bullet$}; 
    \draw[Maroon, ultra thick](0,0.5)--(1,0) node[at start,black] {\Huge$\bullet$};
     \filldraw[white] (1,0) circle (5pt) node[]{};
    \node[black] at (1,0) {\Large$\otimes$};
    \end{tikzpicture}\right]
    &=  8 \,\big(\text{SB}(z,z,z,z)+\text{SB}(z,z,z,1{+}z)\nonumber\\
    &\hspace{1cm}+\text{SB}(z,z,1{+}z,z)-2\,\text{SB}(z,z,1{+}z,1{+}z)\label{depth2hpl}
    \big)\,,
    \end{align}
where the last term of (\ref{depth2hpl}), $\text{SB}(z,z,1{+}z,1{+}z)$, corresponds to $H_{\text{-1,-1,0,0}}$ in eq.~(\ref{L3c}); the latter has (higher) depth two.
%
We will see that this depth property generalizes to more cycles and higher depths. 

\subsection{New three-loop results}

One of the main results of this paper is the computation of the full three-loop Wilson loop ${\cal F}(z)$ using negative geometries. The result was obtained using a traditional method of Feynman diagrams (for Wilson loops) in ref.~\cite{Henn:2019swt}. This expansion contained UV divergent integrals that needed to be regulated, and the finite result was only obtained after performing the whole sum. In our approach, each term is individually finite. The integrand for all contributing negative geometries was provided in eq.~(\ref{eq:omega4}) as the four-loop integrand for the amplitude's logarithm. As before, we need to freeze one of the loops in each diagram and integrate over the others. At the level of diagrams, we have six $L=4$ negative geometry integrands which produce 11 three-loop contributions to ${\cal F}(g,z)$. 
\begin{align}
\centering
   \qquad\qquad \begin{tikzpicture}[baseline=-0.6ex,scale=1.,transform shape]
  \draw[Maroon, ultra thick] (1.4,0)--(2.1,0) node[at end,black]{\Huge$\bullet$};
  \draw[Maroon, ultra thick] (0.7,0)--(1.4,0) node[at end,black]{\Huge$\bullet$};
  \draw[Maroon, ultra thick] (0,0)--(0.7,0) node[at end,black]{\Huge$\bullet$} node[at start,black]{\Huge$\bullet$};
\end{tikzpicture}&\ \to\ \begin{tikzpicture}[baseline=-0.6ex,scale=1.,transform shape]
  \draw[Maroon, ultra thick] (1.4,0)--(2.1,0) node[at end,black]{\Huge$\bullet$};
  \draw[Maroon, ultra thick] (0.7,0)--(1.4,0) node[at end,black]{\Huge$\bullet$};
  \draw[Maroon, ultra thick] (0,0)--(0.7,0) node[at end,black]{\Huge$\bullet$};
   \filldraw[white] (0,0) circle (5pt) node[]{};
  \node[black] at (0,0) {\Large$\otimes$};
\end{tikzpicture}+ \begin{tikzpicture}[baseline=-0.6ex,scale=1.,transform shape]
  \draw[Maroon, ultra thick] (0.7,0.4)--(1.3,0.4) node[at end,black]{\Huge$\bullet$};
  \draw[Maroon, ultra thick] (0,0)--(0.7,0.4) node[at end,black]{\Huge$\bullet$};
  \draw[Maroon, ultra thick] (0,0)--(0.7,-0.4) node[at end,black]{\Huge$\bullet$};
   \filldraw[white] (0,0) circle (5pt) node[]{};
  \node[black] at (0,0) {\Large$\otimes$};
\end{tikzpicture}\nonumber\\
\begin{tikzpicture}[baseline=-0.6ex,scale=1.,transform shape]
  \draw[Maroon, ultra thick] (0,0)--(0.9,0)   node[at end,black]{\Huge$\bullet$};
  \draw[Maroon, ultra thick] (0,0)--(0.8,0.55) node[at end,black]{\Huge$\bullet$};
  \draw[Maroon, ultra thick] (0,0)--(0.8,-0.55) node[at end,black]{\Huge$\bullet$} node[at start,black]{\Huge$\bullet$};
\end{tikzpicture}&\ \to\ \frac{1}{2!}\begin{tikzpicture}[baseline=-0.6ex,scale=1.,transform shape]
  \draw[Maroon, ultra thick] (0.8,0)--(1.4,0.6) node[at end,black]{\Huge$\bullet$};
  \draw[Maroon, ultra thick] (0.8,0)--(1.4,-0.6) node[at end,black]{\Huge$\bullet$};
  \draw[Maroon, ultra thick] (0,0)--(0.8,0) node[at end,black]{\Huge$\bullet$};
   \filldraw[white] (0,0) circle (5pt) node[]{};
  \node[black] at (0,0) {\Large$\otimes$};
\end{tikzpicture}+\frac{1}{3!}\begin{tikzpicture}[baseline=-0.6ex,scale=1.,transform shape]
  \draw[Maroon, ultra thick] (0,0)--(0.9,0)   node[at end,black]{\Huge$\bullet$};
  \draw[Maroon, ultra thick] (0,0)--(0.8,0.55) node[at end,black]{\Huge$\bullet$};
  \draw[Maroon, ultra thick] (0,0)--(0.8,-0.55) node[at end,black]{\Huge$\bullet$};
   \filldraw[white] (0,0) circle (5pt) node[]{};
  \node[black] at (0,0) {\Large$\otimes$};
\end{tikzpicture}\nonumber\\
\begin{tikzpicture}[baseline=1.5ex,scale=1.,transform shape]
  \draw[Maroon, ultra thick] (0.8,0)--(1.8,0);
  \draw[Maroon, ultra thick] (1.8,0)--(1.3,0.9); 
  \draw[Maroon, ultra thick] (0.8,0)--(1.3,0.9) node[at end,black]{\Huge$\bullet$}; 
  \draw[Maroon, ultra thick] (0,0)--(0.8,0) node[at end,black]{\Huge$\bullet$}; 
  \node at (1.8,0) {\Huge$\bullet$};
  \node[black] at (0,0) {\Huge$\bullet$};
\end{tikzpicture}&\ \to\ \begin{tikzpicture}[baseline=1.5ex,scale=1.,transform shape]
  \draw[Maroon, ultra thick] (0.8,0)--(1.8,0);
  \draw[Maroon, ultra thick] (1.8,0)--(1.3,0.9); 
  \draw[Maroon, ultra thick] (0.8,0)--(1.3,0.9);
  \draw[Maroon, ultra thick] (0,0)--(0.8,0) node[at end,black]{\Huge$\bullet$}; 
  \node at (1.8,0) {\Huge$\bullet$};
  \node[black] at (0,0) {\Huge$\bullet$};
   \filldraw[white] (1.3,0.9) circle (5pt) node[]{};
  \node[black] at (1.3,0.9) {\Large$\otimes$};
\end{tikzpicture} +\frac{1}{2!} \begin{tikzpicture}[baseline=1.5ex,scale=1.,transform shape]
  \draw[Maroon, ultra thick] (0.8,0)--(1.8,0);
  \draw[Maroon, ultra thick] (1.8,0)--(1.3,0.9); 
  \draw[Maroon, ultra thick] (0.8,0)--(1.3,0.9) node[at end,black]{\Huge$\bullet$};
  \draw[Maroon, ultra thick] (0,0)--(0.8,0) node[at end,black]{\Huge$\bullet$}; 
  \node at (1.8,0) {\Huge$\bullet$};
  \node[black] at (0,0) {\Huge$\bullet$};
   \filldraw[white] (0.8,0) circle (5pt) node[]{};
  \node[black] at (0.8,0) {\Large$\otimes$};
\end{tikzpicture}+\frac{1}{2!} \begin{tikzpicture}[baseline=1.5ex,scale=1.,transform shape]
  \draw[Maroon, ultra thick] (0.8,0)--(1.8,0);
  \draw[Maroon, ultra thick] (1.8,0)--(1.3,0.9); 
  \draw[Maroon, ultra thick] (0.8,0)--(1.3,0.9) node[at end,black]{\Huge$\bullet$};
  \draw[Maroon, ultra thick] (0,0)--(0.8,0) node[at end,black]{\Huge$\bullet$};
  \node[black] at (1.8,0) {\Huge$\bullet$};
   \filldraw[white] (0,0) circle (5pt) node[]{};
  \node[black] at (0,0) {\Large$\otimes$};
\end{tikzpicture}\nonumber\\
\begin{tikzpicture}[baseline=-0.6ex,scale=1.,transform shape]
  \draw[Maroon, ultra thick] (0,-0.5)--(1.1,-0.5);
  \draw[Maroon, ultra thick] (0,-0.5)--(0,0.5);
  \draw[Maroon, ultra thick] (1.1,-0.5)--(1.1,0.5)
    node[at start,black]{\Huge$\bullet$}; 
  \draw[Maroon, ultra thick] (0,0.5)--(1.1,0.5)
    node[at start,black]{\Huge$\bullet$}
    node[at end,black]{\Huge$\bullet$};
    \node[black] at (0,-0.5) {\Huge$\bullet$};
\end{tikzpicture}&\ \to\ \frac{1}{2!}\begin{tikzpicture}[baseline=-0.6ex,scale=1.,transform shape]
  \draw[Maroon, ultra thick] (0,-0.5)--(0,0.5);
  \draw[Maroon, ultra thick] (1.1,-0.5)--(1.1,0.5)
    node[at start,black]{\Huge$\bullet$}; 
  \draw[Maroon, ultra thick] (0,0.5)--(1.1,0.5)
    node[at start,black]{\Huge$\bullet$}
    node[at end,black]{\Huge$\bullet$};
    \draw[Maroon, ultra thick] (0,-0.5)--(1.1,-0.5)
    node[at start,black]{\Huge$\bullet$}
    node[at end,black]{\Huge$\bullet$};
   \filldraw[white] (1.1,-0.5) circle (5pt) node[]{};
  \node[black] at (1.1,-0.5) {\Large$\otimes$};
\end{tikzpicture}\nonumber\\
\begin{tikzpicture}[baseline=-0.6ex,scale=1.,transform shape]
  \draw[Maroon, ultra thick] (0,-0.5)--(1.1,-0.5);
  \draw[Maroon, ultra thick] (0,-0.5)--(0,0.5); 
  \draw[Maroon, ultra thick] (1.1,0.5)--(0,-0.5);
  \draw[Maroon, ultra thick] (1.1,-0.5)--(1.1,0.5)
    node[at start,black]{\Huge$\bullet$}; 
  \draw[Maroon, ultra thick] (0,0.5)--(1.1,0.5)
    node[at start,black]{\Huge$\bullet$}
    node[at end,black]{\Huge$\bullet$};
  \node[black] at (0,-0.5) {\Huge$\bullet$};
\end{tikzpicture}&\ \to\ \frac{1}{2!}\begin{tikzpicture}[baseline=-0.6ex,scale=1.,transform shape]
  \draw[Maroon, ultra thick] (0,-0.5)--(1.1,-0.5);
  \draw[Maroon, ultra thick] (0,-0.5)--(0,0.5);
  \draw[Maroon, ultra thick] (1.1,0.5)--(0,-0.5);
  \draw[Maroon, ultra thick] (1.1,-0.5)--(1.1,0.5)
    node[at start,black]{\Huge$\bullet$}; 
  \draw[Maroon, ultra thick] (0,0.5)--(1.1,0.5)
    node[at start,black]{\Huge$\bullet$}
    node[at end,black]{\Huge$\bullet$};
   \filldraw[white] (0,-0.5) circle (5pt) node[]{};
  \node[black] at (0,-0.5) {\Large$\otimes$};
\end{tikzpicture} + \frac{1}{2!}\begin{tikzpicture}[baseline=-0.6ex,scale=1.,transform shape]
  \draw[Maroon, ultra thick] (0,-0.5)--(1.1,-0.5);
  \draw[Maroon, ultra thick] (0,-0.5)--(0,0.5);
  \draw[Maroon, ultra thick] (0,0.5)--(1.1,-0.5);
  \draw[Maroon, ultra thick] (1.1,-0.5)--(1.1,0.5)
    node[at start,black]{\Huge$\bullet$}; 
  \draw[Maroon, ultra thick] (0,0.5)--(1.1,0.5)
    node[at start,black]{\Huge$\bullet$}
    node[at end,black]{\Huge$\bullet$};
   \filldraw[white] (0,-0.5) circle (5pt) node[]{};
  \node[black] at (0,-0.5) {\Large$\otimes$};
\end{tikzpicture}\nonumber\\
\begin{tikzpicture}[baseline=-0.6ex,scale=1.,transform shape]
  \draw[Maroon, ultra thick] (0,-0.5)--(1.1,-0.5);
  \draw[Maroon, ultra thick] (0,-0.5)--(0,0.5);
  \draw[Maroon, ultra thick] (0,0.5)--(1.1,-0.5); 
  \draw[Maroon, ultra thick] (1.1,0.5)--(0,-0.5);
  \draw[Maroon, ultra thick] (1.1,-0.5)--(1.1,0.5)
    node[at start,black]{\Huge$\bullet$}; 
  \draw[Maroon, ultra thick] (0,0.5)--(1.1,0.5)
    node[at start,black]{\Huge$\bullet$}
    node[at end,black]{\Huge$\bullet$};
  \node[black] at (0,-0.5) {\Huge$\bullet$};
\end{tikzpicture}&\ \to\ \frac{1}{3!}\begin{tikzpicture}[baseline=-0.6ex,scale=1.,transform shape]
  \draw[Maroon, ultra thick] (0,-0.5)--(1.1,-0.5);
  \draw[Maroon, ultra thick] (0,-0.5)--(0,0.5);
  \draw[Maroon, ultra thick] (0,0.5)--(1.1,-0.5); 
  \draw[Maroon, ultra thick] (1.1,0.5)--(0,-0.5);
  \draw[Maroon, ultra thick] (1.1,-0.5)--(1.1,0.5)
    node[at start,black]{\Huge$\bullet$}; 
  \draw[Maroon, ultra thick] (0,0.5)--(1.1,0.5)
    node[at start,black]{\Huge$\bullet$}
    node[at end,black]{\Huge$\bullet$};
  \filldraw[white] (0,-0.5) circle (5pt) node[]{};
  \node[black] at (0,-0.5) {\Large$\otimes$};
\end{tikzpicture}
\end{align}
where we have included the correct symmetrization factors. 

All zero-cycle (tree) diagrams are easy to evaluate using the differential equation method, 
we obtain for the $\mathcal{G}(z)$ contributions
\begin{align} \label{eq:f3_tree}
\begin{tikzpicture}[baseline=-0.6ex,scale=1.,transform shape]
  \draw[Maroon, ultra thick] (1.4,0)--(2.1,0) node[at end,black]{\Huge$\bullet$};
  \draw[Maroon, ultra thick] (0.7,0)--(1.4,0) node[at end,black]{\Huge$\bullet$};
  \draw[Maroon, ultra thick] (0,0)--(0.7,0) node[at end,black]{\Huge$\bullet$};
   \filldraw[white] (0,0) circle (5pt) node[]{};
  \node[black] at (0,0) {\Large$\otimes$};
\end{tikzpicture}&=8 H_{0,0,0,0,0,0}+4 \pi ^2 H_{0,0,0,0}+\frac{5}{3} \pi ^4 H_{0,0}+\frac{61}{90} \pi ^6\,, \nonumber\\
\begin{tikzpicture}[baseline=-0.6ex,scale=1.,transform shape]
  \draw[Maroon, ultra thick] (0.7,0.4)--(1.3,0.4) node[at end,black]{\Huge$\bullet$};
  \draw[Maroon, ultra thick] (0,0)--(0.7,0.4) node[at end,black]{\Huge$\bullet$};
  \draw[Maroon, ultra thick] (0,0)--(0.7,-0.4) node[at end,black]{\Huge$\bullet$};
   \filldraw[white] (0,0) circle (5pt) node[]{};
  \node[black] at (0,0) {\Large$\otimes$};
\end{tikzpicture}&=120 H_{0,0,0,0,0,0}+28 \pi ^2 H_{0,0,0,0}+\frac{11}{3} \pi ^4 H_{0,0}+\frac{5 \pi ^6}{6}\,,\nonumber\\
\begin{tikzpicture}[baseline=-0.6ex,scale=1.,transform shape]
  \draw[Maroon, ultra thick] (0.8,0)--(1.4,0.6) node[at end,black]{\Huge$\bullet$};
  \draw[Maroon, ultra thick] (0.8,0)--(1.4,-0.6) node[at end,black]{\Huge$\bullet$};
  \draw[Maroon, ultra thick] (0,0)--(0.8,0) node[at end,black]{\Huge$\bullet$};
   \filldraw[white] (0,0) circle (5pt) node[]{};
  \node[black] at (0,0) {\Large$\otimes$};
\end{tikzpicture}&=48 H_{0,0,0,0,0,0}+8 \pi ^2 H_{0,0,0,0}+2 \pi ^4 H_{0,0}+\frac{11 \pi ^6}{15}\,, \nonumber\\
\begin{tikzpicture}[baseline=-0.6ex,scale=1.,transform shape]
  \draw[Maroon, ultra thick] (0,0)--(0.9,0)   node[at end,black]{\Huge$\bullet$};
  \draw[Maroon, ultra thick] (0,0)--(0.8,0.55) node[at end,black]{\Huge$\bullet$};
  \draw[Maroon, ultra thick] (0,0)--(0.8,-0.55) node[at end,black]{\Huge$\bullet$};
  \filldraw[white] (0,0) circle (5pt) node[]{};
  \node[black] at (0,0) {\Large$\otimes$};
\end{tikzpicture}&=720 H_{0,0,0,0,0,0} +72 \pi^2 H_{0,0,0,0}+6 \pi^4 H_{0,0}+\pi^6\,.
\end{align}
As a consistency check, the sum over all tree graphs, weighted by combinatoric factors and the overall minus sign from $(-1)^3$, should reproduce the result from the generating function at order $g^6$~\cite{Arkani-Hamed:2021iya}, {\it i.e.}
\begin{align}
  \mathcal{F}_{\rm tree}(z)\Big|_{g^6}&= -\frac{1}{45}\left[\log ^2 z +\pi ^2\right]\times \left[ 17 \log ^4 z +73 \pi ^2 \log^2 z +92 \pi ^4\right]\nonumber\\
  &= -272 {H}_{0,0,0,0,0,0}-48 \pi ^2 {H}_{0,0,0,0}-\frac{22}{3} \pi ^4 H_{0,0}-\frac{92 \pi ^6}{45}\,,
\end{align}
which is indeed the case. 

\subsubsection*{One-cycle diagrams}

The one-cycle graphs are more involved to evaluate. One of them is just a trivial product of the 2-loop triangle (\ref{L3c}) and the one-loop result (\ref{L2}) using (\ref{tree_fac}),
\begin{align}
&\begin{tikzpicture}[baseline=-0.6ex,scale=1.2,transform shape]
    \draw[Maroon, ultra thick] (0.5,-0.5)--(0.5,0.5);
    \draw[Maroon, ultra thick](0.5,-0.5)--(1,0) node[at start,black] {\huge$\bullet$}; 
    \draw[Maroon, ultra thick](0.5,0.5)--(1,0) node[at start,black] {\huge$\bullet$};
    \draw[Maroon, ultra thick](1,0)--(1.5,0) node[at end,black] {\huge$\bullet$} node[at end, right]{};
    \filldraw[white] (1,0) circle (4pt) node[]{};
    \node[black] at (1,0) {\large$\otimes$};
    \end{tikzpicture}=
16\,\Big(
-2 \zeta_3 H_{-2,0}-2 \zeta_3 H_{-1,0,0}+3 \zeta_3 H_{0,0,0}-\pi ^2 H_{-3,-1}+\frac{3}{2} \pi ^2 H_{-3,0}\nonumber\\&-\pi ^2 H_{-2,-2}-\pi ^2 H_{-1,-3}-\frac{1}{2} \pi ^4 H_{-1,-1}+\frac{1}{4} \pi ^4 H_{-1,0}+\frac{13}{360} \pi ^4 H_{0,0}+3 H_{-4,0,0}-\pi ^2 H_{-2,-1,0}\nonumber\\&+2 \pi ^2 H_{-2,0,0}-\pi ^2 H_{-1,-2,0}-2 H_{-3,-1,0,0}+7 H_{-3,0,0,0}-2 H_{-2,-2,0,0}-2 H_{-1,-3,0,0}\nonumber\\&-2 \pi ^2 H_{-1,-1,0,0}+2 \pi ^2 H_{-1,0,0,0}+\frac{1}{2} \pi ^2 H_{0,0,0,0}-6 H_{-2,-1,0,0,0}+10 H_{-2,0,0,0,0}-6 H_{-1,-2,0,0,0}\nonumber\\&-12 H_{-1,-1,0,0,0,0}+10 H_{-1,0,0,0,0,0}+15 H_{0,0,0,0,0,0}-2 H_{-3} \zeta_3-\pi ^2 H_{-1} \zeta_3+\frac{1}{2} \pi ^2 H_0 \zeta_3\nonumber\\&+\frac{3}{2} \pi ^2 H_{-4}+\frac{1}{4} \pi ^4 H_{-2}+\frac{13 \pi ^6}{720}\Big)\,.   
\end{align}
At the level of the symbol \cite{Goncharov:2010jf} this is
\begin{align} \label{eq:I3,3}
 &{\rm Symbol}
 \left[\begin{tikzpicture}[baseline=-0.6ex,scale=1.2,transform shape]
    \draw[Maroon, ultra thick] (0.5,-0.5)--(0.5,0.5);
    \draw[Maroon, ultra thick](0.5,-0.5)--(1,0) node[at start,black] {\huge$\bullet$}; 
    \draw[Maroon, ultra thick](0.5,0.5)--(1,0) node[at start,black] {\huge$\bullet$};
    \draw[Maroon, ultra thick](1,0)--(1.5,0) node[at end,black] {\huge$\bullet$} node[at end, right]{};
    \filldraw[white] (1,0) circle (4pt) node[]{};
    \node[black] at (1,0) {\large$\otimes$};
    \end{tikzpicture}\right]=16\,\big(15 \,\text{SB}(z,z,z,z,z,z)+10 \,\text{SB}(z,z,z,z,z,1{+}z)\nonumber\\&+10 \,\text{SB}(z,z,z,z,1{+}z,z)-12 \,\text{SB}(z,z,z,z,1{+}z,1{+}z)+7 \,\text{SB}(z,z,z,1{+}z,z,z)\nonumber\\&-6 \,\text{SB}(z,z,z,1{+}z,z,1{+}z)-6 \,\text{SB}(z,z,z,1{+}z,1{+}z,z)+3 \,\text{SB}(z,z,1{+}z,z,z,z)\nonumber\\&-2 \,\text{SB}(z,z,1{+}z,z,z,1{+}z)-2 \,\text{SB}(z,z,1{+}z,z,1{+}z,z)-2 \,\text{SB}(z,z,1{+}z,1{+}z,z,z)\big)\,,
\end{align}
Another one does satisfy the boxing equation and can also be evaluated easily by canonical differential equations \cite{Henn:2013pwa,Henn:2019swt}
\begin{align} \label{eq:I3,1}
   \begin{tikzpicture}[baseline=-0.6ex,scale=1.2,transform shape]
    \draw[Maroon, ultra thick] (0.5,-0.5)--(0.5,0.5);
    \draw[Maroon, ultra thick](0.5,-0.5)--(1,0) node[at start,black] {\huge$\bullet$}; 
    \draw[Maroon, ultra thick](0.5,0.5)--(1,0) node[at start,black] {\huge$\bullet$};
    \draw[Maroon, ultra thick](1,0)--(1.5,0) node[at end,black] {\huge$\bullet$} node[at end, right]{};
    \node[black] at (1.0,0) {\huge$\bullet$};
    \node[white] at (1.5,0) {\huge$\bullet$};
    \filldraw[white] (1.5,0) circle (4pt) node[]{};
    \node[black] at (1.5,0) {$\otimes$};
    \end{tikzpicture}= &2\,\bigg( 8H_{0,0,0,0,0,0}+8H_{0,0,{-}1,0,0,0}{-}16H_{0,0,{-}1,{-}1,0,0}+8H_{0,0,{-}2,0,0} \nonumber\\
    & -8 \zeta_3 (2H_{0,0,{-}1}-H_{0,0,0}) +4 \pi^2(H_{0,0,{-}1,0}-2H_{0,0,{-}1,{-}1}+H_{0,0,{-}2}) \nonumber\\
    &+\frac{13 \pi^4}{90} \log({z})^2 + \left(4 \pi ^2 \zeta_3+8 \zeta_5\right) \log({z})+\left(\frac{7 \pi ^6}{45}-8 \zeta_3^2\right)\bigg)\,,
\end{align}
which reproduces the answer in \cite{Brown:2023mqi} 
\footnote{Here we write the HPL in a partially-uncompressed form, in order to make the boxing equation to (\ref{L3c}) and the comparison with \cite{Brown:2023mqi} manifest. In general, we will often use such partially-uncompressed notation.}
Its symbol is given by
\begin{align}
&{\rm Symbol} 
\left[\begin{tikzpicture}[baseline=-0.6ex,scale=1.2,transform shape]
    \draw[Maroon, ultra thick] (0.5,-0.5)--(0.5,0.5);
    \draw[Maroon, ultra thick](0.5,-0.5)--(1,0) node[at start,black] {\huge$\bullet$}; 
    \draw[Maroon, ultra thick](0.5,0.5)--(1,0) node[at start,black] {\huge$\bullet$};
    \draw[Maroon, ultra thick](1,0)--(1.5,0) node[at end,black] {\huge$\bullet$} node[at end, right]{};
    \node[black] at (1.0,0) {\huge$\bullet$};
    \node[white] at (1.5,0) {\huge$\bullet$};
    \filldraw[white] (1.5,0) circle (4pt) node[]{};
    \node[black] at (1.5,0) {$\otimes$};
    \end{tikzpicture}\right]=2\,\Big(8 \,\text{SB}(z,z,z,z,z,z)+8 \,\text{SB}(z,z,z,1{+}z,z,z)\nonumber\\
&+8 \,\text{SB}(z,z,1{+}z,z,z,z)-16 \,\text{SB}(z,z,1{+}z,1{+}z,z,z)\Big). \end{align}

All the other one-cycle diagrams, as well as both two-cycle and the single three-cycle diagrams, do not satisfy a boxing equation. While these integrals are still very special IR finite integrals, currently there is no technology tailored for their computation. Here we rely on the technology of \cite{Henn:2019swt} and the explicit evaluation using canonical differential equations (not of a boxing type). This generic method involves the expansion of the integrand in terms of 257 uniform transcendental (UT) basis calculated in \cite{Henn:2019swt}. The integral reduction is performed using the {\tt NeatIBP} package \cite{Wu:2023upw}. As a result, we get the following function for the last one-cycle box graph
\begin{align} \label{eq:box_funct}
&\begin{tikzpicture}[baseline=-0.6ex,scale=1.,transform shape]
  \draw[Maroon, ultra thick] (0,-0.5)--(0,0.5);
  \draw[Maroon, ultra thick] (1.1,-0.5)--(1.1,0.5)
    node[at start,black]{\Huge$\bullet$}
    node[at end,black]{\Huge$\bullet$};
  \draw[Maroon, ultra thick] (0,0.5)--(1.1,0.5)
    node[at start,black]{\Huge$\bullet$}
    node[at end,black]{\Huge$\bullet$};
    \draw[Maroon, ultra thick] (0,-0.5)--(1.1,-0.5)
    node[at start,black]{\Huge$\bullet$}
    node[at end,black]{\Huge$\bullet$};
   \filldraw[white] (1.1,-0.5) circle (5pt) node[]{};
  \node[black] at (1.1,-0.5) {\Large$\otimes$};
\end{tikzpicture}=16\,\big(3 H_{-1,-1,0,0,0,0}+2 H_{-1,0,0,0,0,0}+3 H_{0,0,0,0,0,0}
-4 H_{-2,-1,0,0,0}{-}\frac{1}{2} H_{-2,0,0,0,0}\nonumber\\
&-2 H_{-1,-2,0,0,0}+3 H_{-3,0,0,0}+2 H_{-2,-2,0,0}
-5 H_{-1,-3,0,0}+\frac{1}{2} \pi ^2 H_{-1,-1,0,0}
+\frac{7}{6} \pi ^2 H_{-1,0,0,0}\nonumber\\
&+\frac{1}{6} \pi ^2 H_{0,0,0,0}
+4 \zeta_3 H_{-1,0,0}+2 \zeta_3 H_{0,0,0}
+\frac{3}{2} H_{-4,0,0}
-\frac{2}{3} \pi ^2 H_{-2,-1,0}
-\frac{5}{12} \pi ^2 H_{-2,0,0}\nonumber\\
&-\frac{1}{3} \pi ^2 H_{-1,-2,0}-8 \zeta_3 H_{-2,0}
+\frac{1}{2} \pi ^2 H_{-3,0}+\pi ^2 H_{-2,-2}-\frac{5}{2} \pi ^2 H_{-1,-3}+\frac{1}{8} \pi ^4 H_{-1,-1}\nonumber\\&+\frac{11}{45} \pi ^4 H_{-1,0}
+\frac{5}{72} \pi ^4 H_{0,0}-\frac{1}{3} \pi ^2 H_{-1} \zeta_3+\frac{3}{2} \pi ^2 H_0 \zeta_3
-19 H_{-1} \zeta_5+\frac{31 H_0 \zeta_5}{2}\nonumber\\&+\frac{3}{4} \pi ^2 H_{-4}
-\frac{43}{144} \pi ^4 H_{-2}+ {{\frac{505\pi^6}{9072}+11\zeta_3^2}}\big)\,.
\end{align}
The expression in eq.~\eqref{eq:box_funct}, also provided in the ancillary file, has been numerically checked using {\sc AMFlow} \cite{Liu:2022chg} and \verb|pySecDec| \cite{Borowka:2017idc}. The symbol is much more compact and given by
\begin{align}
 &{\rm Symbol}
 \left[\begin{tikzpicture}[baseline=-0.6ex,scale=1.,transform shape]
  \draw[Maroon, ultra thick] (0,-0.5)--(0,0.5);
  \draw[Maroon, ultra thick] (1.1,-0.5)--(1.1,0.5)
    node[at start,black]{\Huge$\bullet$}
    node[at end,black]{\Huge$\bullet$};
  \draw[Maroon, ultra thick] (0,0.5)--(1.1,0.5)
    node[at start,black]{\Huge$\bullet$}
    node[at end,black]{\Huge$\bullet$};
    \draw[Maroon, ultra thick] (0,-0.5)--(1.1,-0.5)
    node[at start,black]{\Huge$\bullet$}
    node[at end,black]{\Huge$\bullet$};
   \filldraw[white] (1.1,-0.5) circle (5pt) node[]{};
  \node[black] at (1.1,-0.5) {\Large$\otimes$};
\end{tikzpicture}\right]=16\,\big(3 \,\text{SB}(z,z,z,z,z,z)+2\,\text{SB}(z,z,z,z,z,1{+}z) -\frac{1}{2} \,\text{SB}(z,z,z,z,1{+}z,z)\nonumber\\
 &+3 \,\text{SB}(z,z,z,z,1{+}z,1{+}z)+3 \,\text{SB}(z,z,z,1{+}z,z,z)-2 \,\text{SB}(z,z,z,1{+}z,z,1{+}z)\nonumber\\
 &-4 \,\text{SB}(z,z,z,1{+}z,1{+}z,z)+\frac{3}{2} \,\text{SB}(z,z,1{+}z,z,z,z)-5 \,\text{SB}(z,z,1{+}z,z,z,1{+}z)\nonumber\\
 &+2 \,\text{SB}(z,z,1{+}z,z,1{+}z,z)\big)\,.
\end{align}

The expression for the remaining one cycle diagram is more complicated and the result is provided in the ancillary file. Here we only present the symbol expression. One new important feature is that there is a \emph{spurious symbol letter} $b(4)$
\begin{align}
    b(4)=\frac{1+i\sqrt{z}}{1-i\sqrt{z}}\,,\quad z=s/t\,.
\end{align}
The letter $b(4)$ shows up in individual negative geometry contributions (starting from the 4$^{\rm th}$ entry), but it drops out of the sum for ${\cal F}^{(3)}$~\cite{Henn:2019swt}.
For the the remaining one cycle graph, the symbol result is 
\begin{align} \label{eq:I3,2}
 &{\rm Symbol}
 \left[\begin{tikzpicture}[baseline=-0.6ex,scale=1.2,transform shape]
    \draw[Maroon, ultra thick] (0.5,-0.5)--(0.5,0.5);
    \draw[Maroon, ultra thick](0.5,-0.5)--(1,0) node[at start,black] {\huge$\bullet$}; 
    \draw[Maroon, ultra thick](0.5,0.5)--(1,0) node[at start,black] {\huge$\bullet$};
    \draw[Maroon, ultra thick](1,0)--(1.5,0) node[at end,black] {\huge$\bullet$} node[at end, right]{};
    \node[black] at (1.0,0) {\huge$\bullet$};
     \filldraw[white] (0.5,0.5) circle (4pt) node[]{};
    \node[black] at (0.5,0.5) {$\otimes$};
    \end{tikzpicture}\right]=2\,\big(8\,\text{SB}(z,z,z,z,b(4),b(4))+2 \,\text{SB}(z,z,z,b(4),z,b(4))\nonumber\\
 &+2 \,\text{SB}(z,z,z,b(4),b(4),z)+28 \,\text{SB}(z,z,z,z,z,z)+8 \,\text{SB}(z,z,z,z,z,1{+}z)+16 \,\text{SB}(z,z,z,z,1{+}z,z)\nonumber\\
 &-8 \,\text{SB}(z,z,z,z,1{+}z,1{+}z)+18 \,\text{SB}(z,z,z,1{+}z,z,z)-2 \,\text{SB}(z,z,z,1{+}z,z,1{+}z)\nonumber\\
 &-2 \,\text{SB}(z,z,z,1{+}z,1{+}z,z)+30 \,\text{SB}(z,z,1{+}z,z,z,z)-6 \,\text{SB}(z,z,1{+}z,z,z,1{+}z)\nonumber\\
 &-22 \,\text{SB}(z,z,1{+}z,z,1{+}z,z)-32 \,\text{SB}(z,z,1{+}z,1{+}z,z,z)\big) \,.
\end{align}
In order to obtain a representation in terms of HPLs~\cite{Remiddi:1999ew} for the full function, we need to rationalize the $b(4)$ letter by defining a new variable $w$, such that
\begin{equation}
    z=-s/t=-w^2 \,.
\end{equation}
The resulting function, in terms of HPLs of $w$, is saved in the ancillary file. 

\subsubsection*{Two-cycle diagrams}

The result for the symbol of the first 2-cycle graph is \\
\begin{align} \label{eq: I_2cyc,1}
 &{\rm Symbol}
 \left[\begin{tikzpicture}[baseline=-0.6ex,scale=1.,transform shape]
  \draw[Maroon, ultra thick] (0,-0.5)--(1.1,-0.5);
  \draw[Maroon, ultra thick] (0,-0.5)--(0,0.5);
  \draw[Maroon, ultra thick] (0,0.5)--(1.1,-0.5);
  \draw[Maroon, ultra thick] (1.1,-0.5)--(1.1,0.5)
    node[at start,black]{\Huge$\bullet$};
  \draw[Maroon, ultra thick] (0,0.5)--(1.1,0.5)
    node[at start,black]{\Huge$\bullet$}
    node[at end,black]{\Huge$\bullet$};
   \filldraw[white] (0,-0.5) circle (5pt) node[]{};
  \node[black] at (0,-0.5) {\Large$\otimes$};
\end{tikzpicture}\right]=8\,\big(4 \,\text{SB}(z,z,z,z,b(4),b(4))+\,\text{SB}(z,z,z,b(4),z,b(4))\nonumber\\&+\text{SB}(z,z,z,b(4),b(4),z)+4 \,\text{SB}(z,z,z,z,z,z)+
 \,\text{SB}(z,z,z,z,z,1{+}z)+\text{SB}(z,z,z,z,1{+}z,z)\nonumber\\&+6 \,\text{SB}(z,z,z,z,1{+}z,1{+}z)+9 \,\text{SB}(z,z,z,1{+}z,z,z)
 -\,\text{SB}(z,z,z,1{+}z,z,1{+}z)\nonumber\\&-5 \,\text{SB}(z,z,z,1{+}z,1{+}z,z)-4 \,\text{SB}(z,z,z,1{+}z,1{+}z,1{+}z)+7 \,\text{SB}(z,z,1{+}z,z,z,z)\nonumber\\&-3\,\text{SB}(z,z,1{+}z,z,z,1{+}z)
 +\text{SB}(z,z,1{+}z,z,1{+}z,z)
 -4 \,\text{SB}(z,z,1{+}z,z,1{+}z,1{+}z)\nonumber\\&-8\,\text{SB}(z,z,1{+}z,1{+}z,z,z)-4 \,\text{SB}(z,z,1{+}z,1{+}z,z,1{+}z)
 -4 \,\text{SB}(z,z,1{+}z,1{+}z,1{+}z,z)\nonumber\\&+8 \,\text{SB}(z,z,1{+}z,1{+}z,1{+}z,1{+}z)\,\big)\,.
\end{align}
\\
The other 2-cycle contribution gives the following symbol result,
\\
\begin{align} \label{eq: I_2cyc,2}
 &{\rm Symbol}
 \left[\begin{tikzpicture}[baseline=-0.6ex,scale=1.,transform shape]
  \draw[Maroon, ultra thick] (0,-0.5)--(0,0.5);
  \draw[Maroon, ultra thick] (0,0.5)--(1.1,-0.5);
  \draw[Maroon, ultra thick] (0,-0.5)--(1.1,-0.5)
    node[at start,black]{\Huge$\bullet$};
  \draw[Maroon, ultra thick] (1.1,-0.5)--(1.1,0.5)
    node[at start,black]{\Huge$\bullet$};
  \draw[Maroon, ultra thick] (0,0.5)--(1.1,0.5)
    node[at end,black]{\Huge$\bullet$};
  \node[white] at (0,0.5) {\Huge$\bullet$};
  \filldraw[white] (0,0.5) circle (5pt) node[]{};
  \node[black] at (0,0.5) {\Large$\otimes$};
\end{tikzpicture}\right]=8\,\big(4 \,\text{SB}(z,z,z,z,b(4),b(4))+\,\text{SB}(z,z,z,b(4),z,b(4))\nonumber\\&+3 \,\text{SB}(z,z,z,b(4),b(4),z)+12 \,\text{SB}(z,z,z,z,z,z)
 +16 \,\text{SB}(z,z,z,z,z,1{+}z)+27 \,\text{SB}(z,z,z,z,1{+}z,z)\nonumber\\&-26 \,\text{SB}(z,z,z,z,1{+}z,1{+}z)+21 \,\text{SB}(z,z,z,1{+}z,z,z)-3 \,\text{SB}(z,z,z,1{+}z,z,1{+}z)\nonumber\\&-19 \,\text{SB}(z,z,z,1{+}z,1{+}z,z)-4 \,\text{SB}(z,z,z,1{+}z,1{+}z,1{+}z)+12 \,\text{SB}(z,z,1{+}z,z,z,z)\nonumber\\&+\text{SB}(z,z,1{+}z,z,z,1{+}z)
 -5 \,\text{SB}(z,z,1{+}z,z,1{+}z,z)-4 \,\text{SB}(z,z,1{+}z,z,1{+}z,1{+}z)\nonumber\\&-16 \,\text{SB}(z,z,1{+}z,1{+}z,z,z)-4 \,\text{SB}(z,z,1{+}z,1{+}z,z,1{+}z)\nonumber\\&-4 \,\text{SB}(z,z,1{+}z,1{+}z,1{+}z,z)+8 \,\text{SB}(z,z,1{+}z,1{+}z,1{+}z,1{+}z)\big)\,,
\end{align}
where the full functional expression in terms of HPLs in $w$ is given in the ancillary file. 

\subsubsection*{Three-cycle diagram}

Finally, the 3-cycle graph evaluates at the level of the symbol to 
\begin{align} \label{eq: I_3cyc}
 &{\rm Symbol}
 \left[\begin{tikzpicture}[baseline=-0.6ex,scale=1.,transform shape]
  \draw[Maroon, ultra thick] (0,-0.5)--(1.1,-0.5);
  \draw[Maroon, ultra thick] (0,-0.5)--(0,0.5);
  \draw[Maroon, ultra thick] (0,0.5)--(1.1,-0.5); 
  \draw[Maroon, ultra thick] (1.1,0.5)--(0,-0.5);
  \draw[Maroon, ultra thick] (1.1,-0.5)--(1.1,0.5)
    node[at start,black]{\Huge$\bullet$}; 
  \draw[Maroon, ultra thick] (0,0.5)--(1.1,0.5)
    node[at start,black]{\Huge$\bullet$}
    node[at end,black]{\Huge$\bullet$};
   \filldraw[white] (0,-0.5) circle (5pt) node[]{};
  \node[black] at (0,-0.5) {\Large$\otimes$};
\end{tikzpicture}\right]=24\,\big(4 \,\text{SB}(z,z,z,z,b(4),b(4))+\,\text{SB}(z,z,z,b(4),z,b(4))\nonumber\\&+3 \,\text{SB}(z,z,z,b(4),b(4),z)+2 \,\text{SB}(z,z,z,z,z,z)+\,\text{SB}(z,z,z,z,z,1{+}z)+11 \,\text{SB}(z,z,z,z,1{+}z,z)\nonumber\\&-10 \,\text{SB}(z,z,z,z,1{+}z,1{+}z)+9 \,\text{SB}(z,z,z,1{+}z,z,z)
+5 \,\text{SB}(z,z,z,1{+}z,z,1{+}z)\nonumber\\&-7 \,\text{SB}(z,z,z,1{+}z,1{+}z,z)-4 \,\text{SB}(z,z,z,1{+}z,1{+}z,1{+}z)+5 \,\text{SB}(z,z,1{+}z,z,z,z)\nonumber\\&+7 \,\text{SB}(z,z,1{+}z,z,z,1{+}z)-\,\text{SB}(z,z,1{+}z,z,1{+}z,z)-4 \,\text{SB}(z,z,1{+}z,z,1{+}z,1{+}z)\nonumber\\&-12 \,\text{SB}(z,z,1{+}z,1{+}z,z,z)-4 \,\text{SB}(z,z,1{+}z,1{+}z,z,1{+}z)-4 \,\text{SB}(z,z,1{+}z,1{+}z,1{+}z,z)\nonumber\\&+8 \,\text{SB}(z,z,1{+}z,1{+}z,1{+}z,1{+}z)\big) \,,
\end{align}
where again the full functional expression in terms of HPLs is provided in the ancillary file.

\subsubsection*{Comparison with Wilson loop}

If we sum all contributions according to \eqref{eq:omega4}, the terms with the spurious symbol letter $b(4)$ cancel and the result is given by 
\begin{align} \label{eq:F3}
-8\,\big(&15 \,\text{SB}(z,z,z,z,z,z)-6 \,\text{SB}(z,z,z,z,z,1{+}z)-5 \,\text{SB}(z,z,z,z,1{+}z,z)\nonumber\\&+6 \,\text{SB}(z,z,z,z,1{+}z,1{+}z)
-5 \,\text{SB}(z,z,z,1{+}z,z,z)+4 \,\text{SB}(z,z,z,1{+}z,z,1{+}z)\nonumber\\&+2 \,\text{SB}(z,z,z,1{+}z,1{+}z,z)
-2 \,\text{SB}(z,z,z,1{+}z,1{+}z,1{+}z)-6 \,\text{SB}(z,z,1{+}z,z,z,z)\nonumber\\&+4 \,\text{SB}(z,z,1{+}z,z,z,1{+}z)
+4 \,\text{SB}(z,z,1{+}z,z,1{+}z,z)-2 \,\text{SB}(z,z,1{+}z,z,1{+}z,1{+}z)\nonumber\\&+6 \,\text{SB}(z,z,1{+}z,1{+}z,z,z){-}2 \,\text{SB}(z,z,1{+}z,1{+}z,z,1{+}z)
{-}2 \,\text{SB}(z,z,1{+}z,1{+}z,1{+}z,z)\nonumber\\&+4 \,\text{SB}(z,z,1{+}z,1{+}z,1{+}z,1{+}z)\big)\,, 
\end{align}
which agrees with the symbol of the Wilson loop calculated in the literature~\cite{Henn:2019swt} (with an overall minus sign defined in \eqref{eq:F_from_geometry}).  An important open question is the cancellation mechanism for the spurious symbol letter $b(4)$. Namely, how exactly is this cancellation encoded in the cuts of the integrands?

\section{Resummation and strong coupling}
\label{sec:resum}

The negative geometries provide a new set of building blocks for the amplitude, and for the Wilson loop ${\cal F}(g,z)$. The important feature of the expansion is its term-wise IR finiteness, which is not manifest in the standard Feynman diagram approach for the Wilson loop. While the integrand is presented in this very special way, the actual process of integration is still cumbersome, as we do not currently have a specific method which would exploit the simplicity of the integrand. In fact, we rely on the canonical differential equation method \cite{Henn:2013pwa,Henn:2019swt} which is very powerful, but also very general, and not tailored to our problem.

However, there are certain classes of negative geometries which are solvable using a differential (boxing) equation method described in section \ref{sec:boxing}. It was shown that this method can be used for any negative geometry with an appropriate topology: one where the marked point sticks out like a lollipop from the graph.  The differential equation
\begin{equation}
    \Box\ \begin{tikzpicture}[baseline={-0.7ex}]
        \draw[Maroon, ultra thick] (0,0)--(1,0);
         \filldraw[white] (0,0) circle (5pt) node[]{};
        \node[black] at (0,0) {\Large$\otimes$};
        \draw[black, fill=gray] (1,0) circle [radius=0.5];
    \end{tikzpicture}\ =\ \begin{tikzpicture}[baseline={-0.7ex}]
        \node[white,left] at (0.63,0) {\Huge$\bullet$};
        \node[black,left] at (0.63,0) {\Large$\otimes$};
        \draw[black, fill=gray] (1,0) circle [radius=0.5];
    \end{tikzpicture} 
\end{equation}
is supplemented by the vanishing boundary condition at $z=-1$. In section \ref{sec:boxing}, we showed that this method can be used to solve for any tree negative geometry. 

\subsection{Tree resummation}

As discussed in ref.~\cite{Arkani-Hamed:2021iya}, this method works not only for any tree graph where the frozen loop $AB$ is connected to the rest of the graph through one link, but it can also be used for resumming all ladder graphs.
\begin{align}\label{eqn:ladder}
    \mathcal{F}_\text{ladder}(g,z) =&\ \begin{tikzpicture}[baseline={(0,-0.1)cm},scale=1.]
         \filldraw[white] (0,0) circle (5pt) node[]{};
        \node[black] at (0,0) {\Large$\otimes$};
    \end{tikzpicture}\ -(g^2)\ \begin{tikzpicture}[baseline={(0,-0.1)cm}]
        \draw[ultra thick, Maroon] (0,0)--(1,0) node[at end, black] {\Huge$\bullet$};
        \node[white] at (0,0) {\Huge$\bullet$};
        \node[black] at (0,0) {\Large$\otimes$};
    \end{tikzpicture}\ +(g^2)^2\ \begin{tikzpicture}[baseline={(0,-0.1)cm},scale=1.]
        \draw[ultra thick, Maroon] (0,0)--(2,0) node[at end, black] {\Huge$\bullet$} node[pos=0.5, black] {\Huge$\bullet$};
         \filldraw[white] (0,0) circle (5pt) node[]{};
        \node[black] at (0,0) {\Large$\otimes$};
    \end{tikzpicture}\nonumber\\
    &-(g^2)^3\ \begin{tikzpicture}[baseline={(0,-0.1)cm},scale=1.]
        \draw[ultra thick, Maroon] (0,0)--(3,0) node[at end, black] {\Huge$\bullet$} node[pos=0.33, black] {\Huge$\bullet$} node[pos=0.66, black] {\Huge$\bullet$};
         \filldraw[white] (0,0) circle (5pt) node[]{};
        \node[black] at (0,0) {\Large$\otimes$};
    \end{tikzpicture}\ +(g^2)^4\ \begin{tikzpicture}[baseline={(0,-0.1)cm},scale=1.]
        \draw[ultra thick, Maroon] (0,0)--(4,0) node[at end, black] {\Huge$\bullet$} node[pos=0.25, black] {\Huge$\bullet$} node[pos=0.5, black] {\Huge$\bullet$} node[pos=0.75, black] {\Huge$\bullet$};
        \filldraw[white] (0,0) circle (5pt) node[]{};
        \node[black] at (0,0) {\Large$\otimes$};
    \end{tikzpicture}\ +\cdots
\end{align}
%
which satisfies
\begin{equation}
\label{eq:box_on_ladder}
\Box \, {\cal F}_{\rm ladder}(g,z) = \tfrac{1}{2} (z\partial_z)^2  {\cal F}_{\rm ladder}(g,z) = - g^2 {\cal F}_{\rm ladder}(g,z).
\end{equation}
Imposing the boundary condition that the $g$-dependent part of the  result must vanish for $z=-1$, which is just a simple consequence of the vanishing of the integrand for $\la AB13\ra = \la AB24\ra = 0$, we get \cite{Arkani-Hamed:2021iya}
\begin{equation} \label{eq:F_ladder}
    {\cal F}_{\rm ladder}(g,z) = \frac{\cos{(\sqrt{2}g\log{z})}}{\cosh{(\sqrt{2}g \pi)}} \,,
\end{equation}
with ${\cal F}_{\rm ladder}(g,-1) = 1$ as required.

For a general tree graph, the marked point can be located inside the graph, and it can be connected to other vertices through multiple links. To resolve that situation, we realize that the sum of all tree graphs is the exponential of all graphs with the marked point sticking out on a link:
\begin{equation}
    {\cal F}_{\rm tree}(g,z) = e^{{\cal H}_{\rm tree}(g,z)}\,,
\end{equation}
where ${\cal F}_{\rm tree}(g,z)$ and ${\cal H}_{\rm tree}(g,z)$ differ in the way that marked point $AB$ can appear, 
\begin{equation}
    \mathcal{F}_{\rm tree}(g,z)\ =\ \begin{tikzpicture}[baseline={-0.6ex},scale=1.]
        \node[white,left] at (0.63,0) {\Huge$\bullet$};
        \node[black,left] at (0.63,0) {\Large$\otimes$};
        \draw[black, fill=gray] (1,0) circle [radius=0.5];
    \end{tikzpicture}\qquad \mathcal{H}_{\rm tree}(g,z)\ =\ \begin{tikzpicture}[baseline={-0.6ex},scale=1.]
        \draw[Maroon, ultra thick] (0,0)--(1,0);
        \filldraw[white] (0,0) circle (5pt) node[]{};
        \node[black] at (0,0) {\Large$\otimes$};
        \draw[black, fill=gray] (1,0) circle [radius=0.5];
    \end{tikzpicture} 
\end{equation}
The generating functional ${\cal H}_{\rm tree}(g,z)$ now satisfies the same Laplace equation because the marked point $AB$ is now connected through a single link to the rest of the graph,
\begin{equation}
    \frac{1}{2}(z\partial_z)^2 {\cal H}_{\rm tree}+g^2 e^{{\cal H}_{\rm tree}}=0\,.
\end{equation}
This was solved in \cite{Arkani-Hamed:2021iya} by
\begin{equation}
\label{eq:tree_final}
    {\cal F}_{\rm tree}(g,z)= \frac{A^2 z^A}{g^2(z^A+1)^2}\qquad \mbox{with}\quad  \frac{A}{2g \cos{\pi A/2}}=1\,.
\end{equation}
The result~\eqref{eq:tree_final} can be used to obtain ${\cal F}^{(n)}_{\rm tree}(g,z)$ for any order $n$ by asymptotically expanding the solution in powers of $g$. We refer the reader to \cite{Arkani-Hamed:2021iya} for more details. 

It is important to note that the Laplacian will collapse a {\it single} propagator connecting the frozen node to the rest of the diagram, even if the rest is not a tree. However, this statement is not true if the unintegrated node is attached to multiple propagators, such as the frozen corner of the 3-loop triangle (\ref{L3c}), since the boxing trick described earlier no longer works. One needs to search for new types of differential operators, perhaps analogous to those in ref.~\cite{Drummond:2010cz}, where the operators acted on the external momentum twistors, or by using methods in ref.~\cite{Henn:2023pkc}.

In ref.~\cite{Arkani-Hamed:2021iya} the resummed result for tree graphs was studied from the point of view of the strong-coupling limit. The leading behavior of $F_{\rm tree}(g,z)$ for $g\gg1$ is\footnote{Note the switch to the $F$ normalization defined in eq.~(\ref{eq:TwoFnormalizations}).} 
\begin{equation}
   F_{\rm tree} =- \frac{z}{(1+z)^2} + {\cal O}\left(\frac{1}{g}\right)\,,
\end{equation}
so we do not reproduce the leading $F_{\rm WLI}\sim g$ behavior, but we do preserve the $1/g$ expansion. See ref.~\cite{Arkani-Hamed:2021iya} for more discussion.

\subsection{Triangle plus ladder}

While it would be desirable to evaluate all one cycle diagrams and resum them to all loop order, this problem is too difficult with our current tools. We can instead turn to a much more modest task: resum a certain class of all one cycle graphs which have the same triangle core solved already in (\ref{L3c}). In particular, we are interested in the infinite set of diagrams of the form 
\begin{equation}
\mathcal{F}_\text{tri-lad}=
g^4\underbrace{\begin{tikzpicture}
[baseline={(0,-0.1)cm},scale=1.,transform shape]
    \draw[Maroon, ultra thick] (0,-0.75)--(0,0.75);
    \draw[Maroon, ultra thick](0,-0.75)--(0.75,0) node[at start,black] {\Huge$\bullet$}; 
    \draw[Maroon, ultra thick](0,0.75)--(0.75,0) node[at start,black] {\Huge$\bullet$};
   \filldraw[white] (0.75,0) circle (5pt) node[]{};
    \node[black] at (0.75,0) {\Large$\otimes$};
\end{tikzpicture}}_{\displaystyle\mathcal{F}_3}-g^6\underbrace{\begin{tikzpicture}[baseline={(0,-0.1)cm}]
    \draw[Maroon, ultra thick] (-0.75,0.75)--(-0.75,-0.75);
    \draw[Maroon, ultra thick] (0,0)--(-0.75,0.75) node[at end, black] {\Huge$\bullet$};
    \draw[Maroon, ultra thick] (0,0)--(-0.75,-0.75) node[at end, black] {\Huge$\bullet$};
    \draw[Maroon, ultra thick] (0,0)--(.75,0) node[at start, black] {\Huge$\bullet$};
   \filldraw[white] (0.75,0) circle (5pt) node[]{};
    \node[black] at (.75,0) {\Large$\otimes$};
\end{tikzpicture}}_{\displaystyle\mathcal{F}_{3,1}} + g^{8}  \underbrace{\begin{tikzpicture}[baseline={(0,-0.1)cm}]
    \draw[Maroon, ultra thick] (-0.75,0.75)--(-0.75,-0.75);
    \draw[Maroon, ultra thick] (0,0)--(-0.75,0.75) node[at end, black] {\Huge$\bullet$};
    \draw[Maroon, ultra thick] (0,0)--(-0.75,-0.75) node[at end, black] {\Huge$\bullet$};
    \draw[Maroon, ultra thick] (0,0)--(1.5,0) node[at start, black] {\Huge$\bullet$} node[pos=0.5, black] {\Huge$\bullet$};
    \filldraw[white] (1.5,0) circle (5pt) node[]{};
    \node[black] at (1.5,0) {\Large$\otimes$};
\end{tikzpicture}}_{\displaystyle\mathcal{F}_{3,2}} + \cdots
\end{equation}


We already saw two of these contributions in the previous section, the first one in eq.~\eqref{L3c} at $L=2$, and the second one in eq.~\eqref{eq:I3,1} as one of the diagrams in the $L=3$ result.
To obtain \eqref{eq:I3,1}, we evaluated the diagram using the differential equation,
\begin{equation}
\Box \,\Bigg(
\begin{tikzpicture}[baseline={(0,-0.1)cm}]
    \draw[Maroon, ultra thick] (-0.75,0.75)--(-0.75,-0.75);
    \draw[Maroon, ultra thick] (0,0)--(-0.75,0.75) node[at end, black] {\Huge$\bullet$};
    \draw[Maroon, ultra thick] (0,0)--(-0.75,-0.75) node[at end, black] {\Huge$\bullet$};
    \draw[Maroon, ultra thick] (0,0)--(.75,0) node[at start, black] {\Huge$\bullet$};
    \filldraw[white] (.75,0) circle (5pt);
    \node[black] at (.75,0) {\Large$\otimes$};
\end{tikzpicture}
\Bigg)
=
\begin{tikzpicture}[baseline={(0,-0.1)cm},scale=1.,transform shape]
    \draw[Maroon, ultra thick] (0,-0.75)--(0,0.75);
    \draw[Maroon, ultra thick] (0,-0.75)--(0.75,0) node[at start,black] {\Huge$\bullet$}; 
    \draw[Maroon, ultra thick] (0,0.75)--(0.75,0) node[at start,black] {\Huge$\bullet$};
    \filldraw[white] (0.75,0) circle (5pt);
    \node[black] at (0.75,0) {\Large$\otimes$};
\end{tikzpicture}
\end{equation}
hence we solve 
\begin{equation}
    \Box\,\mathcal{F}_{3,1}(z) = -\mathcal{F}_3(z) \,,
\end{equation}
where ${\cal F}_3(z)$ denotes the triangle $L=2$ formula (\ref{L3c}), and ${\cal F}_{3,1}(z)$ the triangle diagram with one extra segment. The boundary condition is again the vanishing of the result at $z=-1$ which fixes ${\cal F}_{3,1}(z)$ to (\ref{eq:I3,1}). The same approach continues to work for the next case of the the triangle with two segments ${\cal F}_{3,2}(z)$, 
\begin{equation}
\Box \,\Bigg(
\begin{tikzpicture}[baseline={(0,-0.1)cm}]
    \draw[Maroon, ultra thick] (-0.75,0.75)--(-0.75,-0.75);
    \draw[Maroon, ultra thick] (0,0)--(-0.75,0.75) node[at end, black] {\Huge$\bullet$};
    \draw[Maroon, ultra thick] (0,0)--(-0.75,-0.75) node[at end, black] {\Huge$\bullet$};
    \draw[Maroon, ultra thick] (0,0)--(1.5,0)
        node[at start, black] {\Huge$\bullet$}
        node[pos=0.5, black] {\Huge$\bullet$};
    \filldraw[white] (1.5,0) circle (5pt);
    \node[black] at (1.5,0) {\Large$\otimes$};
\end{tikzpicture}
\Bigg)
=
\begin{tikzpicture}[baseline={(0,-0.1)cm}]
    \draw[Maroon, ultra thick] (-0.75,0.75)--(-0.75,-0.75);
    \draw[Maroon, ultra thick] (0,0)--(-0.75,0.75) node[at end, black] {\Huge$\bullet$};
    \draw[Maroon, ultra thick] (0,0)--(-0.75,-0.75) node[at end, black] {\Huge$\bullet$};
    \draw[Maroon, ultra thick] (0,0)--(.75,0) node[at start, black] {\Huge$\bullet$};
    \filldraw[white] (0.75,0) circle (5pt);
    \node[black] at (0.75,0) {\Large$\otimes$};
\end{tikzpicture}
\end{equation}
with the solution~\cite{Brown:2023mqi}
\begin{align}
\label{eq:F32}
    \mathcal{F}_{3,2}(z) = &4 \Bigl[ 8H_{0,0,0,0,0,0,0,0}(z){+}8H_{0,0,0,0,{-}1,0,0,0}(z){-}16H_{0,0,0,0,{-}1,{-}1,0,0}(z){+}8H_{0,0,0,0,{-}2,0,0}(z) \nonumber\\
    &- 8 \zeta_3 (2H_{0,0,0,0,{-}1}(z)-H_{0,0,0,0,0}(z)) +4 \pi^2(H_{0,0,0,0,{-}1,0}(z)-2H_{0,0,0,0,{-}1,{-}1}(z) \nonumber\\
    &{+}H_{0,0,0,0,{-}2}(z)) {+}\frac{13 \pi^4}{1080} \log({z})^4 {+} \frac{C_{3,1}}{6} \log({z})^3{+} \frac{D_{3,1}}{2}\log({z})^2{+}C_{3,2} \log({z}) {+} D_{3,2} \Bigr] \,,
\end{align}
where
\begin{align}
        C_{3,1} &= 4 \pi ^2 \zeta_3 + 8 \zeta_5\,,\\
        D_{3,1} & = - 8 \zeta^2_3 + \frac{7 \pi ^6}{45}.
\end{align} 
Evaluating ${\cal F}_{3,2}(z {\rightarrow} -1)$ and setting it to 0, we can solve for $C_{3,2}$ and $D_{3,2}$ and get
\begin{align}
       C_{3,2} &=\frac{17 \pi ^4 \zeta_3}{15}+4 \pi ^2 \zeta_5+8 \zeta_7,\\
        D_{3,2} &=-16 \zeta_{5,3}-4\pi^2 \zeta^2_3-96 \zeta_5\zeta_3+\frac{4891 \pi^8}{75600}.
\end{align}
The same method extends to the general case of a triangle plus a ladder of arbitrary length $n$, ${\cal F}_{3,n}(z)$, which satisfies the equation,
\begin{equation}
    \Box\,\mathcal{F}_{3,n}(z) = -\mathcal{F}_{3,n{-}1}(z) \,.
\end{equation}

This equation is easy to solve as the $\Box$ acts easily on the HPLs on the right hand side of the equation. For the harmonic polylogarithms, ``unboxing'' amounts to adding two 0's on the vector argument, namely $(z \partial_z)^2 H_{\{0,0,\Vec{n} \}}(z)= H_{\Vec{n}}(z)$ (remember that $\Box\,=\frac{1}{2}(z \partial_z)^2$). For the constant and $(\log{z})^n$ terms, we can use the fact that $(z \partial_z)^2 (\log{z})^n = n(n-1)(\log{z})^{n-2}$. Finally, there are terms of the form $C \log{z} + D$ that vanish under $(z \partial_z)^2$, which we can fix by the boundary condition that any diagram with branches vanishes as $z \xrightarrow{} -1$. 
The result then reads,

\begin{align}
    \mathcal{F}_{3,n}(z) = (-2)^n \Biggl[ &8H_{\Vec{0}_{2n},0,0,0,0}(z){+}8H_{\Vec{0}_{2n},{-}1,0,0,0}(z){-}16H_{\Vec{0}_{2n},{-}1,{-}1,0,0}(z){+}8H_{\Vec{0}_{2n},0,{-}1,0,0}(z) \nonumber\\
    &
    -16\zeta_3\,H_{\Vec{0}_{2n},{-}1}(z)
    +4 \pi^2\Bigl(H_{\Vec{0}_{2n},{-}1,0}(z)-2 H_{\Vec{0}_{2n},{-}1,{-}1}(z)
    +H_{\Vec{0}_{2n},{-}2}(z)\Bigr) \nonumber\\
    &+ \frac{13 \pi^4}{45} \frac{1}{(2n)!}\log^{2n}(z) 
    +8\zeta_3 \frac{1}{(2n+1)!} \log^{2n+1}(z) \nonumber\\
    &{+} \sum^{n}_{k{=}1} \biggl( \frac{C_k}{[2(n{-}k){+}1]!} 
    \log^{2(n{-}k)+1}(z)
    {+} \frac{D_k}{[2(n{-}k)]!} \log^{2(n{-}k)}(z) \biggr) \Biggr]\,.
\end{align}
where the constants $C_k$ and $D_k$ satisfy recursive formulas as detailed in Appendix \ref{app:boxCoeff}. Let us now consider the following object
\begin{equation}
{\cal F}_\text{tri-lad}(g,z)
=
g^4
\begin{tikzpicture}[baseline={(0,-0.1)cm},scale=1.,transform shape]
    \draw[Maroon, ultra thick] (0,-0.75)--(0,0.75);
    \draw[Maroon, ultra thick] (0,-0.75)--(0.75,0) node[at start,black] {\Huge$\bullet$}; 
    \draw[Maroon, ultra thick] (0,0.75)--(0.75,0) node[at start,black] {\Huge$\bullet$};
    \filldraw[white] (0.75,0) circle (5pt);
    \node[black] at (0.75,0) {\Large$\otimes$};
\end{tikzpicture}
- g^6
\begin{tikzpicture}[baseline={(0,-0.1)cm}]
    \draw[Maroon, ultra thick] (-0.75,0.75)--(-0.75,-0.75);
    \draw[Maroon, ultra thick] (0,0)--(-0.75,0.75) node[at end, black] {\Huge$\bullet$};
    \draw[Maroon, ultra thick] (0,0)--(-0.75,-0.75) node[at end, black] {\Huge$\bullet$};
    \draw[Maroon, ultra thick] (0,0)--(.75,0) node[at start, black] {\Huge$\bullet$};
    \filldraw[white] (0.75,0) circle (5pt);
    \node[black] at (0.75,0) {\Large$\otimes$};
\end{tikzpicture}
+ g^{8}
\begin{tikzpicture}[baseline={(0,-0.1)cm}]
    \draw[Maroon, ultra thick] (-0.75,0.75)--(-0.75,-0.75);
    \draw[Maroon, ultra thick] (0,0)--(-0.75,0.75) node[at end, black] {\Huge$\bullet$};
    \draw[Maroon, ultra thick] (0,0)--(-0.75,-0.75) node[at end, black] {\Huge$\bullet$};
    \draw[Maroon, ultra thick] (0,0)--(1.5,0)
        node[at start, black] {\Huge$\bullet$}
        node[pos=0.5, black] {\Huge$\bullet$};
    \filldraw[white] (1.5,0) circle (5pt);
    \node[black] at (1.5,0) {\Large$\otimes$};
\end{tikzpicture}
+ \cdots
\end{equation}
where the index `tri-lad' refers to the resummed triangle-ladder series. The resummation can be done in two different ways: either the direct resummation, or using the same differential operator applied on the whole series as was done in the earlier case of the ``tree level'' ladder resummation. Denoting the resummed series by ${\cal F}_\text{tri-lad}(g,z)$, we have the differential equation
\begin{equation}
\label{eq:tl_diffeq}
    \Box\ {\cal F}_\text{tri-lad}(g,z)+g^2{\cal F}_\text{tri-lad}(g,z) = g^4 (\Box\,\mathcal{F}_3(z))\,,
\end{equation}
where the right hand side is the differential operator acting on the triangle function ${\cal F}_3(z)$ -- note that ${\cal F}_3(z)$ does not satisfy the Laplace equation, so $(\Box\,{\cal F}_3(z))$ here plays a role of the source term. Explicitly, this source term is 
\begin{equation}
\label{eq:BoxF3}
    \Box \mathcal{F}_3(z) = \frac{4 z \left({-}2 H_{{-}1,0,0}+H_{0,0,0}{-}2 \zeta_3+\pi ^2\right){-}4 \pi ^2 z H_{{-}1}+2 \pi ^2 z H_0+4 (z (z{+}4){+}1) H_{0,0}}{(z+1)^2}\,.
\end{equation}

The differential equation~(\ref{eq:tl_diffeq}) is too complicated for us to solve exactly using the standard toolkit. One approach we can take is to solve the equation in Mellin space, by replacing $f(z) \to f(\nu) = \int_0^1 dz z^\nu f(z)$.  In this space, $\Box$ acts as scalar multiplication by some parameter $\tfrac{1}{2}\nu^2$, and the HPLs turn into harmonic sums. We can then take the inverse transform and express the solution in terms of infinite sums. We can express most of these sums in an analytical form, and take the limit as $g \gg 1$ to get a first order approximation of those we can't. This procedure is outlined in further detail in Appendix \eqref{app:DiffEq}. 

For strong $g$ and $z>0$, we switch to the $F(g,z)$ notation. Recall that $F(g,z)$ differs from ${\cal F}(g,z)$ just by the overall factor of $-g^2$ in eq.~(\ref{eq:TwoFnormalizations}), but the $F$ normalization is used in other literature. The final result then goes like
\begin{align}
 & F_\text{tri-lad}(g,z)|_{g \gg 1}\sim -g^4\frac{\left(2 (z (z+4)+1) \log ^2(z)+2 \pi ^2 z \log (z)-4 \pi ^2 z \log (z+1)\right)}{(z+1)^2} \nonumber\\ 
        &{-}\frac{2 g^4 z \left(12 \text{Li}_3({-}z){-}12 \text{Li}_2({-}z) \log (z){+}\log ^3(z){-}6 \log (z{+}1) \log ^2(z){-}12 \zeta_3{+}6 \pi ^2\right)}{3 (z+1)^2}{+} \mathcal{O}(g^3) \,.
\end{align}

This expression, at the leading order in $g$,
when put into HPL form, turns out to be the source term~(\ref{eq:BoxF3}), suppressed by a factor of $\frac{1}{g^2}$. We defer the discussion of why this is the case to Appendix \eqref{app:DiffEq}. It is worth noting here that the leading behavior at $g\gg 1$ is $F_\text{tri-lad}\sim g^4$, 
which grows much faster than the full Wilson loop behavior $F_{WLI}\sim g$. On the other hand, the series is convergent and the final behavior is weaker than the behavior of the leading term at weak coupling, which is the triangle diagram $\mathcal{F}_3$, which contributes to $F$ as $\sim g^6$. We call such a series \emph{partially summable}.

\subsection{Box plus ladder}

The same procedure can be also used to determine the negative geometries that have the form of a box diagram attached to a ladder:
\begin{equation}
\begin{tikzpicture}[baseline={(0,-0.1)cm}]
    \draw[Maroon, ultra thick] (0,0)--(1,0);
    \draw[Maroon, ultra thick] (1,1)--(1,0);
    \draw[Maroon, ultra thick] (1,1)--(0,1);
    \draw[Maroon, ultra thick] (0,0)--(0,1);
    \draw[Maroon, ultra thick] (1,0)--(3,0);
    \node[Maroon] at (3.3,0) {\small$\bullet$};
    \node[Maroon] at (3.5,0) {\small$\bullet$};
    \node[Maroon] at (3.7,0) {\small$\bullet$};
    \draw[Maroon, ultra thick] (4,0)--(5,0);
    \node at (0,0) {\Huge$\bullet$};
    \node at (1,0) {\Huge$\bullet$};
    \node at (0,1) {\Huge$\bullet$};
    \node at (1,1) {\Huge$\bullet$};
    \node at (2,0) {\Huge$\bullet$};
    \node at (3,0) {\Huge$\bullet$};
    \node at (4,0) {\Huge$\bullet$};
    \filldraw[white] (5,0) circle (5pt);
    \node at (5,0) {\Large$\otimes$};
\end{tikzpicture}
\end{equation}
Just like with the triangle, we are interested in a series of the form
\begin{center}
    \begin{align}
       {\cal F}_{rs,4}(g,z)=g^6\underbrace{\begin{tikzpicture}[baseline={(0,-0.1)cm}]
        \draw[Maroon, ultra thick] (0,-0.5)--(1,-0.5);
        \draw[Maroon, ultra thick] (1,0.5)--(1,-0.5);
        \draw[Maroon, ultra thick] (1,0.5)--(0,0.5);
        \draw[Maroon, ultra thick] (0,-0.5)--(0,0.5);
        \draw[Maroon, ultra thick] (1,-0.5)--(1,-0.5);
        \node at (0,-0.5) {\Huge$\bullet$};
        \node at (0,0.5) {\Huge$\bullet$};
        \node at (1,0.5) {\Huge$\bullet$};
       \filldraw[white] (1,-0.5) circle (5pt) node[]{};
        \node at (1,-0.5) {\Large$\otimes$};
    \end{tikzpicture}}_{\displaystyle\mathcal{F}_4} - g^{8} \underbrace{\begin{tikzpicture}[baseline={(0,-0.1)cm}]
        \draw[Maroon, ultra thick] (0,-0.5)--(1,-0.5);
        \draw[Maroon, ultra thick] (1,0.5)--(1,-0.5);
        \draw[Maroon, ultra thick] (1,0.5)--(0,0.5);
        \draw[Maroon, ultra thick] (0,-0.5)--(0,0.5);
        \draw[Maroon, ultra thick] (1,-0.5)--(2,-0.5);
        \node at (0,-0.5) {\Huge$\bullet$};
        \node at (1,-0.5) {\Huge$\bullet$};
        \node at (0,0.5) {\Huge$\bullet$};
        \node at (1,0.5) {\Huge$\bullet$};
        \filldraw[white] (2,-0.5) circle (5pt) node[]{};
        \node at (2,-0.5) {\Large$\otimes$};
\end{tikzpicture}}_{\displaystyle \mathcal{F}_{4,1}} + g^{10}  \underbrace{\begin{tikzpicture}[baseline={(0,-0.1)cm}]
        \draw[Maroon, ultra thick] (0,-0.5)--(1,-0.5);
        \draw[Maroon, ultra thick] (1,0.5)--(1,-0.5);
        \draw[Maroon, ultra thick] (1,0.5)--(0,0.5);
        \draw[Maroon, ultra thick] (0,-0.5)--(0,0.5);
        \draw[Maroon, ultra thick] (1,-0.5)--(3,-0.5);
        \node at (0,-0.5) {\Huge$\bullet$};
        \node at (1,-0.5) {\Huge$\bullet$};
        \node at (0,0.5) {\Huge$\bullet$};
        \node at (1,0.5) {\Huge$\bullet$};
        \node at (2,-0.5) {\Huge$\bullet$};
       \filldraw[white] (3,-0.5) circle (5pt) node[]{};
        \node at (3,-0.5) {\Large$\otimes$};
\end{tikzpicture}}_{\displaystyle \mathcal{F}_{4,2}}+ \cdots
    \end{align}
\end{center}


We can start with the simplest diagram of the box with one segment, ${\cal F}_{4,1}(z)$ which satisfies a differential equation
\begin{equation}
    \Box\left(\begin{tikzpicture}[baseline={(0,-0.1)cm}]
        \draw[Maroon, ultra thick] (0,-0.5)--(1,-0.5);
        \draw[Maroon, ultra thick] (1,0.5)--(1,-0.5);
        \draw[Maroon, ultra thick] (1,0.5)--(0,0.5);
        \draw[Maroon, ultra thick] (0,-0.5)--(0,0.5);
        \draw[Maroon, ultra thick] (1,-0.5)--(2,-0.5);
        \node at (0,-0.5) {\Huge$\bullet$};
        \node at (1,-0.5) {\Huge$\bullet$};
        \node at (0,0.5) {\Huge$\bullet$};
        \node at (1,0.5) {\Huge$\bullet$};
       \filldraw[white] (2,-0.5) circle (5pt) node[]{};
        \node at (2,-0.5) {\Large$\otimes$};
    \end{tikzpicture}\right) = \begin{tikzpicture}[baseline={(0,-0.1)cm}]
        \draw[Maroon, ultra thick] (0,-0.5)--(1,-0.5);
        \draw[Maroon, ultra thick] (1,0.5)--(1,-0.5);
        \draw[Maroon, ultra thick] (1,0.5)--(0,0.5);
        \draw[Maroon, ultra thick] (0,-0.5)--(0,0.5);
        \draw[Maroon, ultra thick] (1,-0.5)--(1,-0.5);
        \node at (0,-0.5) {\Huge$\bullet$};
        \node at (0,0.5) {\Huge$\bullet$};
        \node at (1,0.5) {\Huge$\bullet$};
        \node[white] at (1,-0.5) {\Huge$\bullet$};
        \node at (1,-0.5) {\Large$\otimes$};
    \end{tikzpicture}
\end{equation}
or $\Box\, \mathcal{F}_{4,1}(z) = -\mathcal{F}_4(z)$ where $\mathcal{F}_4(z)$ is the box $L=3$ diagram evaluated in (\ref{eq:box_funct}). The solution for $\mathcal{F}_{4,1}(z)$ is then, given the same vanishing boundary condition,
\begin{align}
&\mathcal{F}_{4,1}(z)=-\Big(-\frac{32}{3} \pi ^2 \zeta_3 H_{-3}-256 \zeta_3 H_{-4,0}+128 \zeta_3 H_{-3,0,0}+48 \pi ^2 \zeta_3 H_{0,0,0}+64 \zeta_3 H_{0,0,0,0,0} \nonumber \\
&{-}608 \zeta_5 H_{{-}3}+496 \zeta_5 H_{0,0,0}+24 \pi ^2 H_{{-}6}{-}\frac{86}{9} \pi ^4 H_{{-}4}+16 \pi ^2 H_{{-}5,0}+32 \pi ^2 H_{{-}4,{-}2}{-}80 \pi ^2 H_{{-}3,{-}3} \nonumber \\
&+4 \pi ^4 H_{-3,-1}+\frac{352}{45} \pi ^4 H_{-3,0}+48 H_{-6,0,0}-\frac{64}{3} \pi ^2 H_{-4,-1,0}-\frac{40}{3} \pi ^2 H_{-4,0,0}-\frac{32}{3} \pi ^2 H_{-3,-2,0} \nonumber \\
&{+}96 H_{{-}5,0,0,0}{+}64 H_{{-}4,{-}2,0,0}{-}160 H_{{-}3,{-}3,0,0}{+}16 \pi ^2 H_{{-}3,{-}1,0,0}{+}\frac{112}{3} \pi ^2 H_{{-}3,0,0,0}{+}\frac{20}{9} \pi ^4 H_{0,0,0,0}\nonumber \\
&-128 H_{-4,-1,0,0,0}-16 H_{-4,0,0,0,0}-64 H_{-3,-2,0,0,0}+96 H_{-3,-1,0,0,0,0}+64 H_{-3,0,0,0,0,0} \nonumber \\
&+\frac{16}{3} \pi ^2 H_{0,0,0,0,0,0}+96 H_{0,0,0,0,0,0,0,0}+176 \zeta^2_3 \log ^2(z)+\frac{208}{15} \pi^4 \zeta_3 \log (z)+\frac{856}{3} \pi ^2 \zeta_5 \log (z) \nonumber \\
&+1216 \zeta_7 \log (z)+\frac{505}{567} \pi ^6 \log ^2(z)-608\zeta_{5,3}+160 \pi^2 \zeta^2_3-1248 \zeta_3 \zeta_5+\frac{157517 \pi ^8}{170100}
 \Big)
\end{align}
The general diagram ${\cal F}_{4,n}(z)$ again satisfies the same differential equation $\Box\,{\cal F}_{4,n}(z) = -{\cal F}_{4,n{-}1}(z)$ with the solution:
\begin{align}
    \mathcal{F}_{4,n}(z)=&({-}2)^{n{+}4}\bigg[3 H_{\vec{0}_{2 n{+}6}}(z)+ 2 H_{\vec{0}_{2n},{-}1,\vec{0}_5}(z) + 3 H_{\vec{0}_{2n}, {-}1, {-}1, \vec{0}_4}(z)-\frac{1}{2} H_{\vec{0}_{2 n{+}1},{-}1,\vec{0}_4}(z)\nonumber\\
    &{-}4 H_{\vec{0}_{2 n+1},{-}1,{-}1,\vec{0}_3}(z){+}3
   H_{
   \vec{0}_{2 n+2},-1,\vec{0}_3}(z){+}\frac{3}{2} H_{\vec{0}_{2 n+3},-1,0,0}(z){-}2 H_{\vec{0}_{2
   n},-1,-2,\vec{0}_3}(z)\nonumber\\
   &{+}2 H_{\vec{0}_{2 n+1},-1,-2,0,0}(z){+}\frac{1}{6} \pi ^2 H_{\vec{0}_{2
   n+4}}(z){+}\frac{7}{6} \pi ^2 H_{\vec{0}_{2 n},-1,\vec{0}_3}(z){+}\frac{1}{2} \pi ^2 H_{\vec{0}_{2
   n},-1,-1,0,0}(z)\nonumber\\
   &-5 H_{\vec{0}_{2 n},-1,-3,0,0}(z)-\frac{5}{12} \pi ^2 H_{\vec{0}_{2
   n+1},-1,0,0}(z)-\frac{2}{3} \pi ^2 H_{\vec{0}_{2 n+1},-1,-1,0}(z)\nonumber\\
   &+\frac{1}{2} \pi ^2 H_{\vec{0}_{2
   n+2},-1,0}(z)+\frac{3}{4} \pi ^2 H_{\vec{0}_{2 n+3},-1}(z)+2 \zeta_3 H_{\vec{0}_{2 n+3}}(z)+4 \zeta_3 H_{\vec{0}_{2 n},-1,0,0}(z)\nonumber\\
   &-8 \zeta_3 H_{\vec{0}_{2 n+1},-1,0}(z)-\frac{1}{3} \pi ^2 H_{\vec{0}_{2
   n},-1,-2,0}(z)+\pi ^2 H_{\vec{0}_{2 n+1},-1,-2}(z)+\frac{5}{72} \pi ^4 H_{\vec{0}_{2
   n+2}}(z)\nonumber\\
   &{+}\frac{11}{45} \pi ^4 H_{\vec{0}_{2 n},-1,0}(z){+}\frac{1}{8} \pi ^4 H_{\vec{0}_{2
   n},-1,-1}(z){-}\frac{5}{2} \pi ^2 H_{\vec{0}_{2 n},-1,-3}(z){-}\frac{43}{144} \pi ^4 H_{\vec{0}_{2
   n+1},-1}(z)\nonumber\\
   &+\frac{31}{2} \zeta_5 H_{\vec{0}_{2 n+1}}(z)-19 \zeta_5 H_{\vec{0}_{2 n},-1}(z)+\frac{3}{2}
   \pi ^2 \zeta_3 H_{\vec{0}_{2 n+
   1}}(z)-\frac{1}{3} \pi ^2 \zeta_3 H_{\vec{0}_{2 n},-1}(z)\nonumber\\
   &{+}11 \zeta_3^2
   H_{\vec{0}_{2 n}}(z){+}\frac{505 \pi ^6 H_{\vec{0}_{2 n}}(z)}{9072}{+}\sum_{k{=}1}^n \hat{C}_{k} \frac{\log^{2(n{-}k){+}1}(z)}{(2(n{-}k){+}1)!} {+}\sum_{k{=}1}^n \hat{D}_{k} \frac{\log^{2(n{-}k)}(z)}{(2(n{-}k))!}\bigg]\,.
\end{align}
The coefficients $\hat{C}_n$ and $\hat{D}_n$ are determined recursively using the boundary condition $I_{4,n}(z=-1)=0$ in (\ref{eq2}) and are given in Appendix \ref{app:boxCoeff}. In principle, we can use the same procedure to `extend' any $L=3$ diagram by attaching a ladder to the marked point and evaluate the all-loop series, but a more detailed understanding of the infinite series is necessary to write some more explicit formulas for the resummed $\cal F$. 

\section{Leading IR divergence}
\label{sec:cusp}

The Wilson loop with Lagrangian insertion ${\cal F}(g,z)$ is an IR finite object. When we integrate over the insertion point $z$ (the frozen loop momentum), we recover the logarithm of the amplitude which is $1/\epsilon^2$ divergent. The coefficient of this divergence is the cusp anomalous dimension $\Gamma_{\rm cusp}$,
\begin{equation}
\log A_4 = \int_{AB} {\cal F}(g,z)\ \propto \ \frac{\Gamma_{\rm cusp}(g)}{\epsilon^2} + {\cal O}\left(\frac{1}{\epsilon}\right)\,.
\end{equation}
The cusp anomalous dimension is a fascinating object; see refs.~\cite{Beisert:2006ez,Bern:2006ew,Henn:2019swt,Kruczenski:2002fb,Benna:2006nd,Basso:2007wd,Roiban:2007dq} for detailed studies at both weak and strong coupling. In a sense, it is a reduction of the full ${\cal F}(g,z)$ to a function of only the coupling, with no kinematical dependence. Fortunately, we have an access to all-loop order expression for this object from integrability~\cite{Beisert:2006ez}. On the other hand, it is not currently known how to obtain ${\cal F}(g,z)$ from integrability. 

Current amplitude-based methods are not powerful enough to reproduce $\Gamma_{\rm cusp}$ to all loop orders from first principles. That goal would require constructing an all-loop order integrand and performing integration (of the logarithm). It remains a big open question to derive the formula for $\Gamma_{\rm cusp}$ from an amplitudes perspective. We would need some deeper insight into exactly how $\Gamma_{\rm cusp}$ is encoded in the amplitude -- and an ability to extract $\Gamma_{\rm cusp}$ without constructing the all-loop order integrand and integrating it. The negative-geometry expansion, and the construction of the amplitude in momentum-twistor space in general, require a strict $D=4$ framework. Hence we are not allowed to use dimensional regularization, which in principle requires the knowledge of $D\neq 4$ information about the integrand. 

Fortunately, the Wilson loop function ${\cal F}(g,z)$ is IR finite and hence it can be evaluated strictly in $D=4$, for example using our geometry expansion. If we were to integrate over the insertion point, regularization would again be needed. Of course, one is not required to use dimensional-regularization. One can potentially choose another framework such as a deformed amplituhedron which is linked to Coulomb branch amplitudes \cite{Arkani-Hamed:2023epq,Flieger:2025ekn,Alday:2009zm}. Fortunately, there is another way to extract $\Gamma_{\rm cusp}$ from ${\cal F}(g,z)$, without the need for {\it any} regularization. The prescription was first found in refs~\cite{Alday:2013ip,Henn:2019swt}; we use another version of the prescription~\cite{Arkani-Hamed:2021iya}, which introduces an operator ${\cal I}$ that leads to an integration over $z$.\footnote{The operator ${\cal I}$ should not be confused with integrands ${\cal I}$ elsewhere in this paper.}

The first step is to determine the action of the ${\cal I}$ operator on the $\cal F$ function as a contour integral in the complex $z$-plane,
\begin{equation}
    {\cal I}({\cal F}) = -\frac{g^2}{2\pi i}\int_{C} \frac{dz}{z}\cal{F}\,.
\end{equation}
Assuming that ${\cal F}$ only has branch cuts on the negative $z$-axis and is analytic elsewhere, the contour can be deformed to a unit circle around the origin,
\begin{equation}\label{Iintegral}
    {\cal I}({\cal F}) = -\frac{g^2}{2\pi}\int_{-\pi}^\pi d\phi\,{\cal F}(e^{i\phi})\,.
\end{equation}
This function then enters a simple first order differential equation for $\Gamma_{\rm cusp}(g)$,
\begin{equation}
    g\frac{\partial}{\partial g}\Gamma_{\rm cusp}(g) = -8\,{\cal I}( {\cal F}(g,z))\,.
\end{equation}
This equation does not require any regularization: It can be solved for $\Gamma_{\rm cusp}(g)$ without using dimensional regularization, staying entirely in $D=4$. In perturbation theory, we expand
\begin{equation}
\label{eq:cusp_pert}
\Gamma_{\rm cusp}(g) = \sum_{L=1}^\infty g^{2L} \, \Gamma_{\rm cusp}^{(L)}  \,,
\end{equation}
so a contribution to the $L$-loop cusp anomalous dimension is given by dividing by $2L$,
\begin{equation}
\label{eq:cusp_pert2}
\Gamma_{\rm cusp}^{(L)} = - \frac{4}{L}\,{\cal I}( {\cal F}^{(L-1)}(z)) = \frac{2}{\pi L}\int_{-\pi}^\pi d\phi\,{\cal F}^{(L-1)}(e^{i\phi})\,.
\end{equation}
We quickly review below how $\mathcal{I}$ from \eqref{Iintegral} acts on the functions we see in negative geometries. 
\par
All the individual $\mathcal{F}(z)$ we consider here contain HPLs, natural logs, and transcendental constants. For constants, we simply have $\mathcal{I}[c]=-g^2c$. The contour that we use for $\mathcal{I}$ avoids the negative $z-$axis that logs and their powers have branch cuts on, so we can integrate them along the circle as
\begin{equation}
    \mathcal{I}(\log(z)^n) = -\frac{g^2}{2\pi} \int^\pi_{-\pi} (i\phi)^n d\phi = \frac{-g^2\pi^n \cos \left(\frac{\pi  n}{2}\right)}{n+1}
\end{equation}
Finally, we have HPLs that show up in our expressions, starting with single-cycle diagrams. The HPLs we find either have $z$ in their argument, or the spurious letter $b(4)=\frac{1+i\sqrt{z}}{1-i\sqrt{z}}$. Let's first discuss how to deal with the former, for which we use the notation $H_{-m_n,\cdots,-m_1}(z)=H_{-m_n,\cdots,-m_1}$ (the argument is omitted for brevity). Only the HPLs that have 0 as their rightmost index are singular at $z \rightarrow 0$, so we first use the shuffle identities \cite{REMIDDI_2000} to trade them for logs. For example, we can re-write $H_{-m,0}(z)$ as 
\begin{equation}
    H_{-m,0}(z) = \log(z) H_{-m} - m H_{0,-m}(z),
\end{equation}
and $H_{-m,0,0}(z)$ as
\begin{equation}
     H_{-m,0,0}(z) = \frac{1}{2}\log^2(z) H_{-m}(z) - m \log(z) H_{0,-m}(z) + \frac{m(m+1)}{2}H_{0,0,-m}(z),
\end{equation}
and so forth. After getting rid of all the trailing 0's, we have powers of logs multiplied by HPLs that are regular at $z=0$. Thus, all the remaining HPLs have a convergent Taylor series expansion for $z \in (-1,1)$. So we can apply eq.~(5.4) of ref.~\cite{Henn:2019swt} to these terms, obtaining
\begin{equation} \label{eq:hpllogint}
    \mathcal{I}(\log^a(z)H_{-m_n,\cdots ,-m_1}) =  g^2(-1)^{-a} a! \sum_{k=0}^{a} \frac{\pi^{k-1}}{k!} \sin\left(\frac{\pi k}{2}\right) H_{-(a+1-k+m_n),\ldots,-m_1}(-1).
\end{equation}

Starting at four loops, we also get diagrams with the spurious letter $b(4)$, such as \eqref{eq: I_2cyc,1} and \eqref{eq: I_3cyc}. In such cases, we can find a combination of diagrams that is free of the spurious letter. We first apply the above procedure to find how $\mathcal{I}$ acts on the combination. Then we subtract a piece we know, using the independence of $\mathcal{I}$ on which node is frozen. For example, consider \eqref{eq: I_2cyc,1}. The following sum can be expressed in a form that only contains HPLs with $z$ in the argument
\begin{equation}
    -\begin{tikzpicture}[baseline=-0.6ex,scale=1.2,transform shape]
    \draw[Maroon, ultra thick] (0.5,-0.5)--(0.5,0.5);
    \draw[Maroon, ultra thick](0.5,-0.5)--(1,0) node[at start,black] {\huge$\bullet$}; 
    \draw[Maroon, ultra thick](0.5,0.5)--(1,0) node[at start,black] {\huge$\bullet$};
    \draw[Maroon, ultra thick](1,0)--(1.5,0) node[at end,black] {\huge$\bullet$} node[at end, right]{};
    \node[black] at (1.0,0) {\huge$\bullet$};
     \filldraw[white] (0.5,0.5) circle (4pt) node[]{};
    \node[black] at (0.5,0.5) {$\otimes$};
    \end{tikzpicture}+\frac{1}{2!}\begin{tikzpicture}[baseline=-0.6ex,scale=1.,transform shape]
  \draw[Maroon, ultra thick] (0,-0.5)--(1,-0.5);
  \draw[Maroon, ultra thick] (0,-0.5)--(0,0.5);
  \draw[Maroon, ultra thick] (0,0.5)--(1,-0.5);
  \draw[Maroon, ultra thick] (1,-0.5)--(1,0.5)
    node[at start,black]{\Huge$\bullet$};
  \draw[Maroon, ultra thick] (0,0.5)--(1,0.5)
    node[at start,black]{\Huge$\bullet$}
    node[at end,black]{\Huge$\bullet$};
  \filldraw[white] (0,-0.5) circle (5pt) node[]{};
  \node[black] at (0,-0.5) {\Large$\otimes$};
\end{tikzpicture}.
\end{equation}
While the diagram on the left also contains a spurious letter, $\mathcal{I}$ should act on it the same way it does on the graph \eqref{eq:I3,1}, which differs only by the location of the frozen node. We know how to integrate graph \eqref{eq:I3,1} using \eqref{eq:hpllogint}. Then, we get
\begin{equation}
    \mathcal{I}\left(\frac{1}{2!}\begin{tikzpicture}[baseline=-0.6ex,scale=1.,transform shape]
  \draw[Maroon, ultra thick] (0,-0.5)--(1,-0.5);
  \draw[Maroon, ultra thick] (0,-0.5)--(0,0.5);
  \draw[Maroon, ultra thick] (0,0.5)--(1,-0.5);
  \draw[Maroon, ultra thick] (1,-0.5)--(1,0.5)
    node[at start,black]{\Huge$\bullet$};
  \draw[Maroon, ultra thick] (0,0.5)--(1,0.5)
    node[at start,black]{\Huge$\bullet$}
    node[at end,black]{\Huge$\bullet$};
  \filldraw[white] (0,-0.5) circle (5pt) node[]{};
  \node[black] at (0,-0.5) {\Large$\otimes$};
\end{tikzpicture} \right) = \mathcal{I} \left(-\begin{tikzpicture}[baseline=-0.6ex,scale=1.2,transform shape]
    \draw[Maroon, ultra thick] (0.5,-0.5)--(0.5,0.5);
    \draw[Maroon, ultra thick](0.5,-0.5)--(1,0) node[at start,black] {\huge$\bullet$}; 
    \draw[Maroon, ultra thick](0.5,0.5)--(1,0) node[at start,black] {\huge$\bullet$};
    \draw[Maroon, ultra thick](1,0)--(1.5,0) node[at end,black] {\huge$\bullet$} node[at end, right]{};
    \node[black] at (1.0,0) {\huge$\bullet$};
     \filldraw[white] (0.5,0.5) circle (4pt) node[]{};
    \node[black] at (0.5,0.5) {$\otimes$};
    \end{tikzpicture}+\frac{1}{2!}\begin{tikzpicture}[baseline=-0.6ex,scale=1.,transform shape]
  \draw[Maroon, ultra thick] (0,-0.5)--(1,-0.5);
  \draw[Maroon, ultra thick] (0,-0.5)--(0,0.5);
  \draw[Maroon, ultra thick] (0,0.5)--(1,-0.5);
  \draw[Maroon, ultra thick] (1,-0.5)--(1,0.5)
    node[at start,black]{\Huge$\bullet$};
  \draw[Maroon, ultra thick] (0,0.5)--(1,0.5)
    node[at start,black]{\Huge$\bullet$}
    node[at end,black]{\Huge$\bullet$};
  \filldraw[white] (0,-0.5) circle (5pt) node[]{};
  \node[black] at (0,-0.5) {\Large$\otimes$};
\end{tikzpicture}\right) + \mathcal{I} \left( \begin{tikzpicture}[baseline=-0.6ex,scale=1.2,transform shape]
    \draw[Maroon, ultra thick] (0.5,-0.5)--(0.5,0.5);
    \draw[Maroon, ultra thick](0.5,-0.5)--(1,0) node[at start,black] {\huge$\bullet$}; 
    \draw[Maroon, ultra thick](0.5,0.5)--(1,0) node[at start,black] {\huge$\bullet$};
    \draw[Maroon, ultra thick](1,0)--(1.5,0) node[at end,black] {\huge$\bullet$} node[at end, right]{};
    \node[black] at (1.0,0) {\huge$\bullet$};
    \node[white] at (1.5,0) {\huge$\bullet$};
    \filldraw[white] (1.5,0) circle (4pt) node[]{};
    \node[black] at (1.5,0) {$\otimes$};
    \end{tikzpicture}\right)
\end{equation}
Similarly, graph \eqref{eq: I_3cyc} can be combined with the 2-cycle diagram we get from above to form a combination that is free of spurious letters and that we can easily integrate to determine 
\begin{equation}
    \mathcal{I}\left(-\frac{1}{2!}\begin{tikzpicture}[baseline=-0.6ex,scale=1.,transform shape]
  \draw[Maroon, ultra thick] (0,-0.5)--(0,0.5);
  \draw[Maroon, ultra thick] (0,0.5)--(1,-0.5);
  \draw[Maroon, ultra thick] (0,-0.5)--(1,-0.5)
    node[at start,black]{\Huge$\bullet$};
  \draw[Maroon, ultra thick] (1,-0.5)--(1,0.5)
    node[at start,black]{\Huge$\bullet$};
  \draw[Maroon, ultra thick] (0,0.5)--(1,0.5)
    node[at end,black]{\Huge$\bullet$};
  \filldraw[white] (0,0.5) circle (5pt) node[]{};
  \node[black] at (0,0.5) {\Large$\otimes$};
\end{tikzpicture}+\frac{1}{3!} \begin{tikzpicture}[baseline=-0.6ex,scale=1.,transform shape]
  \draw[Maroon, ultra thick] (0,-0.5)--(1,-0.5);
  \draw[Maroon, ultra thick] (0,-0.5)--(0,0.5);
  \draw[Maroon, ultra thick] (0,0.5)--(1,-0.5); 
  \draw[Maroon, ultra thick] (1,0.5)--(0,-0.5);
  \draw[Maroon, ultra thick] (1,-0.5)--(1,0.5)
    node[at start,black]{\Huge$\bullet$}; 
  \draw[Maroon, ultra thick] (0,0.5)--(1,0.5)
    node[at start,black]{\Huge$\bullet$}
    node[at end,black]{\Huge$\bullet$};
  \filldraw[white] (0,-0.5) circle (5pt) node[]{};
  \node[black] at (0,-0.5) {\Large$\otimes$};
\end{tikzpicture} \right) = 40 \zeta^2_3+\frac{4 \pi ^6}{135}.
\end{equation}
Once we determine the 2-cycle piece, we can feed it into the equation above to extract the 3-cycle one.

The purpose of the remainder of this section is to compare how various negative geometries contribute to $\Gamma_{\rm cusp}(g)$, and to see whether any of them dominate over the others.
\subsection{Three- and four-loop \texorpdfstring{$\Gamma_{\rm cusp}$}{ gamma\_cusp}}

From eq.~(\ref{eq:cusp_pert2}), the cusp anomalous dimension at $L+1$ loops can be computed from ${\cal F}(g,z)$ at $L$ loops.
The negative geometry expansion for ${\cal F}(g,z)$ up to three loops (as needed to compute $\Gamma_{\rm cusp}$ through four loops) reads,
\begin{align}
&{\cal F}(g,z)=   \begin{tikzpicture}[baseline=-0.6ex,scale=1.,transform shape]
  \node[white] at (0,0) {\Huge$\bullet$};
  \node[black] at (0,0) {\Large$\otimes$};
\end{tikzpicture}
-g^2\begin{tikzpicture}[baseline=-0.6ex,scale=1.,transform shape]
        \draw[Maroon, ultra thick](1,0)--(2,0) node[at end,black] {\Huge$\bullet$} node[at end, right]{};
        \filldraw[white] (1,0) circle (5pt) node[]{};
        \node[black] at (1,0) {\Large$\otimes$};
    \end{tikzpicture}
\nonumber\\
&+g^4\left\{ 
\begin{tikzpicture}[baseline=-0.6ex,scale=1.,transform shape]
  \draw[Maroon, ultra thick] (1,0)--(3,0) node[pos=0.5,black]{\Huge$\bullet$} node[at end,black]{\Huge$\bullet$};
  \filldraw[white] (1,0) circle (5pt) node[]{};
  \node[black] at (1,0) {\Large$
  \otimes$};
\end{tikzpicture} + \frac{1}{2!} \begin{tikzpicture}[baseline=-0.6ex,scale=1.,transform shape]
  \draw[Maroon, ultra thick] (1,0)--(2,0.6) node[at end,black]{\Huge$\bullet$};
  \draw[Maroon, ultra thick] (1,0)--(2,-0.6) node[at end,black]{\Huge$\bullet$};
  \filldraw[white] (1,0) circle (5pt) node[]{};
  \node[black] at (1,0) {\Large$\otimes$};
\end{tikzpicture}
 - \frac{1}{2!} \begin{tikzpicture}
[baseline=-0.6ex,scale=1.,transform shape]
    \draw[Maroon, ultra thick] (0,-0.5)--(0,0.5);
    \draw[Maroon, ultra thick](0,-0.5)--(1,0) node[at start,black] {\Huge$\bullet$}; 
    \draw[Maroon, ultra thick](0,0.5)--(1,0) node[at start,black] {\Huge$\bullet$};
    \filldraw[white] (1,0) circle (5pt) node[]{};
    \node[black] at (1,0) {\Large$\otimes$};
    \end{tikzpicture}
 \right\} 
\nonumber\\
&-g^6 \Biggl\{\begin{tikzpicture}[baseline=-0.6ex,scale=1.,transform shape]
  \draw[Maroon, ultra thick] (1.4,0)--(2.1,0) node[at end,black]{\Huge$\bullet$};
  \draw[Maroon, ultra thick] (0.7,0)--(1.4,0) node[at end,black]{\Huge$\bullet$};
  \draw[Maroon, ultra thick] (0,0)--(0.7,0) node[at end,black]{\Huge$\bullet$};
  \filldraw[white] (0,0) circle (5pt) node[]{};
  \node[black] at (0,0) {\Large$\otimes$};
\end{tikzpicture} + \begin{tikzpicture}[baseline=-0.6ex,scale=1.,transform shape]
  \draw[Maroon, ultra thick] (0.7,0.4)--(1.3,0.4) node[at end,black]{\Huge$\bullet$};
  \draw[Maroon, ultra thick] (0,0)--(0.7,0.4) node[at end,black]{\Huge$\bullet$};
  \draw[Maroon, ultra thick] (0,0)--(0.7,-0.4) node[at end,black]{\Huge$\bullet$};
  \filldraw[white] (0,0) circle (5pt) node[]{};
  \node[black] at (0,0) {\Large$\otimes$};
\end{tikzpicture} + \frac{1}{2!} \begin{tikzpicture}[baseline=-0.6ex,scale=1.,transform shape]
  \draw[Maroon, ultra thick] (0.8,0)--(1.4,0.6) node[at end,black]{\Huge$\bullet$};
  \draw[Maroon, ultra thick] (0.8,0)--(1.4,-0.6) node[at end,black]{\Huge$\bullet$};
  \draw[Maroon, ultra thick] (0,0)--(0.8,0) node[at end,black]{\Huge$\bullet$};
  \filldraw[white] (0,0) circle (5pt) node[]{};
  \node[black] at (0,0) {\Large$\otimes$};
\end{tikzpicture} \nonumber\\
&+\frac{1}{3!}\begin{tikzpicture}[baseline=-0.6ex,scale=1.,transform shape]
  \draw[Maroon, ultra thick] (0,0)--(0.9,0)   node[at end,black]{\Huge$\bullet$};
  \draw[Maroon, ultra thick] (0,0)--(0.8,0.55) node[at end,black]{\Huge$\bullet$};
  \draw[Maroon, ultra thick] (0,0)--(0.8,-0.55) node[at end,black]{\Huge$\bullet$};
  \filldraw[white] (0,0) circle (5pt) node[]{};
  \node[black] at (0,0) {\Large$\otimes$};
\end{tikzpicture} - \begin{tikzpicture}[baseline=-0.6ex,scale=1.2,transform shape]
    \draw[Maroon, ultra thick] (0.5,-0.5)--(0.5,0.5);
    \draw[Maroon, ultra thick](0.5,-0.5)--(1,0) node[at start,black] {\huge$\bullet$}; 
    \draw[Maroon, ultra thick](0.5,0.5)--(1,0) node[at start,black] {\huge$\bullet$};
    \draw[Maroon, ultra thick](1,0)--(1.5,0) node[at end,black] {\huge$\bullet$} node[at end, right]{};
    \node[black] at (1.0,0) {\huge$\bullet$};
    \filldraw[white] (0.5,0.5) circle (4pt) node[]{};
    \node[black] at (0.5,0.5) {\large$\otimes$};
    \end{tikzpicture} - \frac{1}{2!} \begin{tikzpicture}[baseline=-0.6ex,scale=1.2,transform shape]
    \draw[Maroon, ultra thick] (0.5,-0.5)--(0.5,0.5);
    \draw[Maroon, ultra thick](0.5,-0.5)--(1,0) node[at start,black] {\huge$\bullet$}; 
    \draw[Maroon, ultra thick](0.5,0.5)--(1,0) node[at start,black] {\huge$\bullet$};
    \draw[Maroon, ultra thick](1,0)--(1.5,0) node[at end,black] {\huge$\bullet$} node[at end, right]{};
    \node[black] at (1.0,0) {\huge$\bullet$};
    \filldraw[white] (1.5,0) circle (4pt) node[]{};
    \node[black] at (1.5,0) {\large$\otimes$};
    \end{tikzpicture}-\frac{1}{2!} \begin{tikzpicture}[baseline=-0.6ex,scale=1.2,transform shape]
    \draw[Maroon, ultra thick] (0.5,-0.5)--(0.5,0.5);
    \draw[Maroon, ultra thick](0.5,-0.5)--(1,0) node[at start,black] {\huge$\bullet$}; 
    \draw[Maroon, ultra thick](0.5,0.5)--(1,0) node[at start,black] {\huge$\bullet$};
    \draw[Maroon, ultra thick](1,0)--(1.5,0) node[at end,black] {\huge$\bullet$} node[at end, right]{};
    \filldraw[white] (1,0) circle (4pt) node[]{};
    \node[black] at (1,0) {\large$\otimes$};
    \end{tikzpicture} \nonumber\\
& - \frac{1}{2!} \begin{tikzpicture}[baseline=-0.6ex,scale=1.,transform shape]
  \draw[Maroon, ultra thick] (0,-0.5)--(1,-0.5);
  \draw[Maroon, ultra thick] (0,-0.5)--(0,0.5);
  \draw[Maroon, ultra thick] (1,-0.5)--(1,0.5)
    node[at start,black]{\Huge$\bullet$};
  \draw[Maroon, ultra thick] (0,0.5)--(1,0.5)
    node[at start,black]{\Huge$\bullet$}
    node[at end,black]{\Huge$\bullet$};
  \filldraw[white] (0,-0.5) circle (5pt) node[]{};
  \node[black] at (0,-0.5) {\Large$\otimes$}; 
\end{tikzpicture} + \frac{1}{2!} \begin{tikzpicture}[baseline=-0.6ex,scale=1.,transform shape]
  \draw[Maroon, ultra thick] (0,-0.5)--(1,-0.5);
  \draw[Maroon, ultra thick] (0,-0.5)--(0,0.5);
  \draw[Maroon, ultra thick] (0,0.5)--(1,-0.5);
  \draw[Maroon, ultra thick] (1,-0.5)--(1,0.5)
    node[at start,black]{\Huge$\bullet$};
  \draw[Maroon, ultra thick] (0,0.5)--(1,0.5)
    node[at start,black]{\Huge$\bullet$}
    node[at end,black]{\Huge$\bullet$};
  \filldraw[white] (0,-0.5) circle (5pt) node[]{};
  \node[black] at (0,-0.5) {\Large$\otimes$};
\end{tikzpicture}
+\frac{1}{2!}\begin{tikzpicture}[baseline=-0.6ex,scale=1.,transform shape]
  \draw[Maroon, ultra thick] (0,-0.5)--(0,0.5);
  \draw[Maroon, ultra thick] (0,0.5)--(1,-0.5);
  \draw[Maroon, ultra thick] (0,-0.5)--(1,-0.5)
    node[at start,black]{\Huge$\bullet$};
  \draw[Maroon, ultra thick] (1,-0.5)--(1,0.5)
    node[at start,black]{\Huge$\bullet$};
  \draw[Maroon, ultra thick] (0,0.5)--(1,0.5)
    node[at end,black]{\Huge$\bullet$};
  \filldraw[white] (0,0.5) circle (5pt) node[]{};
  \node[black] at (0,0.5) {\Large$\otimes$};
\end{tikzpicture}- \frac{1}{3!} \begin{tikzpicture}[baseline=-0.6ex,scale=1.,transform shape]
  \draw[Maroon, ultra thick] (0,-0.5)--(1,-0.5);
  \draw[Maroon, ultra thick] (0,-0.5)--(0,0.5);
  \draw[Maroon, ultra thick] (0,0.5)--(1,-0.5); 
  \draw[Maroon, ultra thick] (1,0.5)--(0,-0.5);
  \draw[Maroon, ultra thick] (1,-0.5)--(1,0.5)
    node[at start,black]{\Huge$\bullet$}; 
  \draw[Maroon, ultra thick] (0,0.5)--(1,0.5)
    node[at start,black]{\Huge$\bullet$}
    node[at end,black]{\Huge$\bullet$};
  \filldraw[white] (0,-0.5) circle (5pt) node[]{};
  \node[black] at (0,-0.5) {\Large$\otimes$};
\end{tikzpicture}
 \Biggl\}\,.
\end{align}

At $L=1$, a single diagram represents the whole result for $\Gamma_{\rm cusp}^{(2)}$. At $L=2$, we have three different diagrams, two zero-cycle and one one-cycle, which give the contributions summarized in table \ref{tab:1L}.  We can see from this table that the tree contributions dominate and the one-cycle diagram gives only a small correction to $\Gamma_{\rm cusp}^{(3)}$. 

\begin{table}[ht!]
    \centering
\begin{center}
\begin{tabular}{|c|c|c|c|}
 \hline\rule{0pt}{3ex} 
     Graph &  $\Gamma_{\rm cusp}^{(3)}$ & numerical $\Gamma_{\rm cusp}^{(3)}$ & Fraction $\frac{\Gamma_{\rm diagram}^{(3)}}{\Gamma_{\rm actual}^{(3)}}$ \\[0.4ex]
    \hhline{|=|=|=|=|}\rule{0pt}{3ex} 
    \begin{tikzpicture}[baseline=-0.6ex,scale=.9,transform shape]
  \draw[Maroon, ultra thick] (1,0)--(3,0) node[pos=0.5,black]{\Huge$\bullet$} node[at end,black]{\Huge$\bullet$};
  \filldraw[white] (1,0) circle (5pt) node[]{};
  \node[black] at (1,0) {\Large$
  \otimes$};
\end{tikzpicture}
    &  $\frac{32 \pi ^4}{45}$ & $69.27$ & 0.545  \\[0.4ex]
    \hline
     $\frac{1}{2!}$\begin{tikzpicture}[baseline=-0.6ex,scale=.9,transform shape]
  \draw[Maroon, ultra thick] (1,0)--(2,0.6) node[at end,black]{\Huge$\bullet$};
  \draw[Maroon, ultra thick] (1,0)--(2,-0.6) node[at end,black]{\Huge$\bullet$};
  \filldraw[white] (1,0) circle (5pt) node[]{};
  \node[black] at (1,0) {\Large$\otimes$};
\end{tikzpicture}
    &  $\frac{16 \pi ^4}{45}$ & $34.63$& $0.273$  \\
   \hline\rule{0pt}{3ex} 
     All tree graphs
    &  $\frac{16 \pi ^4}{15}$ & $103.90$ & $1.09$\\[1ex]
    \hhline{|=|=|=|=|}
    \begin{tikzpicture}[baseline=-2.ex,scale=1.1,transform shape]
    \draw[Maroon, ultra thick] (0.5,-0.5)--(0.5,0.5);
    \draw[Maroon, ultra thick](0.5,-0.5)--(1,0) node[at start,black] {\huge$\bullet$}; 
    \draw[Maroon, ultra thick](0.5,0.5)--(1,0) node[at start,black] {\huge$\bullet$};
    \filldraw[white] (1,0) circle (4pt) node[]{};
    \node[black] at (1,0) {\large$\otimes$};
    \node[black] at (0.45,-1) {$\;$\footnotesize Only 1-cycle graph};
    \node[black] at (-0.25,0) {\small$-\frac{1}{2!}$};
    \end{tikzpicture}
    &\rule{0pt}{0.5ex}  $-\frac{4 \pi ^4}{45}$ & ${-}8.66$ & $-0.09$ \\
    \hhline{|=|=|=|=|}\rule{0pt}{3ex} 
     Full result
    & $\frac{44\pi ^4}{45}$ & $95.24$ & 1 \\[1ex]
    \hline
\end{tabular}
\end{center}
    \caption{Contributions of individual negative geometries to the three-loop coefficient of the cusp anomalous dimension, $\Gamma_{\rm cusp}^{(3)}$. We see that the one-cycle geometry contribution to $\Gamma_{\rm cusp}$ is numerically suppressed in comparison to the tree graphs contribution.}
    \label{tab:1L}
\end{table}

At $L=3$, we have 11 diagrams in table \ref{tab:2L}. While numerically the tree diagrams still dominate overall, the one-cycle graphs give a smaller contribution than the two-cycle graphs. 
\begin{table}[H]
 \centering   
\begin{tabular}{|c|c|c|c|}
\hline\rule{0pt}{3ex} 
    Graph &  $\Gamma_{\rm cusp}^{(4)}$ & Numerical $\Gamma_{\rm cusp}^{(4)}$ & Fraction $\frac{\Gamma_{\rm graph}^{(4)}}{\Gamma_{\rm actual}^{(4)}}$\\[0.4ex]
 \hhline{|=|=|=|=|}\rule{0pt}{3ex} 
   $-$\begin{tikzpicture}[baseline=-0.6ex,scale=.8,transform shape]
  \draw[Maroon, ultra thick] (1.4,0)--(2.1,0) node[at end,black]{\Huge$\bullet$};
  \draw[Maroon, ultra thick] (0.7,0)--(1.4,0) node[at end,black]{\Huge$\bullet$};
  \draw[Maroon, ultra thick] (0,0)--(0.7,0) node[at end,black]{\Huge$\bullet$};
 \filldraw[white] (0,0) circle (5pt) node[]{};
  \node[black] at (0,0) {\Large$\otimes$};
\end{tikzpicture}
    &  $-\frac{136}{315} \pi^6$ & $-415.075$ &0.443 \\[0.4ex]
    \hline
   $-$\begin{tikzpicture}[baseline=-0.6ex,scale=.8,transform shape]
  \draw[Maroon, ultra thick] (0.7,0.4)--(1.3,0.4) node[at end,black]{\Huge$\bullet$};
  \draw[Maroon, ultra thick] (0,0)--(0.7,0.4) node[at end,black]{\Huge$\bullet$};
  \draw[Maroon, ultra thick] (0,0)--(0.7,-0.4) node[at end,black]{\Huge$\bullet$};
  \filldraw[white] (0,0) circle (5pt) node[]{};
  \node[black] at (0,0) {\Large$\otimes$};
\end{tikzpicture}
    &   $-\frac{136}{315}\pi^6$ & $-415.075$ &0.443 \\
    \hline
    $-\frac{1}{2!}$\begin{tikzpicture}[baseline=-0.6ex,scale=.8,transform shape]
  \draw[Maroon, ultra thick] (0.8,0)--(1.4,0.6) node[at end,black]{\Huge$\bullet$};
  \draw[Maroon, ultra thick] (0.8,0)--(1.4,-0.6) node[at end,black]{\Huge$\bullet$};
  \draw[Maroon, ultra thick] (0,0)--(0.8,0) node[at end,black]{\Huge$\bullet$};
  \filldraw[white] (0,0) circle (5pt) node[]{};
  \node[black] at (0,0) {\Large$\otimes$};
\end{tikzpicture} 
    &  $-\frac{8}{35}\pi^6$ & $-219.746 $ & 0.234 \\
    \hline
   $-\frac{1}{3!}$\begin{tikzpicture}[baseline=-0.6ex,scale=.8,transform shape]
  \draw[Maroon, ultra thick] (0,0)--(0.9,0)   node[at end,black]{\Huge$\bullet$};
  \draw[Maroon, ultra thick] (0,0)--(0.8,0.55) node[at end,black]{\Huge$\bullet$};
  \draw[Maroon, ultra thick] (0,0)--(0.8,-0.55) node[at end,black]{\Huge$\bullet$};
  \filldraw[white] (0,0) circle (5pt) node[]{};
  \node[black] at (0,0) {\Large$\otimes$};
\end{tikzpicture}
    &  $-\frac{8}{105} \pi^6$ & $-73.249 $ &0.078 \\
 \hline\rule{0pt}{3ex} 
    All tree graphs
    &  $-\frac{368}{315} \pi^6$ & $-1123.15 $ & 1.1981 \\[1ex]
    \hhline{|=|=|=|=|}
    $\frac{1}{2!}$\begin{tikzpicture}[baseline=-0.6ex,scale=1.,transform shape]
    \draw[Maroon, ultra thick] (0.5,-0.5)--(0.5,0.5);
    \draw[Maroon, ultra thick](0.5,-0.5)--(1,0) node[at start,black] {\huge$\bullet$}; 
    \draw[Maroon, ultra thick](0.5,0.5)--(1,0) node[at start,black] {\huge$\bullet$};
    \draw[Maroon, ultra thick](1,0)--(1.5,0) node[at end,black] {\huge$\bullet$} node[at end, right]{};
    \filldraw[white] (1,0) circle (4pt) node[]{};
    \node[black] at (1,0) {\large$\otimes$};
    \end{tikzpicture}
    & $\frac{176 \pi ^6}{2835}-24\zeta_3^2$ & 25.005 & $-0.0267$\\
    \hline
    \phantom{ $\frac{1}{2!}$}\begin{tikzpicture}[baseline=-0.6ex,scale=1.,transform shape]
    \draw[Maroon, ultra thick] (0.5,-0.5)--(0.5,0.5);
    \draw[Maroon, ultra thick](0.5,-0.5)--(1,0) node[at start,black] {\huge$\bullet$}; 
    \draw[Maroon, ultra thick](0.5,0.5)--(1,0) node[at start,black] {\huge$\bullet$};
    \draw[Maroon, ultra thick](1,0)--(1.5,0) node[at end,black] {\huge$\bullet$} node[at end, right]{};
    \node[black] at (1.0,0) {\huge$\bullet$};
   \filldraw[white] (0.5,0.5) circle (4pt) node[]{};
    \node[black] at (0.5,0.5) {\large$\otimes$};
    \end{tikzpicture}
    & $ \frac{352 \pi ^6}{2835}-48\zeta_3^2$ & 50.01 & $-0.0533$\\
    \hline
     $\frac{1}{2!}$ \begin{tikzpicture}[baseline=-0.6ex,scale=1.,transform shape]
    \draw[Maroon, ultra thick] (0.5,-0.5)--(0.5,0.5);
    \draw[Maroon, ultra thick](0.5,-0.5)--(1,0) node[at start,black] {\huge$\bullet$}; 
    \draw[Maroon, ultra thick](0.5,0.5)--(1,0) node[at start,black] {\huge$\bullet$};
    \draw[Maroon, ultra thick](1,0)--(1.5,0) node[at end,black] {\huge$\bullet$} node[at end, right]{};
    \node[black] at (1.0,0) {\huge$\bullet$};
   \filldraw[white] (1.5,0) circle (4pt) node[]{};
    \node[black] at (1.5,0) {\large$\otimes$};
    \end{tikzpicture}
    & $\frac{176 \pi ^6}{2835}-24\zeta_3^2$ & 25.005 & $-0.0267$\\
    \hline
\hspace{-0.4cm}\phantom{-}$\frac{1}{2!}$ \begin{tikzpicture}[baseline=-0.6ex,scale=1.,transform shape]
  \draw[Maroon, ultra thick] (0,-0.375)--(0.75,-0.375);
  \draw[Maroon, ultra thick] (0,-0.375)--(0,0.375);
  \draw[Maroon, ultra thick] (0.75,-0.375)--(0.75,0.375)
    node[at start,black]{\huge$\bullet$};
  \draw[Maroon, ultra thick] (0,0.375)--(0.75,0.375)
    node[at start,black]{\huge$\bullet$}
    node[at end,black]{\huge$\bullet$};
  \filldraw[white] (0,-0.375) circle (4pt) node[]{};
  \node[black] at (0,-0.375) {\large$\otimes$};
\end{tikzpicture}& $-\frac{164}{2835}\pi^6$ & $-55.61$ & 0.0593\\
 \hline\rule{0pt}{3ex} 
    All 1-cycle graphs   &  $\frac{540 \pi^6}{2835}-96\zeta_3^2$ & 44.41 & $-0.047$ \\[1ex]
    \hhline{|=|=|=|=|}
  \hspace{-0.4cm}$-\frac{1}{2!}$ \begin{tikzpicture}[baseline=-0.6ex,scale=1.,transform shape]
  \draw[Maroon, ultra thick] (0,-0.375)--(0,0.375);
  \draw[Maroon, ultra thick] (0,0.375)--(0.75,-0.375);
  \draw[Maroon, ultra thick] (0,-0.375)--(0.75,-0.375)
    node[at start,black]{\huge$\bullet$};
  \draw[Maroon, ultra thick] (0.75,-0.375)--(0.75,0.375)
    node[at start,black]{\huge$\bullet$};
  \draw[Maroon, ultra thick] (0,0.375)--(0.75,0.375)
    node[at end,black]{\huge$\bullet$};
 \filldraw[white] (0,0.375) circle (4pt) node[]{};
  \node[black] at (0,0.375) {\large$\otimes$};
\end{tikzpicture}
   &  $\frac{4 \pi ^6}{189}+24\zeta_3^2$ & 55.025 & $-0.0587$\\
    \hline
    \hspace{-0.4cm}$-\frac{1}{2!}$ \begin{tikzpicture}[baseline=-0.6ex,scale=1.,transform shape]
  \draw[Maroon, ultra thick] (0,-0.375)--(0.75,-0.375);
  \draw[Maroon, ultra thick] (0,-0.375)--(0,0.375);
  \draw[Maroon, ultra thick] (0,0.375)--(0.75,-0.375);
  \draw[Maroon, ultra thick] (0.75,-0.375)--(0.75,0.375)
    node[at start,black]{\huge$\bullet$};
  \draw[Maroon, ultra thick] (0,0.375)--(0.75,0.375)
    node[at start,black]{\huge$\bullet$}
    node[at end,black]{\huge$\bullet$};
   \filldraw[white] (0,-0.375) circle (4pt) node[]{};
  \node[black] at (0,-0.375) {\large$\otimes$};
\end{tikzpicture}
  & $\frac{4\pi ^6}{189}+24\zeta_3^2$ & 55.025 & $-0.0587$\\
    \hline\rule{0pt}{3.5ex} 
    All 2-cycle graphs
        &  $\frac{8 \pi ^6}{189}+48\zeta_3^2$ & 110.05 & $-0.117$ \\[1.5ex]
    \hhline{|=|=|=|=|}
    \hspace{0.2cm}\begin{tikzpicture}[baseline=-2.ex,scale=1.,transform shape]
  \draw[Maroon, ultra thick] (0,-0.375)--(0.75,-0.375);
  \draw[Maroon, ultra thick] (0,-0.375)--(0,0.375);
  \draw[Maroon, ultra thick] (0,0.375)--(0.75,-0.375); 
  \draw[Maroon, ultra thick] (0.75,0.375)--(0,-0.375); 
  \draw[Maroon, ultra thick] (0.75,-0.375)--(0.75,0.375)
    node[at start,black]{\huge$\bullet$};
  \draw[Maroon, ultra thick] (0,0.375)--(0.75,0.375)
    node[at start,black]{\huge$\bullet$}
    node[at end,black]{\huge$\bullet$};
   \filldraw[white] (0,-0.375) circle (4pt) node[]{};
  \node[black] at (0,-0.375) {\large$\otimes$};
  \node[black] at (-0.65,0) {$\frac{1}{3!}$};
  \node[black] at (0.25,-0.8) {Only 3-cycle graph};
\end{tikzpicture}
 &$\frac{8 \pi ^6}{945}+16\zeta_3^2$ & 31.26 & $-0.0333$\\
    \hhline{|=|=|=|=|}\rule{0pt}{3ex} 
     Full result
    &  $-8 \left(\frac{73 \pi ^6}{630}+4 \zeta_3^2\right)$ & $-937.43 $ & 1 \\[1ex]
    \hline
\end{tabular}
\caption{Contributions of individual negative geometries to $\Gamma_{\rm cusp}^{(4)}$. The tree graphs give the dominant contribution. The contribution of one-cycle, two-cycle and three-cycle graphs are numerically similar.}\label{tab:2L}
\end{table}

\begin{table}[H]
    \centering
\begin{tabular}{|c|c|c|c|c|}
\hline
    Graph &  $\pi^6$ coefficient & Fraction $\frac{\text{graph}}{\text{actual}}$ & $\zeta^2_3$ coefficient & Fraction $\frac{\text{graph}}{\text{actual}}$\\
 \hhline{|=|=|=|=|=|}
    \rule{0pt}{3ex} $-$\begin{tikzpicture}[baseline=-0.6ex,scale=.8,transform shape]
  \draw[Maroon, ultra thick] (1.4,0)--(2.1,0) node[at end,black]{\Huge$\bullet$};
  \draw[Maroon, ultra thick] (0.7,0)--(1.4,0) node[at end,black]{\Huge$\bullet$};
  \draw[Maroon, ultra thick] (0,0)--(0.7,0) node[at end,black]{\Huge$\bullet$};
  \filldraw[white] (0,0) circle (5pt) node[]{};
  \node[black] at (0,0) {\Large$\otimes$};
\end{tikzpicture}
    &  $-\frac{136}{315}$ & $0.466$ &0 & 0\\[0.4ex]
    \hline
   $-$\begin{tikzpicture}[baseline=-0.6ex,scale=.8,transform shape]
  \draw[Maroon, ultra thick] (0.7,0.4)--(1.3,0.4) node[at end,black]{\Huge$\bullet$};
  \draw[Maroon, ultra thick] (0,0)--(0.7,0.4) node[at end,black]{\Huge$\bullet$};
  \draw[Maroon, ultra thick] (0,0)--(0.7,-0.4) node[at end,black]{\Huge$\bullet$};
  \filldraw[white] (0,0) circle (5pt) node[]{};
  \node[black] at (0,0) {\Large$\otimes$};
\end{tikzpicture}
    &   $-\frac{136}{315}$ &  0.466 &0 & 0 \\[0.4ex]
    \hline
    $-\frac{1}{2}$ \begin{tikzpicture}[baseline=-0.6ex,scale=.8,transform shape]
  \draw[Maroon, ultra thick] (0.8,0)--(1.4,0.6) node[at end,black]{\Huge$\bullet$};
  \draw[Maroon, ultra thick] (0.8,0)--(1.4,-0.6) node[at end,black]{\Huge$\bullet$};
  \draw[Maroon, ultra thick] (0,0)--(0.8,0) node[at end,black]{\Huge$\bullet$};
  \filldraw[white] (0,0) circle (5pt) node[]{};
  \node[black] at (0,0) {\Large$\otimes$};
\end{tikzpicture} 
    &  $-\frac{8}{35}$ & $0.247$ & 0 & 0 \\
    \hline
   $-\frac{1}{3!}\;$\begin{tikzpicture}[baseline=-0.6ex,scale=.8,transform shape]
  \draw[Maroon, ultra thick] (0,0)--(0.9,0)   node[at end,black]{\Huge$\bullet$};
  \draw[Maroon, ultra thick] (0,0)--(0.8,0.55) node[at end,black]{\Huge$\bullet$};
  \draw[Maroon, ultra thick] (0,0)--(0.8,-0.55) node[at end,black]{\Huge$\bullet$};
  \filldraw[white] (0,0) circle (5pt) node[]{};
  \node[black] at (0,0) {\Large$\otimes$};
\end{tikzpicture}
    &  $-\frac{8}{105}$ & $0.082$ &0 & 0 \\
 \hline\rule{0pt}{3ex} 
    All tree graphs 
    &  $-\frac{368}{315}$ & $1.26$ & 0 & 0 \\[1ex]
    \hhline{|=|=|=|=|=|}
     $\frac{1}{2!}$\begin{tikzpicture}[baseline=-0.6ex,scale=1.,transform shape]
    \draw[Maroon, ultra thick] (0.5,-0.5)--(0.5,0.5);
    \draw[Maroon, ultra thick](0.5,-0.5)--(1,0) node[at start,black] {\huge$\bullet$}; 
    \draw[Maroon, ultra thick](0.5,0.5)--(1,0) node[at start,black] {\huge$\bullet$};
    \draw[Maroon, ultra thick](1,0)--(1.5,0) node[at end,black] {\huge$\bullet$} node[at end, right]{};
    \filldraw[white] (1,0) circle (4pt) node[]{};
    \node[black] at (1,0) {\large$\otimes$};
    \end{tikzpicture}
    & $ \frac{176 }{2835}$ & $-0.067$ & $-24$ & $0.75$\\
    \hline
    \phantom{$\frac{1}{2!}$}\begin{tikzpicture}[baseline=-0.6ex,scale=1.,transform shape]
    \draw[Maroon, ultra thick] (0.5,-0.5)--(0.5,0.5);
    \draw[Maroon, ultra thick](0.5,-0.5)--(1,0) node[at start,black] {\huge$\bullet$}; 
    \draw[Maroon, ultra thick](0.5,0.5)--(1,0) node[at start,black] {\huge$\bullet$};
    \draw[Maroon, ultra thick](1,0)--(1.5,0) node[at end,black] {\huge$\bullet$} node[at end, right]{};
    \node[black] at (1.0,0) {\huge$\bullet$};
   \filldraw[white] (0.5,0.5) circle (4pt) node[]{};
    \node[black] at (0.5,0.5) {\large$\otimes$};
    \end{tikzpicture}
     & $ \frac{352 }{2835}$ & $-0.134$ & $-48$ & $1.5$\\
    \hline
  $\frac{1}{2!}$ \begin{tikzpicture}[baseline=-0.6ex,scale=1.,transform shape]
    \draw[Maroon, ultra thick] (0.5,-0.5)--(0.5,0.5);
    \draw[Maroon, ultra thick](0.5,-0.5)--(1,0) node[at start,black] {\huge$\bullet$}; 
    \draw[Maroon, ultra thick](0.5,0.5)--(1,0) node[at start,black] {\huge$\bullet$};
    \draw[Maroon, ultra thick](1,0)--(1.5,0) node[at end,black] {\huge$\bullet$} node[at end, right]{};
    \node[black] at (1.0,0) {\huge$\bullet$};
   \filldraw[white] (1.5,0) circle (4pt) node[]{};
    \node[black] at (1.5,0) {\large$\otimes$};
    \end{tikzpicture} 
    & $ \frac{176 }{2835}$ & $-0.067$ & $-24$ & $0.75$\\
    \hline
 \hspace{-0.6cm}\phantom{$-$}$\frac{1}{2!}$ \begin{tikzpicture}[baseline=-0.6ex,scale=1.,transform shape]
  \draw[Maroon, ultra thick] (0,-0.375)--(0.75,-0.375);
  \draw[Maroon, ultra thick] (0,-0.375)--(0,0.375);
  \draw[Maroon, ultra thick] (0.75,-0.375)--(0.75,0.375)
    node[at start,black]{\huge$\bullet$};
  \draw[Maroon, ultra thick] (0,0.375)--(0.75,0.375)
    node[at start,black]{\huge$\bullet$}
    node[at end,black]{\huge$\bullet$};
  \filldraw[white] (0,-0.375) circle (4pt) node[]{};
  \node[black] at (0,-0.375) {\large$\otimes$};
\end{tikzpicture}& $-\frac{164}{2835}$ & $0.062$ & 0 & 0\\
 \hline\rule{0pt}{3ex} 
    All 1-cycle graphs
    &  $\frac{540}{2835}$ & $-0.205$ & $-96$ & $3$ \\[1ex]
    \hhline{|=|=|=|=|=|}
 \hspace{-0.6cm}$-\frac{1}{2!}$ \begin{tikzpicture}[baseline=-0.6ex,scale=1.,transform shape]
  \draw[Maroon, ultra thick] (0,-0.375)--(0,0.375);
  \draw[Maroon, ultra thick] (0,0.375)--(0.75,-0.375);
  \draw[Maroon, ultra thick] (0,-0.375)--(0.75,-0.375)
    node[at start,black]{\huge$\bullet$};
  \draw[Maroon, ultra thick] (0.75,-0.375)--(0.75,0.375)
    node[at start,black]{\huge$\bullet$};
  \draw[Maroon, ultra thick] (0,0.375)--(0.75,0.375)
    node[at end,black]{\huge$\bullet$};
  \filldraw[white] (0,0.375) circle (4pt) node[]{};
  \node[black] at (0,0.375) {\large$\otimes$};
\end{tikzpicture}
   &  $\frac{4}{189}$ & $-0.023$ & $24$ & $-0.75$\\
    \hline
    \hspace{-0.6cm}$-\frac{1}{2!}$ \begin{tikzpicture}[baseline=-0.6ex,scale=1.,transform shape]
  \draw[Maroon, ultra thick] (0,-0.375)--(0.75,-0.375);
  \draw[Maroon, ultra thick] (0,-0.375)--(0,0.375);
  \draw[Maroon, ultra thick] (0,0.375)--(0.75,-0.375);
  \draw[Maroon, ultra thick] (0.75,-0.375)--(0.75,0.375)
    node[at start,black]{\huge$\bullet$};
  \draw[Maroon, ultra thick] (0,0.375)--(0.75,0.375)
    node[at start,black]{\huge$\bullet$}
    node[at end,black]{\huge$\bullet$};
 \filldraw[white] (0,-0.375) circle (4pt) node[]{};
  \node[black] at (0,-0.375) {\large$\otimes$};
\end{tikzpicture}
 &  $\frac{4 }{189}$ & $-0.023$ & 24 & $-0.75$\\
    \hline\rule{0pt}{2.5ex} 
    All 2-cycle graphs
    &  $\frac{8}{189}$ & $-0.046$ & 48 & $-1.5$ \\[0.5ex]
    \hhline{|=|=|=|=|=|}
 \begin{tikzpicture}[baseline=-2.ex,scale=1.,transform shape]
  \draw[Maroon, ultra thick] (0,-0.375)--(0.75,-0.375);
  \draw[Maroon, ultra thick] (0,-0.375)--(0,0.375);
  \draw[Maroon, ultra thick] (0,0.375)--(0.75,-0.375); 
  \draw[Maroon, ultra thick] (0.75,0.375)--(0,-0.375); 
  \draw[Maroon, ultra thick] (0.75,-0.375)--(0.75,0.375)
    node[at start,black]{\huge$\bullet$};
  \draw[Maroon, ultra thick] (0,0.375)--(0.75,0.375)
    node[at start,black]{\huge$\bullet$}
    node[at end,black]{\huge$\bullet$};
 \filldraw[white] (0,-0.375) circle (4pt) node[]{};
  \node[black] at (0,-0.375) {\large$\otimes$};
  \node[black] at (-0.65,0) {$\frac{1}{3!}$};
  \node[black] at (0.25,-0.8) {Only 3-cycle graph};
\end{tikzpicture}
 & $\frac{8}{945}$ & $-0.009$ & 16 & $-0.5$\\
    \hhline{|=|=|=|=|=|}\rule{0pt}{3.5ex}
     Full result
    &  $-\frac{292}{315}$ &1 & $-32$ & 1 \\[1ex]
    \hline
\end{tabular}
    \caption{When dividing the contributions to $\Gamma_{\rm cusp}$ into $\pi^6$ and $\zeta_3^2$ parts, the hierarchy becomes more clear. In both sectors separately, the higher-cycle graphs are suppressed.}
    \label{tab:3L_2}
\end{table}

In table~\ref{tab:3L_2}, we see that in the even-zeta sector, i.e.~the $\pi^6$ terms, the tree graphs dominate, one-cycle graphs are sub-leading, two-cycle graphs are sub-sub-leading and the three-cycle graph gives the smallest contribution. In the $\zeta_3^2$ sector, the tree graphs do not contribute at all, and the one-cycle graphs dominate. The two-cycle graphs are sub-leading and the three-cycle graph again gives a smallest contribution. 

\begin{tcolorbox}[colback=white]
\begin{center}
\vspace{-0.1cm}
Through $L=3$, lower-cycle graphs dominate over higher-cycle graphs in their contribution to $\Gamma_{\rm cusp}$, if we separate contributions into even and odd zeta sectors. 
\vspace{-0.1cm}
\end{center}
\end{tcolorbox} 

For $L=4$ and higher, we do not have a systematic way to evaluate ${\cal F}(g,z)$ and extract the cusp anomalous dimension. We can only do it for diagrams which can be solved using the differential equation discussed in section \ref{sec:integrated_neg_geom}. For one of these solvable diagrams, we find
\begin{align} 
\label{eq:I_3,2}
    \Gamma_{\rm cusp}\left(\begin{tikzpicture}[baseline={(0,0)cm},scale=0.8]
    \draw[Maroon, ultra thick] (-0.75,0.75)--(-0.75,-0.75);
    \draw[Maroon, ultra thick] (0,0)--(-0.75,0.75) node[at end, black] {\Huge$\bullet$};
    \draw[Maroon, ultra thick] (0,0)--(-0.75,-0.75) node[at end, black] {\Huge$\bullet$};
    \draw[Maroon, ultra thick] (0,0)--(1.5,0) node[at start, black] {\Huge$\bullet$} node[pos=0.5, black] {\Huge$\bullet$};
    \node[white] at (1.5,0) {\Huge$\bullet$};
    \filldraw[white] (1.5,0) circle (6pt) node[]{};
    \node[black] at (1.5,0) {\Large$\otimes$};
\end{tikzpicture}\right) =  \frac{4 g^{10}}{5} \left(-\frac{448}{5} \zeta_{5,3}-\frac{32}{3}  \pi ^2 \zeta^2_3-640 \zeta_5\zeta_3+\frac{9944 \pi ^8}{70875}\right).
\end{align}

We will discuss this diagram, which contributes to the full $\Gamma_{\rm cusp}$ with a factor of $\frac{1}{2!}$, in more detail as a part of the fully solvable series in the next section. But here we just note that the contribution to $\Gamma_{\rm cusp}$ contains a {\bf multiple-zeta value} (MZV) $\zeta_{5,3}$ for the first time. For this object, we use the convention that a multiple-zeta value of depth $k$ is equal to 
\begin{equation}
    \zeta_{m_1,m_2, \ldots, m_k} = \sum_{n_1>n_2 > \cdots > n_k>0} \frac{1}{n^{m_1}_1 \cdots n^{m_k}_k} \,.
\end{equation}
For $n$ ladders attached to the triangle, the $\Gamma_{\rm cusp}$ contribution coming from ${\cal F}_{3,n}$ can be seen in eq.~\eqref{gammaF3n}. For $n\geq2$, irreducible MZVs appear. From the point of view of the full $\Gamma_{\rm cusp}$, the term with $\zeta_{5,3}$ is spurious and must cancel in the sum over all negative geometries at $L=4$. Understanding the mechanism of the cancellation of MZVs might be an important clue to how integrability could be made more manifest in the context of (integrated) amplitudes, and how the presence of MZVs might be seen as violating manifest integrability. (See e.g.~ref.~\cite{Drummond:2009fd} for the manifest Yangian symmetry for tree-level amplitudes and leading singularities.)

\subsection{\texorpdfstring{$\Gamma_{\rm cusp}$}{ gamma\_cusp} of infinite classes}

The task of calculating the $\Gamma_{\rm cusp}$ contribution for certain infinite classes of negative geometries is simpler than computing the full $F(g,z)$ function. The main reason is that for the purpose of $\Gamma_{\rm cusp}$, it does not matter where the marked point is located in the graph. This is obvious from the fact that $\Gamma_{\rm cusp}$ can be also recovered from the integration over all of the loops of the integrand (with no marked points) as the leading IR divergence, although this is not the method we use here. Hence for this purpose, labeling a marked point in the graph is just an intermediate step (in the end all loops are integrated over), and the result for $\Gamma_{\rm cusp}$ from the given graph does not depend on where this marked point was located.

This also explains why different diagrams contribute to $\Gamma_{\rm cusp}$ in the same way in $L=2$ and $L=3$ tables (modulo symmetry factors). In particular,

\begin{equation}
    \Gamma_{\rm cusp} \left(\begin{tikzpicture}[baseline=-0.6ex,scale=1.,transform shape]
  \draw[Maroon, ultra thick] (1,0)--(3,0) node[pos=0.5,black]{\Huge$\bullet$} node[at end,black]{\Huge$\bullet$};
  \filldraw[white] (1,0) circle (5pt) node[]{};
  \node[black] at (1,0) {\Large$
  \otimes$};
\end{tikzpicture}\right) = \Gamma_{\rm cusp} \left(\begin{tikzpicture}[baseline=-0.6ex,scale=1.,transform shape]
  \draw[Maroon, ultra thick] (1,0)--(2,0.6) node[at end,black]{\Huge$\bullet$};
  \draw[Maroon, ultra thick] (1,0)--(2,-0.6) node[at end,black]{\Huge$\bullet$};
  \filldraw[white] (1,0) circle (5pt) node[]{};
  \node[black] at (1,0) {\Large$\otimes$};
\end{tikzpicture}\right) = \frac{32\pi^4 g^6}{45}\,,
\end{equation}
\begin{center}
   \begin{equation}
    \Gamma_{\rm cusp} \left(\begin{tikzpicture}[baseline=-0.6ex,scale=1.5,transform shape]
  \draw[Maroon, ultra thick] (0,-0.375)--(0,0.375);
  \draw[Maroon, ultra thick] (0,0.375)--(0.75,-0.375);
  \draw[Maroon, ultra thick] (0,-0.375)--(0.75,-0.375)
    node[at start,black]{\Large$\bullet$};
  \draw[Maroon, ultra thick] (0.75,-0.375)--(0.75,0.375)
    node[at start,black]{\Large$\bullet$};
  \draw[Maroon, ultra thick] (0,0.375)--(0.75,0.375)
    node[at end,black]{\Large$\bullet$};
  \filldraw[white] (0,0.375) circle (4pt) node[]{};
  \node[black] at (0,0.375) {$\otimes$};
\end{tikzpicture}\right) = \Gamma_{\rm cusp} \left(\begin{tikzpicture}[baseline=-0.6ex,scale=1.5,transform shape]
  \draw[Maroon, ultra thick] (0,-0.375)--(0.75,-0.375);
  \draw[Maroon, ultra thick] (0,-0.375)--(0,0.375);
  \draw[Maroon, ultra thick] (0,0.375)--(0.75,-0.375);
  \draw[Maroon, ultra thick] (0.75,-0.375)--(0.75,0.375)
    node[at start,black]{\Large$\bullet$};
  \draw[Maroon, ultra thick] (0,0.375)--(0.75,0.375)
    node[at start,black]{\Large$\bullet$}
    node[at end,black]{\Large$\bullet$};
  \filldraw[white] (0,-0.375) circle (4pt) node[]{};
  \node[black] at (0,-0.375) {$\otimes$};
\end{tikzpicture}\right) = -\left(48 \zeta^2_3+\frac{8 \pi ^6}{189}\right)g^8\,.
\end{equation}
\end{center}
Similarly, we can apply this method to the triangle plus ladder diagrams in the form 
\begin{center}
  \begin{equation}
     \Gamma_{\rm cusp} \left( \begin{tikzpicture}[baseline=-0.5ex,]
    \draw[Maroon, ultra thick] (-0.75,0.75)--(-0.75,-0.75);
    \draw[Maroon, ultra thick] (0,0)--(-0.75,0.75) node[at end, black] {\Huge$\bullet$};
    \draw[Maroon, ultra thick] (0,0)--(-0.75,-0.75) node[at end, black] {\Huge$\bullet$};
    \draw[Maroon, ultra thick] (0,0)--(1.5,0) node[at end, black] {\Huge$\bullet$} node[pos=0.5, black] {\Huge$\bullet$};
   \filldraw[white] (0,0) circle (5pt) node[]{};
    \node[black] at (0,0) {\Large$\otimes$};
\end{tikzpicture} \right) = \Gamma_{\rm cusp} \left( \begin{tikzpicture}[baseline=-0.5ex,]
    \draw[Maroon, ultra thick] (-0.75,0.75)--(-0.75,-0.75);
    \draw[Maroon, ultra thick] (0,0)--(-0.75,0.75) node[at end, black] {\Huge$\bullet$};
    \draw[Maroon, ultra thick] (0,0)--(-0.75,-0.75) node[at end, black] {\Huge$\bullet$};
    \draw[Maroon, ultra thick] (0.75,0)--(1.5,0) node[at end, black] {\Huge$\bullet$};
    \draw[Maroon, ultra thick] (0,0)--(0.75,0) node[at end, black] {\Huge$\bullet$} node[at start, black] {\Huge$\bullet$};
    \filldraw[white] (1.5,0) circle (5pt) node[]{};
    \node[black] at (1.5,0) {\Large$\otimes$};
\end{tikzpicture} \right)\,.
  \end{equation}
\end{center}
We can also apply the same relation for the full sum of all graphs,
\begin{center}
  \begin{align}
        &\Gamma_{\rm cusp} \left( g^4\begin{tikzpicture}
[baseline=-0.5ex,scale=1.,transform shape]
    \draw[Maroon, ultra thick] (0,-0.75)--(0,0.75);
    \draw[Maroon, ultra thick](0,-0.75)--(.75,0) node[at start,black] {\Huge$\bullet$}; 
    \draw[Maroon, ultra thick](0,0.75)--(.75,0) node[at start,black] {\Huge$\bullet$};
    \filldraw[white] (0.75,0) circle (5pt) node[]{};
    \node[black] at (.75,0) {\Large$\otimes$};
\end{tikzpicture}\quad - g^6 \begin{tikzpicture}[baseline=-0.5ex]
    \draw[Maroon, ultra thick] (-0.75,0.75)--(-0.75,-0.75);
    \draw[Maroon, ultra thick] (0,0)--(-0.75,0.75) node[at end, black] {\Huge$\bullet$};
    \draw[Maroon, ultra thick] (0,0)--(-0.75,-0.75) node[at end, black] {\Huge$\bullet$};
    \draw[Maroon, ultra thick] (0,0)--(.75,0) node[at start, black] {\Huge$\bullet$};
    \filldraw[white] (0.75,0) circle (5pt) node[]{};
    \node[black] at (.75,0) {\Large$\otimes$};
\end{tikzpicture} + g^{8}  \begin{tikzpicture}[baseline=-0.5ex,]
    \draw[Maroon, ultra thick] (-0.75,0.75)--(-0.75,-0.75);
    \draw[Maroon, ultra thick] (0,0)--(-0.75,0.75) node[at end, black] {\Huge$\bullet$};
    \draw[Maroon, ultra thick] (0,0)--(-0.75,-0.75) node[at end, black] {\Huge$\bullet$};
    \draw[Maroon, ultra thick] (0,0)--(1.5,0) node[at start, black] {\Huge$\bullet$} node[pos=0.5, black] {\Huge$\bullet$};
    \filldraw[white] (1.5,0) circle (5pt) node[]{};
    \node[black] at (1.5,0) {\Large$\otimes$};
\end{tikzpicture} + \cdots\right)= \\
& \Gamma_{\rm cusp}\left( g^4 \begin{tikzpicture}[baseline={(0,-0.1)cm}]
    \draw[Maroon, ultra thick] (-0.75,0.75)--(-0.75,-0.75);
    \draw[Maroon, ultra thick] (0,0)--(-0.75,0.75) node[at end, black] {\Huge$\bullet$};
    \draw[Maroon, ultra thick] (0,0)--(-0.75,-0.75) node[at end, black] {\Huge$\bullet$};
    \filldraw[white] (0,0) circle (5pt) node[]{};
    \node[black] at (0,0) {\Large$\otimes$};
\end{tikzpicture} \times  \begin{tikzpicture}[baseline={(0,-0.1)cm},scale=1.]
        \node[ellipse, draw, fill=almond, minimum width=4cm, minimum height=1.5cm] {Sum of ladders};
        \filldraw[white] (-2.17,0) circle (5pt) node[]{};
        \node[black] at (-2.17,0) {\Large$\otimes$};
    \end{tikzpicture} \right) \nonumber
  \end{align}
\end{center}
where the ``Ladder'' blob corresponds to the series in \eqref{eqn:ladder}. 

The $F(g,z)$ function for both the triangle and the resummed ladder diagrams are known, eqs.~(\ref{L3c}), (\ref{eq:F_ladder}), hence the differential equation for the $\Gamma_{\rm cusp}$ contribution is
\begin{equation}
g\frac{\partial}{\partial g}\Gamma_{\rm cusp} = -8\mathcal{I}[g^4\mathcal{F}_3(z){\cal F}_{\rm ladder}(g,z)]\,,
\end{equation}
where the $\cal{I}$ operation acts on the product of the triangle, given by ${\cal F}_3(z)$ and the resummed ladder function ${\cal F}_{\rm ladder}(g,z)$. There is an additional factor of $g^4$ which originates from the $g^6$ factor of the triangle, but divided by $1/g^2$ to avoid double counting the marked point (which is already counted in the resummed ladder). We leave the details of this computation to Appendix \ref{sec: Gamma-cups-resum}. In the end, we find that $\Gamma_{\rm cusp}(g^4\mathcal{F}_3(z){\cal F}_{\rm ladder}(g,z)) \ \propto \ g^5+\mathcal{O}(g^4)$ at $g \gg 1$ (see \eqref{eq:gammacusptriladder} for the full result). While this is suppressed by a factor of $\frac{1}{g}$ compared to the leading term in the weak coupling expansion, namely $F_{3}(g,z) \ \propto \  g^6$, it still shows polynomial growth in $g$ and thus isn't resummable. As mentioned before, the triangle-plus-ladder is an example of a partially summable class. Attaching the ladder class to the single-cycle box, we again find suppression only by $\frac{1}{g}$ compared to the leading term. Namely, we find  $\Gamma_{\rm cusp}(g^6\mathcal{F}_4(z){\cal F}_{\rm ladder}(g,z)) \ \propto \ g^7+\mathcal{O}(g^6)$ at $g \gg 1$ (see \eqref{eq:gammacuspboxladder} for the full result). These results suggest that one might need to embed the triangle-plus-ladder and box-plus-ladder classes into an even larger class of diagrams -- one starting at a lower perturbative order -- before resumming, in order to obtain sufficient suppression in $g$ at strong coupling.

\subsection{Geometric Series}

At the moment, we have examples of \emph{summable} negative geometries: these are ladders or all tree graphs, where the $g\gg1$ limit of the contribution to $\Gamma_{\rm cusp}$ is linear in $g$, as in the full Wilson loop. We also have the {\it partially summable} geometries that we discussed above. There are also classes of \emph{super-summable} geometries which are suppressed at large $g$: these are geometric series. In fact, any negative geometry participates in such a series. For example, we can consider a following sequence,
\begin{equation}
 \begin{tikzpicture}[baseline=-0.6ex]
        \node[black] at (0,0) {\Large$\otimes$};
    \end{tikzpicture}\,-\, g^2\, \begin{tikzpicture}[baseline=-0.6ex,scale=1.,transform shape]
  \draw[Maroon, ultra thick] (1,0)--(2,0) node[at end,black]{\Huge$\bullet$};
   \filldraw[white] (1,0) circle (5pt) node[]{};
  \node[black] at (1,0) {\Large$
  \otimes$}; 
\end{tikzpicture} \, + \,\frac{g^4}{2!}  \begin{tikzpicture}[baseline=-0.6ex,scale=1.,transform shape]
  \draw[Maroon, ultra thick] (0.8,0)--(1.4,0.6) node[at end,black]{\Huge$\bullet$};
  \draw[Maroon, ultra thick] (0.8,0)--(1.4,-0.6) node[at end,black]{\Huge$\bullet$};
   \filldraw[white] (0.8,0) circle (5pt) node[]{};
  \node[black] at (0.8,0) {\Large$\otimes$};
\end{tikzpicture} - \frac{g^6}{3!} \begin{tikzpicture}[baseline=-0.6ex,scale=1.,transform shape]
  \draw[Maroon, ultra thick] (0.8,0)--(1.4,0.6) node[at end,black]{\Huge$\bullet$};
  \draw[Maroon, ultra thick] (0.8,0)--(1.4,-0.6) node[at end,black]{\Huge$\bullet$};
  \draw[Maroon, ultra thick] (0,0)--(0.8,0) node[at end,black]{\Huge$\bullet$};
   \filldraw[white] (0.8,0) circle (5pt) node[]{};
  \node[black] at (0.8,0) {\Large$\otimes$};
  \node[pos=0.0,black]{\Huge$\bullet$}; 
\end{tikzpicture}  \, + \, \cdots + \, \frac{g^{2n}}{n!} \begin{tikzpicture}[baseline=-0.6ex,scale=1.,transform shape]
 \draw[Maroon, ultra thick] (0.8,0)--(1.645,0) node[at end,black]{\Huge$\bullet$};
 \node at (0.8,0.5){$\cdots$};
 \draw[Maroon, ultra thick] (0.8,0)--(0.8,-0.845) node[at end,black]{\Huge$\bullet$};
  \draw[Maroon, ultra thick] (0.8,0)--(1.4,0.6) node[at end,black]{\Huge$\bullet$};
  \draw[Maroon, ultra thick] (0.8,0)--(0.2,0.6)
  node[at end,black]{\Huge$\bullet$};
  \draw[Maroon, ultra thick] (0.8,0)--(0.2,-0.6)
  node[at end,black]{\Huge$\bullet$};
  \draw[Maroon, ultra thick] (0.8,0)--(1.4,-0.6) node[at end,black]{\Huge$\bullet$};
  \draw[Maroon, ultra thick] (0.8,0)--(1.4,-0.6) node[at end,black]{\Huge$\bullet$};
  \draw[Maroon, ultra thick] (0,0)--(0.8,0) node[at end,black]{\Huge$\bullet$};
   \filldraw[white] (0.8,0) circle (5pt) node[]{};
  \node[black] at (0.8,0) {\Large$\otimes$};
  \node[pos=0.0,black]{\Huge$\bullet$}; 
\end{tikzpicture}
\end{equation}
with the ${\cal F}(g,z)$ function 
\begin{equation}
    {\cal F}_{\text{sun}}(g,z) = \sum^{\infty}_{n=0} \frac{(-g^2)^n}{n!}(\pi^2+\log(z)^2)^n\,.
    \end{equation}
This sums into an exponential, namely 
\begin{equation}
\label{eq:Fsunexp}
{\cal F}_{\text{sun}}(g,z) = e^{-g^2 \left(\log ^2(z)+\pi ^2\right)}\,.
\end{equation}
It is easy to see that at $g \gg 1$, ${\cal F}$ is exponentially suppressed. We can easily take the $\mathcal{I}$ integral of ${\cal F}_{\text{sun}}(g,z)$ and compute
\begin{equation}
    \Gamma_{\rm cusp}\left({\cal F}_{\text{sun}}(g,z) \right)_{g \gg 1} \sim \frac{4\log \left(\pi g\right)}{\pi ^2}\,.
\end{equation}
This contribution is not exponentially suppressed. It comes from the region of integration near $z=-1$, where $\log^2(z)+\pi^2$ vanishes. That is, the exponential suppression of the integrand~\eqref{eq:Fsunexp} is not uniform.

We can also attach 1-cycle diagrams together in a similar fashion and get another series that sums into an exponential. The simplest example is the following
\begin{equation}
   \,\begin{tikzpicture}[baseline=-0.6ex]
        \node[black] at (0,0) {\Large$\otimes$};
    \end{tikzpicture}\,- \, \frac{g^4}{2!}\,\begin{tikzpicture}[baseline={(0,0)cm}]
        \draw[Maroon, ultra thick] (-1,0.5) -- (-1,-0.5);
        \draw[Maroon, ultra thick] (0,0) -- (-1,0.5) node[at end, black] {\Huge$\bullet$};
        \draw[Maroon, ultra thick] (0,0) -- (-1,-0.5) node[at end, black] {\Huge$\bullet$};
        \filldraw[white] (0,0) circle (5pt) node[]{};
        \node[black] at (0,0) {\Large$\otimes$};
    \end{tikzpicture}\, + \frac{g^{8}}{8}\,\begin{tikzpicture}[baseline={(0,0)cm}]
        \draw[Maroon, ultra thick] (0.5,1) -- (-0.5,1);
        \draw[Maroon, ultra thick] (0,0) -- (0.5,1) node[at end, black] {\Huge$\bullet$};
        \draw[Maroon, ultra thick] (0,0) -- (-0.5,1) node[at end, black] {\Huge$\bullet$};
        \draw[Maroon, ultra thick] (0.5,-1) -- (-0.5,-1);
        \draw[Maroon, ultra thick] (0,0) -- (0.5,-1) node[at end, black] {\Huge$\bullet$};
        \draw[Maroon, ultra thick] (0,0) -- (-0.5,-1) node[at end, black] {\Huge$\bullet$};
        \filldraw[white] (0,0) circle (5pt) node[]{};
        \node[black] at (0,0) {\Large$\otimes$};
    \end{tikzpicture}\, - \frac{g^{12}}{48}\,\begin{tikzpicture}[baseline={(0,0)cm}]
        \draw[Maroon, ultra thick] (0.25,1) -- (-0.75,0.75);
        \draw[Maroon, ultra thick] (0,0) -- (0.25,1) node[at end, black] {\Huge$\bullet$};
        \draw[Maroon, ultra thick] (0,0) -- (-0.75,0.75) node[at end, black] {\Huge$\bullet$};
        \draw[Maroon, ultra thick] (0.25,-1) -- (-0.75,-0.75);
        \draw[Maroon, ultra thick] (0,0) -- (0.25,-1) node[at end, black] {\Huge$\bullet$};
        \draw[Maroon, ultra thick] (0,0) -- (-0.75,-0.75) node[at end, black] {\Huge$\bullet$};
        \draw[Maroon, ultra thick] (1,0.5) -- (1,-0.5);
        \draw[Maroon, ultra thick] (0,0) -- (1,0.5) node[at end, black] {\Huge$\bullet$};
        \draw[Maroon, ultra thick] (0,0) -- (1,-0.5) node[at end, black] {\Huge$\bullet$};
         \filldraw[white] (0,0) circle (5pt) node[]{};
        \node[black] at (0,0) {\Large$\otimes$};
    \end{tikzpicture}\, + \frac{g^{16}}{384}\,\begin{tikzpicture}[baseline={(0,0)cm}]
        \draw[Maroon, ultra thick] (0.5,1) -- (-0.5,1);
        \draw[Maroon, ultra thick] (0,0) -- (0.5,1) node[at end, black] {\Huge$\bullet$};
        \draw[Maroon, ultra thick] (0,0) -- (-0.5,1) node[at end, black] {\Huge$\bullet$};
        \draw[Maroon, ultra thick] (0.5,-1) -- (-0.5,-1);
        \draw[Maroon, ultra thick] (0,0) -- (0.5,-1) node[at end, black] {\Huge$\bullet$};
        \draw[Maroon, ultra thick] (0,0) -- (-0.5,-1) node[at end, black] {\Huge$\bullet$};
        \draw[Maroon, ultra thick] (1,0.5) -- (1,-0.5);
        \draw[Maroon, ultra thick] (0,0) -- (1,0.5) node[at end, black] {\Huge$\bullet$};
        \draw[Maroon, ultra thick] (0,0) -- (1,-0.5) node[at end, black] {\Huge$\bullet$};
        \draw[Maroon, ultra thick] (-1,0.5) -- (-1,-0.5);
        \draw[Maroon, ultra thick] (0,0) -- (-1,0.5) node[at end, black] {\Huge$\bullet$};
        \draw[Maroon, ultra thick] (0,0) -- (-1,-0.5) node[at end, black] {\Huge$\bullet$};
         \filldraw[white] (0,0) circle (5pt) node[]{};
        \node[black] at (0,0) {\Large$\otimes$};
    \end{tikzpicture}\,\cdots
\end{equation}
which contains the triangle function ${\cal F}_3(z)$ and its powers. We can think about the first term as the zero power and the it sums to 
\begin{equation}
  {\cal F}(g,z) = \sum_{m=0}^\infty \frac{(-1)^n}{2^n n!}\left(g^4 \mathcal{F}_3(z)\right)^n =  e^{-\frac{1}{2} \left( g^4 \mathcal{F}_{3}(z)\right)}\,.
\end{equation}
The extraction of $\Gamma_{\rm cusp}$ is trickier. We note that for all points in the integration domain $\phi \in [-\pi,\pi]$ relevant for the $\Gamma_{\rm cusp}$ contribution, $e^{-\frac{g^4}{2}{\cal F}_3(e^{i\phi})}$ is exponentially suppressed. $\mathcal{F}_3(e^{i\phi})$ has a global minimum at $\phi = 0$ in this domain, so we can do a saddle point approximation and compute (for large $g$)
\begin{equation}
    \mathcal{I}[ e^{-\frac{1}{2} \left( g^4 \mathcal{F}_{3}(z)\right)}]_{g\gg1} \sim -\frac{e^{-\frac{1}{2} g^4 \mathcal{F}_3(z=1)}}{\sqrt{- \pi  \mathcal{F}''_3(z=1)}} + \mathcal{O}(\frac{e^{-g^4}}{g^4})\,.
\end{equation}
This gives us the $\Gamma_{\rm cusp}$ contribution
\begin{equation}
    \Gamma_{\rm cusp}\left(\sum_{m=0}^\infty \frac{(-1)^n}{2^n n!}\left(g^4 \mathcal{F}_3(z)\right)^n \right)_{g \gg 1} \sim -\frac{0.61 e^{-1.2 g^4}}{g^4}+\mathcal{O}\left(\frac{e^{-1.2g^4}}{g^8}\right)\,,
\end{equation}
which is heavily suppressed in large $g$. We can also compute the individual terms in this expansion using the HPL shuffle identities. Let's see how this works for the bow-tie diagram.
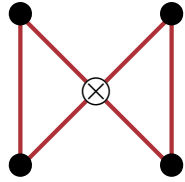
\begin{center}
\begin{tikzpicture}
  \draw[Maroon, ultra thick] (2,0)--(2,2);
    \draw[Maroon, ultra thick] (0,0)--(0,2);
    \draw[Maroon, ultra thick](0,0)--(1,1) node[at start,black] {\Huge$\bullet$}; 
    \draw[Maroon, ultra thick](0,2)--(1,1) node[at start,black] {\Huge$\bullet$};
    \draw[Maroon, ultra thick](1,1)--(2,0) node[at end,black] {\Huge$\bullet$}; 
    \draw[Maroon, ultra thick](1,1)--(2,2) node[at end,black] {\Huge$\bullet$};
    \filldraw[white] (1,1) circle (5pt) node[]{};
    \node[black] at (1,1) {\Large$\otimes$};
\end{tikzpicture}
\end{center}
Clearly this simply evaluates to $\mathcal{F}_3(z)^2$. In order to calculate $\mathcal{I}[\mathcal{F}_3(z)^2]$ we must use the shuffle identities for harmonic polylogarithms:
\begin{align}
    H_{\vec{p}}(z) H_{\vec{q}}(z) =\sum_{\vec{r}=\vec{p} \bigsqcup\hspace{-1mm}\bigsqcup \vec{q}} H_{\vec{r}}(z)\,,
\end{align}
where $\bigsqcup\hspace{-1.25mm}\bigsqcup$ denotes the shuffle product of the two ordered lists $\vec{p}$ and $\vec{q}$. We can then re-write $\mathcal{F}_3(z)^2$ to be linear in HPLs and calculate $\Gamma_{\rm cusp}(\mathcal{F}_3(z)^2)$
\begin{align}
   &\frac{4g^{10}}{5}\Big[ {-}512 \zeta_{3}(\zeta_{3,2}{-}2 \zeta_{2,1,2}{-}3 \zeta_{2,2,1})+\frac{64}{3} \pi ^2 (4 (\zeta_{2,4}+\zeta_{3,3}+3 \zeta_{2,3,1}+4 \zeta_{3,1,2}+6 \zeta_{3,2,1}+9 \zeta_{4,1,1})\nonumber\\
   &+\zeta^2_{3})
   {-}256 \zeta_{5,3}{-}1024 \zeta_{6,2}{-}512 \zeta_{2,2,4}{-}1536 \zeta_{2,3,3}{-}3584 \zeta_{2,4,2}{-}7168 \zeta_{2,5,1}{-}1024 \zeta_{3,1,4}{-}2560 \zeta_{3,2,3}\nonumber\\
   &{-}5632 \zeta_{3,3,2}{-}11264 \zeta_{3,4,1}{-}3072 \zeta_{4,1,3}{-}6144 \zeta_{4,2,2}{-}12288 \zeta_{4,3,1}{-}5632 \zeta_{5,1,2}{-}10752 \zeta_{5,2,1}\nonumber\\
   &{-}7680 \zeta_{6,1,1}
    {-}1024 \zeta_{2,1,2,3}{-}3072 \zeta_{2,1,3,2}{-}6144 \zeta_{2,1,4,1}{-}1536 \zeta_{2,2,1,3}{-}6656 \zeta_{2,2,3,1}{-}4608 \zeta_{2,3,1,2}\nonumber\\
    &{-}3072 \zeta_{3,1,1,3}
    {-}7168 \zeta_{3,1,2,2}{-}9216 \zeta_{3,2,1,2}{-}15360 \zeta_{3,2,2,1}{-}18432 \zeta_{3,3,1,1}{-}13824 \zeta_{4,1,1,2}
    \nonumber\\
    &{-}9216 \zeta_{2,4,1,1}{-}7680 \zeta_{2,3,2,1}{-}23040 \zeta_{4,1,2,1}{-}27648 \zeta_{4,2,1,1}
    {-}36864 \zeta_{5,1,1,1}+5120 \zeta_{3} \zeta_{5}{-}\frac{10196 \pi ^8}{14175} \Big]\,.
\end{align}
Note that while the $\Gamma_{\rm cusp}$ contribution of the single cycle diagrams we have considered above had at most depth-2 multiple zeta values, here we have depth-4. We can see from the shuffle identities that each cycle we add increases the maximum depth of MZVs that appear in the $\Gamma_{\rm cusp}$ contribution by 2. In general, a diagram with $L$ 1-cycles glued together will contribute to $\Gamma_{\rm cusp}$ with at most $n^L$ depth multiple zeta values, where $n$ is the maximum depth of the HPLs that appear in the $\mathcal{F}(z)$ function of the 1-cycle diagram at hand. As these multiple zetas do not appear in the full $\Gamma_{\rm cusp}$, there must be some mechanism that cancels them out. It is possible to re-write these higher depth zetas using relations between multiple-zeta values. For example, the expression above can be written as
\begin{equation}
    \Gamma_{\rm cusp}(\mathcal{F}_3(z)^2) \sim \frac{4 g^{10}}{5}\left(\frac{406888}{45}\zeta_8+384\zeta_2\zeta^2_3-7680\zeta_3\zeta_5-\frac{13056}{5}\zeta_{5,3}\right)\,,
\end{equation}
in terms of lower depth zetas. However, we still have the depth-2 $\zeta_{5,3}$ that we can't reduce further, which doesn't appear in the full cusp.
\section{Conclusion}
\label{sec:conclusion}

In this paper, we studied scattering amplitudes in planar ${\cal N}=4$ SYM amplitudes in the Amplituhedron framework. The logarithm of the amplitude is dual to the Wilson loop with a Lagrangian insertion (WLI), an IR finite object, a perfect laboratory to explore the analytic structure of amplitudes. The negative geometry expansion is a new basis of objects, analogous to the Amplituhedron, which makes the IR properties manifest. 

In this paper, we pushed the exploration of these objects further by computing all negative geometries that contribute to the three-loop WLI. We showed that the number of internal cycles of the associated graphs is closely related to the depth of polylogarithms. We also discussed how these negative geometries contribute to the leading IR divergence given by the cusp anomalous dimension, and found some evidence of low-cycle dominance at weak coupling. Furthermore, we resummed certain classes of one-cycle negative geometries to all loop orders, and provided strong-coupling expansions of these formulas.

There are many natural questions and directions to explore further. The computation of ${\cal F}(g,z)$ at the next loop order, four loops, would require new methods, tailored to IR finite objects and negative geometries; with more generic integration technology the four-loop calculation is not currently within reach. One option is to find new differential equations, beyond the Laplace equation, that hold for more general negative geometries. Another direction is to consider more types of resummations. Our work only scratched the surface of this important question, and we only considered special classes of one-cycle geometries. The real breakthrough would be to sum \emph{all} one-cycle geometries and study the properties of this object, mainly the strong coupling expansion. We know that the summation of all zero-cycle (tree) negative geometries has good strong coupling behavior, and it would be very interesting to see if this extends to the set of all geometries with fixed number of cycles. This would also necessarily require a new integration method.

The most important open problem is related to the cusp anomalous dimension $\Gamma_{\rm cusp}(g)$. This object is predicted from integrability via the BES equation \cite{Beisert:2006ez}, but how to derive it from amplitudes is a big challenge, which would very likely uncover an imprint of integrability in ${\cal N}=4$ SYM amplitudes. While this seems like an extremely difficult problem, even in the context of negative geometries, we have one particular technical aspect to study and understand -- the presence of only single zeta values in $\Gamma_{\rm cusp}(g)$. As we have seen in Section \ref{sec:cusp}, generic individual negative geometries (even with only one cycle) do contribute  multiple zeta values to $\Gamma_{\rm cusp}$. These MZVs must cancel while summing over multiple (or all) negative geometries. Exploring the precise mechanism of how these cancellations happen would be an important first step towards a better understanding of integrability constraints on negative geometries and amplitudes in general. 

\section*{Acknowledgments}

We thank Nima Arkani-Hamed, Johannes Henn, Chia-Kai Kuo, Artyom Lisitsyn, Elia Mazzucchelli, Melvyn Nabavi, Dan Romik, Marcelo Augusto Ferreira Dos Santos and Qinglin Yang for very useful discussions. LD thanks the Max-Planck-Institut f\"{u}r Physik for hospitality while this paper was completed.  This work was supported by the US Department of Energy under contract 
DE--AC02--76SF00515 and grant No.~SC0009999, and by the funds of the University of California.
YX is supported by the grants from the NNSF of China with Grant No:12247103 and 12575078.

\vspace{0.2cm}

\appendix

\section{Integrands of four-loop negative geometries}
\label{app:3-cycle}

The integrand for $L=4$ negative geometries is known for all tree, one-cycle and two-cycle graphs and is provided in \cite{Arkani-Hamed:2021iya,Brown:2023mqi}. The only missing term is the integrand for the three-cycle diagram
\begin{equation}
    \tilde{\Omega}^{3\text{-cycle}}_4 = \begin{tikzpicture}[baseline={(0,0.75)cm}]
        \draw[Maroon, ultra thick] (0,0)--(1.75,0);
        \draw[Maroon, ultra thick] (0,0)--(0.87,1.5);
        \draw[Maroon, ultra thick] (0.87,1.5)--(1.75,0);
        \draw[Maroon, ultra thick] (0,0)--(0.87,0.65) node[at start, black] {\Huge$\bullet$};
        \draw[Maroon, ultra thick] (1.75,0)--(0.87,0.65) node[at start, black] {\Huge$\bullet$};
        \draw[Maroon, ultra thick] (0.87,1.5)--(0.87,0.65) node[at start, black] {\Huge$\bullet$} node[at end, black] {\Huge$\bullet$};
    \node[below left] at (0,0) {$AB_1$};
    \node[below right] at (1.75,0) {$AB_4$};
    \node[above] at (0.87,1.6) {$AB_2$};
    \node[below] at (0.87,0.55) {$AB_3$};
    \end{tikzpicture}
\end{equation}
Following the procedure in \cite{Brown:2023mqi}, the numerator of the integrand can be written as
\begin{align}
\cal{N} &= n_{12}n_{23}n_{34}n_{14}n_{24}n_{13}  - \sum_{\pi_1} R_{123}^\text{1-cycle} n_{14} n_{24}n_{34} + \sum_{\pi_2} R_{1234}^\text{1-cycle} n_{13}n_{24}\nonumber\\ 
&\hspace{0.3cm} - \sum_{\pi_3} R_{1234,13}^\text{2-cycle}n_{24} + R_{1234}^\text{3-cycle} \,.
\end{align}
Graphically, this is
\begin{align}
    \mathcal{N}=&\,\begin{tikzpicture}[baseline={(0,0.75)cm}]
        \draw[Maroon, ultra thick, dashed] (0,0)--(1.75,0);
        \draw[Maroon, ultra thick, dashed] (0,0)--(0.87,1.5);
        \draw[Maroon, ultra thick, dashed] (0.87,1.5)--(1.75,0);
        \draw[Maroon, ultra thick, dashed] (0,0)--(0.87,0.65) node[at start, black] {\Huge$\bullet$};
        \draw[Maroon, ultra thick, dashed] (1.75,0)--(0.87,0.65) node[at start, black] {\Huge$\bullet$};
        \draw[Maroon, ultra thick, dashed] (0.87,1.5)--(0.87,0.65) node[at start, black] {\Huge$\bullet$} node[at end, black] {\Huge$\bullet$};
    \node[below left] at (0,0) {$AB_1$};
    \node[below right] at (1.75,0) {$AB_4$};
    \node[above] at (0.87,1.6) {$AB_2$};
    \node[below] at (0.87,0.55) {$AB_3$};
    \end{tikzpicture}\,+\sum_{\pi_1}\,\begin{tikzpicture}[baseline={(0,0.75)cm}]
        \draw[chalkdust] (0,0)--(1.75,0);
        \draw[chalkdust] (0,0)--(0.87,1.5);
        \draw[chalkdust] (0.87,1.5)--(1.75,0);
        \draw[Maroon, ultra thick, dashed] (0,0)--(0.87,0.65) node[at start, black] {\Huge$\bullet$};
        \draw[Maroon, ultra thick, dashed] (1.75,0)--(0.87,0.65) node[at start, black] {\Huge$\bullet$};
        \draw[Maroon, ultra thick, dashed] (0.87,1.5)--(0.87,0.65) node[at start, black] {\Huge$\bullet$} node[at end, black] {\Huge$\bullet$};
    \node[below left] at (0,0) {$AB_1$};
    \node[below right] at (1.75,0) {$AB_4$};
    \node[above] at (0.87,1.6) {$AB_2$};
    \node[below] at (0.87,0.55) {$AB_3$};
    \end{tikzpicture}\,+\sum_{\pi_2}\,\begin{tikzpicture}[baseline={(0,0.75)cm}]
        \draw[chalkdust] (0,0)--(1.75,0);
        \draw[chalkdust] (0,0)--(0.87,1.5);
        \draw[Maroon, ultra thick, dashed] (0.87,1.5)--(1.75,0);
        \draw[Maroon, ultra thick, dashed] (0,0)--(0.87,0.65) node[at start, black] {\Huge$\bullet$};
        \draw[chalkdust] (1.75,0)--(0.87,0.65) node[at start, black] {\Huge$\bullet$};
        \draw[chalkdust] (0.87,1.5)--(0.87,0.65) node[at start, black] {\Huge$\bullet$} node[at end, black] {\Huge$\bullet$};
    \node[below left] at (0,0) {$AB_1$};
    \node[below right] at (1.75,0) {$AB_4$};
    \node[above] at (0.87,1.6) {$AB_2$};
    \node[below] at (0.87,0.55) {$AB_3$};
    \end{tikzpicture}\nonumber\\
    &+\sum_{\pi_3}\,\begin{tikzpicture}[baseline={(0,0.75)cm}]
        \draw[chalkdust] (0,0)--(1.75,0);
        \draw[chalkdust] (0,0)--(0.87,1.5);
        \draw[chalkdust] (0.87,1.5)--(1.75,0);
        \draw[Maroon, ultra thick, dashed] (0,0)--(0.87,0.65) node[at start, black] {\Huge$\bullet$};
        \draw[chalkdust] (1.75,0)--(0.87,0.65) node[at start, black] {\Huge$\bullet$};
        \draw[chalkdust] (0.87,1.5)--(0.87,0.65) node[at start, black] {\Huge$\bullet$} node[at end, black] {\Huge$\bullet$};
    \node[below left] at (0,0) {$AB_1$};
    \node[below right] at (1.75,0) {$AB_4$};
    \node[above] at (0.87,1.6) {$AB_2$};
    \node[below] at (0.87,0.55) {$AB_3$};
    \end{tikzpicture}\,+\,\begin{tikzpicture}[baseline={(0,0.75)cm}]
        \draw[chalkdust] (0,0)--(1.75,0);
        \draw[chalkdust] (0,0)--(0.87,1.5);
        \draw[chalkdust] (0.87,1.5)--(1.75,0);
        \draw[chalkdust] (0,0)--(0.87,0.65) node[at start, black] {\Huge$\bullet$};
        \draw[chalkdust] (1.75,0)--(0.87,0.65) node[at start, black] {\Huge$\bullet$};
        \draw[chalkdust] (0.87,1.5)--(0.87,0.65) node[at start, black] {\Huge$\bullet$} node[at end, black] {\Huge$\bullet$};
    \node[below left] at (0,0) {$AB_1$};
    \node[below right] at (1.75,0) {$AB_4$};
    \node[above] at (0.87,1.6) {$AB_2$};
    \node[below] at (0.87,0.55) {$AB_3$};
    \end{tikzpicture}\,.
\end{align}
%
The permutation $\pi_1$ is over all embedded one-loop triangles, there are 4 of them $(123)$, $(124)$, $(134)$, $(234)$. The permutation $\pi_2$ is over all embedded one-loop "squares" (where the square is marked by the non-dashed line, given a particular ordering), there are 3 of them $(1234)$, $(1243)$, $(1324)$. The permutation $\pi_3$ is over all embedded two-loop sub-topologies (where we ignore one of the links), this is 6 terms (ignoring one of the links each). The final purely three-loop piece is again universal for any link sign assignment and satisfies
\begin{equation}
\boxed{R^\text{3-cycle}_{1234} = 0 \quad \mbox{on any cut when $\la AB_i AB_j\ra$ has a definite sign}}
\end{equation}
Here we use the remainders $R_{1234}^\text{1-cycle}$ and $R_{1234,13}^\text{2-cycle}$ from \cite{Brown:2023mqi}, and determine the remainder $R_{1234}^\text{3-cycle}$. We follow the same procedure, ie. first to write the ansatz of all terms which satisfy singlet and doublet conditions -- only involving cuts in two of the loops, whenever $\la AB_iAB_j\ra$ has definite sign, this cut is not allowed. This is a subset of all cuts that render a definite sign, but we make this subset of conditions manifest term-by-term in our basis. 

We use the short-hand notation
\begin{align}
n_{ij}^{(a)} &\equiv \la AB_i12\ra \la AB_j34\ra + \la AB_i34\ra\la AB_j12\ra\,,\\
n_{ij}^{(b)} &\equiv \la AB_i23\ra \la AB_j14\ra + \la AB_i14\ra\la AB_j23\ra\,,\\
n_{ij}^{(c)} &\equiv \la AB_i13\ra \la AB_j24\ra + \la AB_i24\ra\la AB_j13\ra\,.
\end{align}
and 
\begin{align}
n_{i}^{(a)} &\equiv \frac12 n_{ii}^{(a)} = \la AB_i12\ra \la AB_i34\ra\,, \\
n_{i}^{(b)} &\equiv \frac12 n_{ii}^{(b)} = \la AB_i23\ra \la AB_i14\ra\,,
\end{align}
and finally
\begin{align}
n^{(a)}_{1234} &= \la AB_112\ra\la AB_234\ra\la AB_312\ra\la AB_434\ra +\la AB_134\ra\la AB_212\ra\la AB_334\ra\la AB_412\ra\,,\\
n^{(b)}_{1234} &= \la AB_123\ra\la AB_214\ra\la AB_323\ra\la AB_414\ra +\la AB_114\ra\la AB_223\ra\la AB_314\ra\la AB_423\ra\,,\\
n^{(c)}_{1234} &= \la AB_113\ra\la AB_224\ra\la AB_313\ra\la AB_424\ra +\la AB_124\ra\la AB_213\ra\la AB_324\ra\la AB_413\ra\,.
\end{align}
We start with $n_1^{(a)}n_2^{(a)}n_3^{(a)}n_4^{(a)}$ in the $a$-sector which already takes care of all conditions, and there are following terms to complete it:
\begin{align}
   & {\cal B}_a^{(1)} = n_1^{(a)}n_2^{(a)}n_3^{(a)}n_4^{(a)}n_{12}^{(a)}n_{34}^{(a)}, \quad   {\cal B}_a^{(2)} = n_1^{(a)}n_2^{(a)}n_3^{(a)}n_4^{(a)}n_{12}^{(a)}n_{34}^{(c)},\\
   & {\cal B}_a^{(3)} = n_1^{(a)}n_2^{(a)}n_3^{(a)}n_4^{(a)}n_{12}^{(c)}n_{34}^{(c)},
\end{align}
where each term is implicitly summed over permutations (modulo the symmetry factor). For example, 
\begin{equation}
    {\cal B}_a^{(1)} = n_1^{(a)}n_2^{(a)}n_3^{(a)}n_4^{(a)}(n_{12}^{(a)}n_{34}^{(a)}+n_{13}^{(a)}n_{24}^{(a)}+n_{14}^{(a)}n_{23}^{(a)})\,.
\end{equation}
Note that in the ansatz there are no terms of the form $n_1^{(a)}n_2^{(a)}n_3^{(a)}n_4^{(a)}n^{(a)}_{1234}$, because after permutation
\begin{equation}
    n^{(a)}_{1234}+n^{(a)}_{1324}+n^{(a)}_{1243} = \frac{1}{2}\left(n^{(a)}_{12}n^{(a)}_{34}+n^{(a)}_{14}n^{(a)}_{23}+n^{(a)}_{13}n^{(a)}_{24}\right)\,.
\end{equation}
In the ${\cal B}_{ab}$ sector, we again start with $n_1^{(a)}n_2^{(a)}n_3^{(a)}n_4^{(a)}$ and complete it with 
\begin{align}
   & {\cal B}_{ab}^{(1)} = n_1^{(a)}n_2^{(a)}n_3^{(a)}n_4^{(a)}n_{12}^{(a)}n_{34}^{(b)}, \quad   {\cal B}_{ab}^{(2)} = n_1^{(a)}n_2^{(a)}n_3^{(a)}n_4^{(a)}n_{12}^{(b)}n_{34}^{(b)},\\
   & {\cal B}_{ab}^{(3)} = n_1^{(a)}n_2^{(a)}n_3^{(a)}n_4^{(a)}n_{12}^{(b)}n_{34}^{(c)}\,.
\end{align}
The next set of four $n_1$ terms has a $ab$ mixing in the $n_1$ terms with 3-1 and 2-2 mixings. For the 3-1 mixing we have following terms 
\begin{align}
   & {\cal B}_{ab}^{(4)} = n_1^{(a)}n_2^{(a)}n_3^{(a)}n_4^{(b)}n_{12}^{(b)}n_{34}^{(b)}, \quad   {\cal B}_{ab}^{(5)} = n_1^{(a)}n_2^{(a)}n_3^{(a)}n_4^{(b)}n_{12}^{(b)}n_{34}^{(a)},\\
   & {\cal B}_{ab}^{(6)} = n_1^{(a)}n_2^{(a)}n_3^{(a)}n_4^{(b)}n_{12}^{(c)}n_{34}^{(a)}, \quad  {\cal B}_{ab}^{(7)} = n_1^{(a)}n_2^{(a)}n_3^{(a)}n_4^{(b)}n_{12}^{(a)}n_{34}^{(a)}.
\end{align}
For the 2-2 mixing we have 
\begin{align}
   & {\cal B}_{ab}^{(8)} = n_1^{(a)}n_2^{(a)}n_3^{(b)}n_4^{(b)}n_{12}^{(b)}n_{34}^{(a)}, \quad   {\cal B}_{ab}^{(9)} = n_1^{(a)}n_2^{(a)}n_3^{(b)}n_4^{(b)}n_{12}^{(a)}n_{34}^{(a)},\\
   & {\cal B}_{ab}^{(10)} = n_1^{(a)}n_2^{(a)}n_3^{(b)}n_4^{(b)}n_{12}^{(c)}n_{34}^{(a)}, \quad  {\cal B}_{ab}^{(11)} = n_1^{(a)}n_2^{(a)}n_3^{(b)}n_4^{(b)}n_{13}^{(a)}n_{24}^{(a)}\,.
\end{align}
now considering only 3 insertions of $n_1$ we have in $a$-sector,
\begin{align}
   & {\cal B}_{a}^{(4)} = n_1^{(a)}n_2^{(a)}n_3^{(a)}n_{24}^{(a)}n_{14}^{(c)}n_{34}^{(a)}, \quad {\cal B}_{a}^{(5)} = n_1^{(a)}n_2^{(a)}n_3^{(a)}n_{24}^{(a)}n_{14}^{(c)}n_{34}^{(c)},\\
\end{align}
and in $a-b$ sector,
\begin{align}
   & {\cal B}_{ab}^{(12)} = n_1^{(a)}n_2^{(a)}n_3^{(a)}n_{24}^{(a)}n_{14}^{(c)}n_{34}^{(b)}, \quad {\cal B}_{ab}^{(13)} = n_1^{(a)}n_2^{(a)}n_3^{(b)}n_{14}^{(c)}n_{24}^{(a)}n_{34}^{(a)},\\
   & {\cal B}_{ab}^{(14)} = n_1^{(a)}n_2^{(a)}n_3^{(b)}n_{14}^{(c)}n_{24}^{(b)}n_{34}^{(a)}, \quad {\cal B}_{ab}^{(15)} = n_1^{(a)}n_2^{(a)}n_3^{(b)}n_{14}^{(c)}n_{24}^{(c)}n_{34}^{(a)},\\
  & {\cal B}_{ab}^{(16)} = n_1^{(a)}n_2^{(a)}n_3^{(b)}n_{14}^{(b)}n_{24}^{(b)}n_{34}^{(c)}\,.
\end{align}
next we have insertion of two $n_1$'s. The term $n_1^{(a)}n_2^{(a)}n_{34}^{(a)}n_{34}^{(c)}$ already takes care of all links, so we can complete it with arbitrary collection of other terms. In the $a$-sector this is 
\begin{align}
   & {\cal B}_{a}^{(6)} = n_1^{(a)}n_2^{(a)}n_{34}^{(a)}n_{34}^{(c)}n_{12}^{(a)}n_{34}^{(a)}, \quad {\cal B}_{a}^{(7)} = n_1^{(a)}n_2^{(a)}n_{34}^{(a)}n_{34}^{(c)}n_{12}^{(a)}n_{34}^{(c)},\\
   & {\cal B}_{a}^{(8)} = n_1^{(a)}n_2^{(a)}n_{34}^{(a)}n_{34}^{(c)}n_{12}^{(c)}n_{34}^{(a)}, \quad {\cal B}_{a}^{(9)} = n_1^{(a)}n_2^{(a)}n_{34}^{(a)}n_{34}^{(c)}n_{12}^{(c)}n_{34}^{(c)},\\
 & {\cal B}_{a}^{(10)} = n_1^{(a)}n_2^{(a)}n_{34}^{(a)}n_{34}^{(c)}n_{13}^{(a)}n_{24}^{(a)}, \quad {\cal B}_{a}^{(11)} = n_1^{(a)}n_2^{(a)}n_{34}^{(a)}n_{34}^{(c)}n_{13}^{(a)}n_{24}^{(c)},\\
 & {\cal B}_{a}^{(12)} = n_1^{(a)}n_2^{(a)}n_{34}^{(a)}n_{34}^{(c)}n_{13}^{(c)}n_{24}^{(c)}\,,
\end{align}
and in the $ab$ sector, it is 
\begin{align}
   & {\cal B}_{ab}^{(17)} = n_1^{(a)}n_2^{(a)}n_{34}^{(a)}n_{34}^{(c)}n_{12}^{(a)}n_{34}^{(b)}, \quad {\cal B}_{ab}^{(18)} = n_1^{(a)}n_2^{(a)}n_{34}^{(a)}n_{34}^{(c)}n_{12}^{(b)}n_{34}^{(a)},\\
   & {\cal B}_{ab}^{(19)} = n_1^{(a)}n_2^{(a)}n_{34}^{(a)}n_{34}^{(c)}n_{12}^{(b)}n_{34}^{(b)}, \quad {\cal B}_{ab}^{(20)} = n_1^{(a)}n_2^{(a)}n_{34}^{(a)}n_{34}^{(c)}n_{12}^{(b)}n_{34}^{(c)},\\
 & {\cal B}_{ab}^{(21)} = n_1^{(a)}n_2^{(a)}n_{34}^{(a)}n_{34}^{(c)}n_{12}^{(c)}n_{34}^{(b)}, \quad {\cal B}_{ab}^{(22)} = n_1^{(a)}n_2^{(a)}n_{34}^{(a)}n_{34}^{(c)}n_{13}^{(b)}n_{24}^{(b)},\\
 & {\cal B}_{ab}^{(23)} = n_1^{(a)}n_2^{(a)}n_{34}^{(a)}n_{34}^{(c)}n_{13}^{(a)}n_{24}^{(b)}, \quad {\cal B}_{ab}^{(24)} = n_1^{(a)}n_2^{(a)}n_{34}^{(a)}n_{34}^{(c)}n_{13}^{(c)}n_{24}^{(b)}\,.
\end{align}
The next option is to have $n_1^{(a)}n_2^{(a)}n_{34}^{(c)}$, the link $n_{34}^{(c)}$ is needed, there is no way to replace it. Because we do not want $n_{34}^{(a)}$ the only option is to connect through $n_{34}^{(b)}$ and we get 
\begin{align}
     {\cal B}_{ab}^{(25)} = n_1^{(a)}n_2^{(a)}n_{34}^{(c)}n_{34}^{(b)}n_{13}^{(a)}n_{24}^{(a)}\,.
\end{align}
We can also have $n_1^{(a)}n_1^{(b)}$ which has three options,
\begin{align}
   & {\cal B}_{ab}^{(26)} = n_1^{(a)}n_2^{(b)}n_{34}^{(b)}n_{34}^{(c)}n_{12}^{(a)}n_{34}^{(a)}, \quad {\cal B}_{ab}^{(27)} = n_1^{(a)}n_2^{(b)}n_{34}^{(b)}n_{34}^{(c)}n_{13}^{(a)}n_{24}^{(a)},\\
   & {\cal B}_{ab}^{(28)} = n_1^{(a)}n_2^{(b)}n_{34}^{(b)}n_{34}^{(c)}n_{13}^{(b)}n_{24}^{(a)}\,,
\end{align}
and hence we can set its coefficients to zero. If we solve for the double pole and spurious cuts cancellations, we get an unique solution. As it turns out there are still two parameters left in the ansatz but that is just because the basis is actually not independent. In particular, after symmetrization
\begin{align}
  &{\cal B}_a^{(2)}-{\cal B}_a^{(4)}-\frac14{\cal B}_a^{(6)}+\frac12{\cal B}_a^{(10)} = 0\,,\\
  &{\cal B}_a^{(2)}-{\cal B}_a^{(5)}-\frac14{\cal B}_a^{(9)}+\frac12{\cal B}_a^{(12)}+2{\cal B}_{ab}^{(6)}+{\cal B}_{ab}^{(10)}-{\cal B}_{ab}^{(15)} = 0\,.
\end{align}
We can then eliminate two of the terms, e.g. $B_a^{(10)}$ and $B_{a}^{(15)}$, write the ansatz,
\begin{equation} \label{eq:3-loop_remainder}
    A = \sum_j c_a^{(j)} {\cal B}_a^{(j)} + \sum_k c_{ab}^{(k)}{\cal B}_{ab}^{(k)}\,.
\end{equation}
In the ansatz we assume each term to be completely permuted (we do not factor out the symmetry factor). As a solution, we get
\begin{align}
& c_a^{(1)} = \frac12, \quad c_a^{(2)} = 2,\quad c_a^{(3)} = \frac12, \quad c_a^{(4)} = 0, \quad c_a^{(5)} = 1, \quad c_a^{(6)} = -\frac14, \quad c_a^{(7)} = -\frac14, \nonumber\\ & c_a^{(8)} = -\frac14, \quad c_a^{(9)} = 0, \quad c_a^{(10)} = 0,\quad c_a^{(11)} = -1,\quad c_a^{(12)} = -\frac12\,,
\end{align}
and 
\begin{align}
& c_{ab}^{(1)} = 0, \quad c_{ab}^{(2)} = -\frac12, \quad c_{ab}^{(3)} = 0, \quad c_{ab}^{(4)} = 1, \quad c_{ab}^{(5)} = 1, \quad c_{ab}^{(6)} = 0, \quad c_{ab}^{(7)} = 0, \quad c_{ab}^{(8)} = 0, \nonumber\\ & c_{ab}^{(9)} = 0, \quad c_{ab}^{(10)} = -1, \quad c_{ab}^{(11)} = 0, \quad c_{ab}^{(12)} = 0, \quad c_{ab}^{(13)} = 0, \quad c_{ab}^{(14)} = -1, \quad c_{ab}^{(15)} = 0, \nonumber\\ & c_{ab}^{(16)} = 1, \quad c_{ab}^{(17)} = 0, \quad c_{ab}^{(18)} = 0, \quad c_{ab}^{(19)} = 0, \quad c_{ab}^{(20)} = 0, \quad c_{ab}^{(21)} = 0, \quad c_{ab}^{(22)} = \frac12,\nonumber\\ & c_{ab}^{(23)} = 0, \quad c_{ab}^{(24)} = 0, \quad c_{ab}^{(25)} = 0, \quad c_{ab}^{(26)} = -\frac14, \quad c_{ab}^{(27)} = -\frac12, \quad c_{ab}^{(28)} = -1 \,.
\end{align}
\par
With the 3-cycle integrand solved, we can now put the full $ \tilde{\Omega}_4$ together. Starting with trees, we have
\begin{equation}
     \tilde{\Omega}^{\text{tree}}_4 = \frac{\la 1234\ra^5 n_{12} n_{23}(n_{34}\la AB_2 AB_4\ra+n_{24} \la AB_3 AB_4\ra)}{D_1D_2D_3D_4\la AB_1AB_2\ra \la AB_2 AB_3 \ra \la AB_3 AB_4\ra \la AB_2 AB_4\ra}\,.
\end{equation}
For 1-cycle and 2-cycle, we have
\begin{equation}
 \tilde{\Omega}^{1-\text{cycle}}_4 = \frac{(n_{12}n_{23}n_{34}n_{14}+R^{1-\text{cycle}}_{1234})\la AB_2 AB_4\ra+ \la AB_4 AB_1 \ra n_{12}(n_{23}n_{34}n_{24}-R^{1-\text{cycle}}_{234})}{D_1D_2D_3 D_4 \la AB_1AB_2\ra \la AB_2 AB_3 \ra\la AB_3 AB_4 \ra \la AB_4 AB_1 \ra \la AB_2 AB_4\ra} \,,
\end{equation}

\begin{equation}
 \tilde{\Omega}^{2-\text{cycle}}_4 = \frac{n_{12}n_{23}n_{34}n_{41}n_{13}-\sum^{}_{\pi_1}R^{1-\text{cycle}}_{123}n_{14}n_{34}+R^{1-\text{cycle}}_{1234}n_{13}-R^{2-\text{cycle}}_{1234,13}}{D_1D_2D_3 D_4 \la AB_1AB_2\ra \la AB_2 AB_3 \ra\la AB_3 AB_4 \ra \la AB_4 AB_1 \ra \la AB_1 AB_3\ra} \,,
\end{equation}
respectively. Finally, for the $3-$cycle integrand, putting everything together gives us
\begin{align}
 &\tilde{\Omega}^{3-\text{cycle}}_4 = \frac{n_{12}n_{23}n_{34}n_{14}n_{24}n_{13}  - \sum_{\pi_1} R_{123}^\text{1-cycle} n_{14} n_{24}n_{34} }{D_1D_2D_3 D_4 \la AB_1AB_2\ra \la AB_2 AB_3 \ra\la AB_3 AB_4 \ra \la AB_4 AB_1 \ra \la AB_1 AB_3\ra \la AB_2 AB_4\ra} \nonumber \\ 
 & +\frac{\sum_{\pi_2} R_{1234}^\text{1-cycle} n_{13}^{(-)}n_{24}^{(-)}- \sum_{\pi_3} R_{1234,13}^\text{2-cycle}n_{24}^{(-)} + R_{1234}^\text{3-cycle}}{D_1D_2D_3 D_4 \la AB_1AB_2\ra \la AB_2 AB_3 \ra\la AB_3 AB_4 \ra \la AB_4 AB_1 \ra \la AB_1 AB_3\ra \la AB_2 AB_4\ra}\,,
\end{align}
where the remainder functions are listed below
\begin{align}
R^{1-\text{cycle}}_{123} = \left\{ 4 n_1^{(a)} n_2^{(a)} n_3^{(a)} - \left( n_1^{(a)} n_{23}^{(a)} n_{23}^{(c)} + n_2^{(a)} n_{13}^{(a)} n_{13}^{(c)} + n_3^{(a)} n_{12}^{(a)} n_{12}^{(c)} \right) \right\} + (a \to b),
\end{align}

\begin{align}
R^{1-\text{cycle}}_{1234} &= 12\,n_1^{(a)}n_2^{(a)}n_3^{(a)}n_4^{(a)} -\bigg\{n_1^{(a)} n_3^{(a)} n_{24}^{(a)} n_{24}^{(c)} + n_2^{(a)} n_4^{(a)} n_{13}^{(a)} n_{13}^{(c)}\bigg\}  \label{R1234ans2} \nonumber\\
&\hspace{-0.5cm}- \bigg\{n_1^{(a)} n_2^{(a)} n_{34}^{(a)} n_{34}^{(c)} {+} n_2^{(a)} n_3^{(a)} n_{14}^{(a)} n_{14}^{(c)} {+} n_3^{(a)} n_4^{(a)} n_{12}^{(a)} n_{12}^{(c)} {+} n_1^{(a)} n_4^{(a)} n_{23}^{(a)} n_{23}^{(c)} \Bigg\}\nonumber\\
&\hspace{-0.5cm}-\, n^{(a)}_{1234}n^{(c)}_{1234}\hspace{8.3cm} + (a\rightarrow b)\,. 
\end{align}

\begin{equation}
R_{1234,13}^{\rm 2-\text{cycle}} = \sum_{j=1}^{15} c_j\left({\cal B}_a^{(j)} + {\cal B}_b^{(j)}\right) + \sum_{k=1}^{12} d_k\,{\cal B}_{ab}^{(k)}\,,
\end{equation}
where ${\cal B}_a^{(j)}$, ${\cal B}_b^{(j)}$, and ${\cal B}_{ab}^{(k)}$ are defined in the appendix B in \cite{Brown:2023mqi} and the c's are 
\begin{align}
&c_1 = c_2 = -8,\,\,\, c_3 = -2,\,\,\, c_8 = c_9 = c_{12} = c_{13} = 0,\\
&c_4 = c_5 = c_6 = c_7 = c_{10} = c_{11} = c_{14} = c_{15} = 1, \nonumber\\
& d_1 = d_3 = d_4 = d_5 = 0, \,\,\, d_2 = -4,\,\,\,d_6 = d_7 = d_8 = 0, \,\,\, d_9 = d_{10} = d_{11} = 1,\,\,\, d_{12} = -2\,.\nonumber
\end{align}
and $R^{3-\text{cycle}}_{1234}$ is given in \eqref{eq:3-loop_remainder}.
\section{Results for the $\Gamma_{\text{cusp}}$ Contribution of Infinite Classes}
\label{sec: Gamma-cups-resum}
In this section, we provide detailed results for how the classes of diagrams where a ladder is attached to the triangle and box diagrams contributes to $\Gamma_{\rm cusp}$ after resummation. Recall that the $\mathcal{I}$ operation is the integral \eqref{Iintegral}, and we can perform it in our case explicitly. We can use the shuffle identities to put any HPL that appear in the ${\cal F}_3(z)$ in a form
\begin{equation}
    \log^a(z) H_{-m_n,...,-m_1}(z)\,,
\end{equation}
where $m_i$ are non-zero. Then the integrals we have to perform to get the $\Gamma_{\rm cusp}$ contribution are of the form
\begin{equation}
    \frac{1}{2\pi} \int^\pi_{-\pi} H_{-m_2,-m_1}(e^{i\phi})(i\phi)^a\frac{\cosh(\sqrt{2}g\phi)}{\cosh{\sqrt{2}g\pi}} d\phi \,,
\end{equation}
which result in the following types of sums
\begin{align}
    \sum_{j_2>j_1>0}^\infty \frac{j_2^a}{j_1^b(j^2_2+2g^2)^3}\,,\ \ -3\le a\le 4\,,\ \ 1\le b\le 3\,.
\end{align}
The precise list of $a$'s and $b$'s needed is 
\begin{center}
\begin{tabular}{c|c|c|c|c|c|c|c|c|c|c|c|}
    $a$ & -3 & -2 & -1 & -1 & 0 & 1 & 1 & 2 & 3 & 3 & 4\\
    \hline
    $b$  & 1 & 2 & 1 & 3 & 2 & 1 & 3 & 2 & 1 & 3 &2 \\
\end{tabular}
\end{center}
All of these sums can be performed analytically and be written in terms of Gamma functions and their derivatives (as explained in Appendix \ref{app:sums}). Note that the $\frac{\Gamma'(2-i\sqrt{2}g)}{\Gamma(2-i\sqrt{2}g)} = \psi^0(2-i\sqrt{2}g)$ terms that appear go like $\log g$ at strong coupling, but these terms end up canceling out between each other. While the leading $\log{g}$ terms that come from the HPLs with -1 indices cancel out between $H_{-1,-1}(z)$, $H_{-1,-1,0,0}(z)$, and $H_{-1}(z)$, we find that the lower order terms that have $\log(g)$ also cancel or come in pairs that go like $i\pi$ (such as $\log(-g)-\log(g)=-i\pi$). The full result is too long to include here (put in ancillary file {\tt IFladdertriangle.m}) but at $g \gg 1$ it becomes a polynomial in $g$ as seen below 
\begin{align}
    \mathcal{I}[g^4\mathcal{F}_{3}(g,z){\cal F}_{\rm ladder}(g,z)]_{g>>1} &\sim \frac{11}{45} \pi ^3 \sqrt{2} g^5 - 8 g^4+\frac{\sqrt{2} \left(1+\pi ^2\right)}{ \pi }g^3
    \\ \nonumber
    &-\frac{19}{9} g^2+\frac{9\sqrt{2} }{8\pi}g \,.
  \end{align}
Using $g\partial_g \Gamma_{\rm cusp}=-8\mathcal{I}[{\cal F}(g,z)]$, we can extract the contribution to $\Gamma_{\rm cusp}$ as
\begin{align} \label{eq:gammacusptriladder}
    \Gamma_{\rm cusp} &\sim -\frac{88}{225} \pi ^3 \sqrt{2} g^5 +16 g^4-\frac{8\sqrt{2} \left(1+\pi ^2\right)}{ 3\pi }g^3
    \\ \nonumber
    &+\frac{76}{9} g^2-\frac{9\sqrt{2} }{\pi}g  + {\cal O}(1) \label{triladder}\,.
\end{align}
We see that the leading order scales like $\Gamma_{\rm cusp}\sim g^5$ which is in contrast with the correct behavior $\Gamma_{\rm cusp}\sim g$ and it is also faster that the resummed Wilson loop function ${\cal F}_{rs}(g,z)\sim g^4$. Note that this is similar to what happens with the resummation of all tree diagrams where ${\cal F}_{\rm tree}(g,z) \sim {\cal O}(\frac{1}{g^2})$ for $g\gg 1$ but the contribution to Gamma cusp is $\Gamma_{\rm tree}\sim g$. On the other hand, even with the behavior $\sim g^5$, this is better than the leading $\sim g^6$ behavior of a triangle diagram at weak coupling. Hence we can refer to this sum also a partially summable. Note that the appearance of Euler constant $\gamma$ in (\ref{triladder}) is spurious from the point of view of the full result where only the single zeta values appear. 

Similar result can be obtained for the box and ladder series, 
\begin{center}
  \begin{align}
        &\Gamma_{\rm cusp} \left( g^6\begin{tikzpicture}[baseline={(0,-0.1)cm}]
        \draw[Maroon, ultra thick] (0,-0.5)--(1,-0.5);
        \draw[Maroon, ultra thick] (1,0.5)--(1,-0.5);
        \draw[Maroon, ultra thick] (1,0.5)--(0,0.5);
        \draw[Maroon, ultra thick] (0,-0.5)--(0,0.5);
        \draw[Maroon, ultra thick] (1,-0.5)--(1,-0.5);
        \node at (0,-0.5) {\Huge$\bullet$};
        \node at (0,0.5) {\Huge$\bullet$};
        \node at (1,0.5) {\Huge$\bullet$};
        \filldraw[white] (1,-0.5) circle (5pt) node[]{};
        \node at (1,-0.5) {\Large$\otimes$};
    \end{tikzpicture} - g^{8} \begin{tikzpicture}[baseline={(0,-0.1)cm}]
        \draw[Maroon, ultra thick] (0,-0.5)--(1,-0.5);
        \draw[Maroon, ultra thick] (1,0.5)--(1,-0.5);
        \draw[Maroon, ultra thick] (1,0.5)--(0,0.5);
        \draw[Maroon, ultra thick] (0,-0.5)--(0,0.5);
        \draw[Maroon, ultra thick] (1,-0.5)--(2,-0.5);
        \node at (0,-0.5) {\Huge$\bullet$};
        \node at (1,-0.5) {\Huge$\bullet$};
        \node at (0,0.5) {\Huge$\bullet$};
        \node at (1,0.5) {\Huge$\bullet$};
       \filldraw[white] (2,-0.5) circle (5pt) node[]{};
        \node at (2,-0.5) {\Large$\otimes$};
\end{tikzpicture} + g^{10}  \begin{tikzpicture}[baseline={(0,-0.1)cm}]
        \draw[Maroon, ultra thick] (0,-0.5)--(1,-0.5);
        \draw[Maroon, ultra thick] (1,0.5)--(1,-0.5);
        \draw[Maroon, ultra thick] (1,0.5)--(0,0.5);
        \draw[Maroon, ultra thick] (0,-0.5)--(0,0.5);
        \draw[Maroon, ultra thick] (1,-0.5)--(3,-0.5);
        \node at (0,-0.5) {\Huge$\bullet$};
        \node at (1,-0.5) {\Huge$\bullet$};
        \node at (0,0.5) {\Huge$\bullet$};
        \node at (1,0.5) {\Huge$\bullet$};
        \node at (2,-0.5) {\Huge$\bullet$};
       \filldraw[white] (3,-0.5) circle (5pt) node[]{};
        \node at (3,-0.5) {\Large$\otimes$};
\end{tikzpicture}+ \cdots\right)= \\
& \Gamma_{\rm cusp}\left( g^6 \begin{tikzpicture}[baseline={(0,-0.1)cm}]
        \draw[Maroon, ultra thick] (0,-0.5)--(1,-0.5);
        \draw[Maroon, ultra thick] (1,0.5)--(1,-0.5);
        \draw[Maroon, ultra thick] (1,0.5)--(0,0.5);
        \draw[Maroon, ultra thick] (0,-0.5)--(0,0.5);
        \draw[Maroon, ultra thick] (1,-0.5)--(1,-0.5);
        \node at (0,-0.5) {\Huge$\bullet$};
        \node at (0,0.5) {\Huge$\bullet$};
        \node at (1,0.5) {\Huge$\bullet$};
        \filldraw[white] (1,-0.5) circle (5pt) node[]{};
        \node at (1,-0.5) {\Large$\otimes$};
    \end{tikzpicture} \times  \begin{tikzpicture}[baseline={(0,-0.1)cm},scale=1.]
        \node[ellipse, draw, fill=almond, minimum width=4cm, minimum height=1.5cm] {Sum of ladders};
        \filldraw[white] (-2.17,0) circle (5pt) node[]{};
        \node[black] at (-2.17,0) {\Large$\otimes$};
    \end{tikzpicture} \right) \nonumber \nonumber
  \end{align}
\end{center}
and
\begin{align}
      \mathcal{I}[g^6\mathcal{F}_{4}(z) {\cal F}_{\rm ladder}(g,z)]_{g>>1}& \sim \frac{394}{945} \pi ^5 \sqrt{2}  \tanh \left(\sqrt{2} \pi  g\right)g^7+\left(\frac{4 \pi ^4}{15}-192 \zeta_3\right)g^6\\ \nonumber
      &-\frac{\left(2 \sqrt{2} \left(990 \zeta_3+13 \pi ^4-255 \pi ^2\right)\right) \tanh \left(\sqrt{2} \pi  g\right)}{15 \pi }g^5 + {\cal O}(g^4)\,,
\end{align}
\begin{equation} \label{eq:gammacuspboxladder}
  \Gamma_{\rm cusp}\sim  -\frac{3152\sqrt{2} \pi ^5}{6615}g^7-\frac{4}{3} \left(\frac{4 \pi ^4}{15}-192 \zeta_3\right)g^6-\frac{16 \sqrt{2} \left(990 \zeta_3+13 \pi ^4-255 \pi ^2\right)}{75 \pi } g^5 + {\cal O}(g^4)\,,
\end{equation}
where we don't include the powers of $g$ lower than $g^5$. We note that this sum is also partially summable (the leading term contributes $\sim g^8$ in weak coupling).

Just like the three-loop case, the $\Gamma_{\rm cusp}$ contribution from the resummed result is the same as the leading diagram in the weak coupling expansion suppressed by $\frac{1}{g}$. Furthermore, even though individual terms give $\log(g)$ contributions, they again end up canceling out.
We can actually see the same phenomenon even for tree diagrams, we consider 
\begin{equation}
    g^4\begin{tikzpicture}[baseline={(0,-0.1)cm}]
    \draw[Maroon, ultra thick] (0,0)--(-0.75,0.75) node[at end, black] {\Huge$\bullet$};
    \draw[Maroon, ultra thick] (0,0)--(-0.75,-0.75) node[at end, black] {\Huge$\bullet$};
     \filldraw[white] (0,0) circle (5pt) node[]{};
    \node[black] at (0,0) {\Large$\otimes$};
\end{tikzpicture}\,-\, g^6\begin{tikzpicture}[baseline={(0,-0.1)cm}]
    \draw[Maroon, ultra thick] (0,0)--(-0.75,0.75) node[at end, black] {\Huge$\bullet$};
    \draw[Maroon, ultra thick] (0,0)--(-0.75,-0.75) node[at end, black] {\Huge$\bullet$};
    \draw[Maroon, ultra thick] (0,0)--(0.75,0) node[at end, black] {\Huge$\bullet$};
     \filldraw[white] (0,0) circle (5pt) node[]{};
    \node[black] at (0,0) {\Large$\otimes$};
\end{tikzpicture}\,+\,g^{8}\begin{tikzpicture}[baseline={(0,-0.1)cm}]
    \draw[Maroon, ultra thick] (0,0)--(-0.75,0.75) node[at end, black] {\Huge$\bullet$};
    \draw[Maroon, ultra thick] (0,0)--(-0.75,-0.75) node[at end, black] {\Huge$\bullet$};
    \draw[Maroon, ultra thick] (0,0)--(1.5,0) node[at end, black] {\Huge$\bullet$} node[pos=0.5, black] {\Huge$\bullet$};
   \filldraw[white] (0,0) circle (5pt) node[]{};
    \node[black] at (0,0) {\Large$\otimes$};
\end{tikzpicture}\,+\,\cdots
\end{equation}
\begin{equation}
    \mathcal{I}[{\cal F}_{3l,\text{ladder}}] = \sqrt{8} \pi  g^3 \tanh \left(\sqrt{2} \pi  g\right)-6 g^2+\frac{3 \sqrt{2} g \tanh \left(\sqrt{2} \pi  g\right)}{\pi}\,.
\end{equation}
In this case, the leading behavior is $\sim g^3$, with even a larger suppression from the first term in the sum that contributes to $\sim g^6$, still short of the correct $\Gamma_{\rm cusp}\sim g$ behavior. 

We can also attach the whole series of tree graphs to the triangle core and compute the contribution to the $\Gamma_{\rm cusp}$ of the following series, 
\begin{center}
  \begin{align}
    &\Gamma_{\rm cusp}\Bigg( g^4 \begin{tikzpicture}[baseline={(0,-0.1)cm}]
    \draw[Maroon, ultra thick] (-0.75,0.75)--(-0.75,-0.75);
    \draw[Maroon, ultra thick] (0,0)--(-0.75,0.75) node[at end, black] {\Huge$\bullet$};
    \draw[Maroon, ultra thick] (0,0)--(-0.75,-0.75) node[at end, black] {\Huge$\bullet$};
    \filldraw[white] (0,0) circle (5pt) node[]{};
    \node[black] at (0,0) {\Large$\otimes$};
\end{tikzpicture} \times \Big( \begin{tikzpicture}[baseline={(0,-0.1)cm},scale=1.]
        \node[white] at (0,0) {\Huge$\bullet$};
        \node[black] at (0,0) {\Large$\otimes$};
    \end{tikzpicture}\ -g^2\ \begin{tikzpicture}[baseline={(0,-0.1)cm}]
        \draw[ultra thick, Maroon] (0,0)--(1,0) node[at end, black] {\Huge$\bullet$};
        \filldraw[white] (0,0) circle (5pt) node[]{};
        \node[black] at (0,0) {\Large$\otimes$};
    \end{tikzpicture}\ +g^4\ \begin{tikzpicture}[baseline={(0,-0.1)cm},scale=1.]
        \draw[ultra thick, Maroon] (0,0)--(2,0) node[at end, black] {\Huge$\bullet$} node[pos=0.5, black] {\Huge$\bullet$};
        \filldraw[white] (0,0) circle (5pt) node[]{};
        \node[black] at (0,0) {\Large$\otimes$};
    \end{tikzpicture} \\
    &+\frac{g^4}{2!} \begin{tikzpicture}[baseline=-0.6ex,scale=1.,transform shape]
  \draw[Maroon, ultra thick] (1,0)--(2,0.6) node[at end,black]{\Huge$\bullet$};
  \draw[Maroon, ultra thick] (1,0)--(2,-0.6) node[at end,black]{\Huge$\bullet$};
  \node[white] at (1,0) {\Huge$\bullet$};
  \node[black] at (1,0) {\Large$\otimes$};
\end{tikzpicture}-g^6\begin{tikzpicture}[baseline=-0.6ex,scale=1.,transform shape]
  \draw[Maroon, ultra thick] (1.4,0)--(2.1,0) node[at end,black]{\Huge$\bullet$};
  \draw[Maroon, ultra thick] (0.7,0)--(1.4,0) node[at end,black]{\Huge$\bullet$};
  \draw[Maroon, ultra thick] (0,0)--(0.7,0) node[at end,black]{\Huge$\bullet$};
  \filldraw[white] (0,0) circle (5pt) node[]{};
  \node[black] at (0,0) {\Large$\otimes$};
\end{tikzpicture} - g^6 \begin{tikzpicture}[baseline=-0.6ex,scale=1.,transform shape]
  \draw[Maroon, ultra thick] (0.7,0.4)--(1.3,0.4) node[at end,black]{\Huge$\bullet$};
  \draw[Maroon, ultra thick] (0,0)--(0.7,0.4) node[at end,black]{\Huge$\bullet$};
  \draw[Maroon, ultra thick] (0,0)--(0.7,-0.4) node[at end,black]{\Huge$\bullet$};
  \filldraw[white] (0,0) circle (5pt) node[]{};
  \node[black] at (0,0) {\Large$\otimes$};
\end{tikzpicture} - \frac{g^6}{2!} \begin{tikzpicture}[baseline=-0.6ex,scale=1.,transform shape]
  \draw[Maroon, ultra thick] (0.8,0)--(1.4,0.6) node[at end,black]{\Huge$\bullet$};
  \draw[Maroon, ultra thick] (0.8,0)--(1.4,-0.6) node[at end,black]{\Huge$\bullet$};
  \draw[Maroon, ultra thick] (0,0)--(0.8,0) node[at end,black]{\Huge$\bullet$};
  \filldraw[white] (0,0) circle (5pt) node[]{};
  \node[black] at (0,0) {\Large$\otimes$};
\end{tikzpicture} \cdots \Big) \Bigg) \\
&=\Gamma_{\rm cusp}\left(g^4\begin{tikzpicture}[baseline={(0,-0.1)cm}]
    \draw[Maroon, ultra thick] (-0.75,0.75)--(-0.75,-0.75);
    \draw[Maroon, ultra thick] (0,0)--(-0.75,0.75) node[at end, black] {\Huge$\bullet$};
    \draw[Maroon, ultra thick] (0,0)--(-0.75,-0.75) node[at end, black] {\Huge$\bullet$};
    \filldraw[white] (0,0) circle (5pt) node[]{};
    \node[black] at (0,0) {\Large$\otimes$};
\end{tikzpicture} \times \begin{tikzpicture}[baseline={(0,-0.1)cm},scale=1.]
        \node[ellipse, draw, fill=almond, minimum width=4cm, minimum height=1.5cm] {Sum of trees};
        \filldraw[white] (-2.17,0) circle (5pt) node[]{};
        \node[black] at (-2.17,0) {\Large$\otimes$};
    \end{tikzpicture} \right)
\nonumber
  \end{align}
\end{center}
We cannot solve the full result analytically, but the leading contribution for $g\gg1$ is again $\Gamma_{\rm cusp} \sim g^5$ as shown in Appendix \eqref{app:tretriangle}. This all demonstrates that we have to consider a larger set of one-cycle negative geometries to get the right behavior for $g\gg1$. In fact, it is an important open question if the sum of \emph{all} one-cycle geometries actually goes like $\sim g$ for large $g$ or if we have to combine it with higher cycle geometries.
\section{Details on the triangle-ladder computation}
\label{sec:tri-ladder}

\subsection*{Differential Equation for \texorpdfstring{${\cal F}_\text{tri-lad}(g,z)$}{F\_tri-lad(g,z)}}\label{app:DiffEq}
We can solve the differential equation for ${\cal F}_\text{tri-lad}(g,z)$ in terms of infinite series, by taking the Mellin transform of both sides. Since the source ${\cal F}_3(z)$ is linear in HPLs, the solution ${\cal F}_\text{tri-lad}(g,z)$ will be a linear combination of functions $f_i(g,z)$ that solve this differential equation for each source term. Taking the Mellin Transform, we have
\begin{equation}
    \tilde{{\cal F}}(\nu,g) = \int^1_0 z^{\nu-1}{\cal F}(g,z) dz\,.
\end{equation}
Note that the integration domain is reduced from $(0,\infty)$ to $(0,1)$ since we are working with harmonic polylogarithms which remain invariant under $z \rightarrow \frac{1}{z}$. Then in Mellin space, the action of the differential box operator is given by
\begin{equation}
 \frac{1}{2}\int^1_0 (z\partial_z)^2z^{\nu-1}f(z) dz = \frac{\nu^2}{2} \tilde{f}(\nu)\,.
\end{equation}
Using this for the full differential equation, the Laplace equation from before \eqref{eq:tl_diffeq} conveniently becomes
\begin{equation}
    \tilde{{\cal F}}(\nu,g) = \frac{g^4\nu^2}{2g^2+\nu^2} \tilde{\mathcal{F}_3}(\nu,g)\,,
\end{equation}
where $\tilde{{\cal F}}(\nu,g)$ is the Mellin transform of $\mathcal{F}_\text{tri-lad}(g,z)$ and $\tilde{\mathcal{F}_3}(\nu)$ is the Mellin transform of $\mathcal{F}_3(z)$. The RHS is a sum over different harmonic sums, so the task at hand is to take the inverse Mellin transform of terms of the form $\frac{\nu^2}{\nu^2+g^2} \sum^\infty_{i=\nu}\frac{-1}{i^p}$. Some pieces are easier than others, so let's start with the easiest term to see how it works for the general case. In Mellin space, $H_{0,0,0,0}(z)=\frac{\log^4({z})}{24}$ is given by
\begin{equation}
    \mathcal{M}(H_{0,0,0,0}(z),\nu)= \frac{1}{\nu^5}\,.
\end{equation}
Then the inverse transform of $\frac{8\nu^2}{2g^2+\nu^2}\mathcal{M}(H_{0,0,0,0}(z),\nu)$ is given by
\begin{equation}
    f_1(g,z) = \int_C z^{-\nu} \frac{8}{(\nu^2+2g^2)\nu^3} d\nu\,.
\end{equation}
If we choose a contour that only has real poles, we find
\begin{equation}
    f_1(g,z) = \frac{2(g^2 \log ^2(z)-1)}{ g^4}\,.
\end{equation}
This indeed satisfies the equation
\begin{equation}
    \frac{1}{2}(z \partial_z)^2 f_1(g,z) + g^2 f_1(g,z) = \frac{1}{2}(z \partial_z)^2\left( 8 H_{0,0,0,0}(z) \right)= 2 \log^2({z})\,.
\end{equation}
where we have suppressed the $g^4$ on the RHS for convenience and will restore it at the end after all terms are gathered. Note that we can also add terms of the form $C(g) \cos{\left( \sqrt{2}g \log{z}\right)}$ and $D(g) \sin{\left( \sqrt{2}g \log{z}\right)}$ which vanish under the operator $\frac{1}{2}(z\partial_z)^2+g^2$.  This corresponds to the homogeneous solution of the differential equation. We will discuss how to fix the coefficients of the homogeneous solution later and focus first on the particular solution, which we write as ${\cal F}_p(g,z)$.
\par
Using the same logic as above, we can similarly solve rest of the terms in $\Box {\cal F}_{p}(g,z)+g^2{\cal F}_{p}(g,z)=g^4\Box \mathcal{F}_{3}(z)$ for each HPL that the $\Box$ on the RHS acts on. Recalling that 
\begin{align}
    \mathcal{F}_{3}(z)&= 8\displaystyle H_{\text{0,0,0,0}}+8 H_{\text{-1,0,0,0}}-16 H_{\text{-1,-1,0,0}}+8 H_{\text{-2,0,0}} - 8 \zeta_3 \left(2 H_{-1}-H_0\right)\nonumber \\
&+ 4\pi ^2 \left(H_{\text{-1,0}}- 2\,H_{\text{-1,-1}}+H_{-2}\right) +\frac{13 \pi ^4}{45}\,.
\end{align}
We denote the partial solutions $f_i(g,z)$ such that ${\cal F}_{p}(g,z)=g^4 \sum^9_i f_i(g,z)$ and each solves
\begin{align}
    &\Box f_1(g,z)+g^2f_1(g,z)=\Box( 8\displaystyle H_{\text{0,0,0,0}}(z)) \nonumber\\
    & \Box f_2(g,z)+g^2f_2(g,z)=\Box( 8\displaystyle H_{\text{-1,0,0,0}}(z)) \nonumber \\
    & \Box f_3(g,z)+g^2f_3(g,z)=\Box( -16\displaystyle H_{\text{-1,-1,0,0}}(z)) \nonumber \\
     & \Box f_4(g,z)+g^2f_4(g,z)=\Box( 8\displaystyle H_{\text{-2,0,0}}(z)) \nonumber \\
      & \Box f_5(g,z)+g^2f_5(g,z)=\Box( -16 \zeta_3\displaystyle H_{-1}(z)) \nonumber \\
      & \Box f_6(g,z)+g^2f_6(g,z)=\Box( 8 \zeta_3\displaystyle H_{0}(z)) \nonumber \\
       & \Box f_7(g,z)+g^2f_7(g,z)=\Box( 4 \pi^2 \displaystyle H_{-1,0}(z)) \nonumber \\
        & \Box f_8(g,z)+g^2f_8(g,z)=\Box( -8\pi^2 \displaystyle H_{-1,-1}(z)) \nonumber \\
           & \Box f_9(g,z)+g^2f_9(g,z)=\Box( 4\pi^2 \displaystyle H_{-2}(z)) \nonumber \\
\end{align}
These functions are written below
\begin{align}
    f_2(g,z) =& \sum^\infty_{n=1}-\frac{4 e^{i \pi  n} n z^n \log ^3(z)}{6 g^2+3 n^2}+\frac{4 e^{i \pi  n} \left(n^2-2 g^2\right) z^n \log ^2(z)}{\left(2 g^2+n^2\right)^2} \nonumber\\
    &-\frac{8 e^{i \pi  n} n \left(n^2-6 g^2\right) z^n \log (z)}{\left(2 g^2+n^2\right)^3}+\frac{8 e^{i \pi  n} \left(4 g^4-12 g^2 n^2+n^4\right) z^n}{\left(2 g^2+n^2\right)^4}\,
\end{align}

\begin{align}
   f_3(g,z) =& \sum^\infty_{i,j{=}1}{-}\frac{8 e^{i \pi  (i{+}j)} (i{+}j) \log ^2(z) z^{i{+}j}}{j \left(2 g^2+(i+j)^2\right)}+\frac{16 e^{i \pi  (i{+}j)} \log (z) \left(2 g^2 i{+}(i{+}j)^2 (i{+}2 j)\right) z^{i{+}j}}{j^2 \left(2 g^2+(i+j)^2\right)^2} \nonumber\\
   &-\frac{16 e^{i \pi  (i+j)} \left(4 g^4 i+2 g^2 (i+j) \left(2 i^2+4 i j-j^2\right)+(i+j)^3 \left(i^2+3 i j+3 j^2\right)\right) z^{i+j}}{j^3 \left(2 g^2+(i+j)^2\right)^3}
\end{align}

\begin{align}
    f_4(g,z) = \sum^\infty_{n=1}\frac{8 e^{i \pi  n} \left(2 g^2-3 n^2\right) z^n}{\left(2 g^2+n^2\right)^3}-\frac{4 e^{i \pi  n} z^n \log ^2(z)}{2 g^2+n^2}+\frac{16 e^{i \pi  n} n z^n \log (z)}{\left(2 g^2+n^2\right)^2}
\end{align}

\begin{equation}
    f_5(g,z) = 16 \zeta_3 \sum^\infty_{n=1} \frac{e^{i \pi n} nz^n}{2g^2+n^2}\,,
\end{equation}

\begin{equation}
    f_6(g,z) = f_h,
\end{equation}

\begin{equation}
    f_7(g,z) = -4\pi^2 \sum^\infty_{n=1}\frac{z^n e^{i \pi n} \left(g^2 n \log (z)+g^2+n^3 \log (z)-n^2\right)}{\left(2g^2+n^2\right)^2}\,,
\end{equation}

\begin{align}
    f_8(g,z) = -\sum^\infty_{n=1}\frac{4 \pi ^2 e^{i \pi  n} z^n \left(2 g^2 n \log (z)+2 g^2+n^3 \log (z)-n^2\right)}{\left(2 g^2+n^2\right)^2},
\end{align}

\begin{equation}
    f_9(g,z) = -4\pi^2 \sum^\infty_{n=1} \frac{e^{i\pi n}z^n}{2g^2+n^2}\,,
\end{equation}
and thus
\begin{equation}
    {\cal F}_{p}(g,z) = g^4\sum^9_{i=1}f_i(g,z)\,.
\end{equation}
where $f_h$ denotes terms that go into the homogenous solution. Most of the sums here can be analytically expressed in terms of hypergeometric functions (put in the ancillary file {\tt Ftriangleladder.m}), but the sums $f_3(g,z)$ and $f_8(g,z)$ don't have a nice form. These terms however have an integral representation, which we can extract the large $g$ asymptotics of using integration by parts. For $g \gg 1$, these integral terms go like (to leading order in $g$) 

\begin{align}
   {\cal F}_{\rm IT}(g,z)_{g\gg1}\sim&\frac{4z g^2}{(z+1)^2}\left(2 \text{Li}_3(-z)-2 \text{Li}_2(-z) \log (z)- \log (z+1) \log ^2(z) \right)+\mathcal{O}(g)\,
\end{align}
 We can then expand the full result for the particular solution at strong coupling and get
\begin{align}
        &{\cal F}_{3,p}(g,z)_{g \gg 1} \sim \frac{g^2 \left(4 (z (z+4)+1) \log ^2(z)+4 \pi ^2 z \log (z)-8 \pi ^2 z \log (z+1)\right)}{2 (z+1)^2}\\ \nonumber
        &+\frac{2 g^2 z \left(12 \text{Li}_3(-z)-12 \text{Li}_2(-z) \log (z)+\log ^3(z)-6 \log (z+1) \log ^2(z)-12 \zeta_3+6 \pi ^2\right)}{3 (z+1)^2}\\ \nonumber
        &+ \mathcal{O}(g) \,.
\end{align}
Note that this is only the particular solution to the equation 
\begin{equation}
    \Box {\cal F}_\text{tri-lad}(g,z)+g^2{\cal F}_\text{tri-lad}(g,z)=g^4\Box\mathcal{F}_{3}(z)\,,
\end{equation}
but we also have to add the homogeneous solution that solves
\begin{equation}
    \Box {\cal F}_h(g,z)+g^2 {\cal F}_h(g,z)=0\,,
\end{equation}
namely ${\cal F}_{h}(g,z)=c_1\sin(\sqrt{2}g\log(z))+c_2\cos(\sqrt{2}g\log(z))$. While this looks identical to the ladder example, this time the boundary conditions are slightly different. The full solution is then ${\cal F}_{rs}(g,z)={\cal F}_{p}(g,z)+{\cal F}_{h}(g,z)$ and the coefficients $c_1,c_2$ must be fixed using the boundary conditions. The first boundary condition, namely $\frac{d}{dz}{\cal F}_\text{tri-lad}(g,z)|_{z=1}=0$, sets $c_1$ to 0. Fixing $c_2$ is more subtle. For this, we need the value of the full solution at $z=-1$, which we know must equal $g^4 \mathcal{F}_3(-1)$. However, the source term of the differential equation has a branch point at $z=0$ and the solution is only analytic for $z>0$. So we can essentially treat our ${\cal F}_\text{tri-lad}(g,z)$ as being defined on positive real z, and analytically continue it through the upper half plane and evaluate it at $z=-1$. Then we have the following condition
\begin{equation}
   {\cal F}_\text{tri-lad}(z=e^{i\pi},g) = g^4\mathcal{F}_3(z=e^{i\pi})= \frac{22 \pi^4 g^4}{45}
\end{equation}
This fixes the $c_2$ term and gives us the homogeneous solution
\begin{align}
   {\cal F}_h(g,z)|_{g\gg1} &\sim \frac{1}{45} g^2 \cos \left(\sqrt{2} g \log (z)\right) \Big( \left(22 \pi ^4 g^2+90 \pi ^2+90\right) \text{sech}\left(\sqrt{2} \pi  g\right) + \mathcal{O}(g e^{-\sqrt{2}g \pi} )\,.
\end{align}
Then at strong $g$ and $z>0$, the full solution $F_\text{tri-lad}(g,z)=F_p(g,z)+F_h(g,z)$ to leading order is
\begin{align}
        &F_\text{tri-lad}(g,z)_{g \gg 1} \sim -\frac{g^4 \left(2 (z (z+4)+1) \log ^2(z)+2 \pi ^2 z \log (z)-4 \pi ^2 z \log (z+1)\right)}{(z+1)^2} \nonumber \\ 
        &{-}\frac{2 g^4 z \left(12 \text{Li}_3({-}z){-}12 \text{Li}_2({-}z) \log (z){+}\log ^3(z){-}6 \log (z{+}1) \log ^2(z){-}12 \zeta_3{+}6 \pi ^2\right)}{3 (z{+}1)^2} {+} \mathcal{O}(g^3)\,. \nonumber
\end{align}
where we have written $F(g,z)$ instead of $\mathcal{F}(g,z)=-\frac{1}{g^2} F(g,z)$. Note that this is simply the source term divided by $g^2$, which amounts to ignoring the differential operator to leading order when $g$ is large. This is only true for $z>0$, where the negative geometries live, because the homogeneous solution is exponentially suppressed in this region. In particular, the leading behavior of ${\cal F}_h(z>0,g)$ is $g^4 e^{-\sqrt{2}g\pi}$, which is exponentially suppressed for large $g$. However, the numerator has a $\cos(\sqrt{2}g\log(z))$ factor, which contributes a $e^{\sqrt{2}g \pi}$ that cancels this suppression upon analytical continuation to $z=-1$. This explains how we can match the triangle diagram's $g^4$ power at $z=-1$ despite having the full $\mathcal{F}_\text{tri-lad}(g,z)$ behave like $g^2$ for large $g$ in the region we are interested in. Indeed, the $\Gamma_{\rm cusp}$ contribution of $\mathcal{F}_\text{tri-lad}(g,z)$ is not given by the leading order term we quote because the homogeneous solution $\mathcal{F}_h$ picks up a $\cosh{\sqrt{2}g\pi}$ along the contour we integrate. Thus, the part that was sub-leading for positive and real $z$ becomes relevant.

For well behaved source terms\footnote{Recent work \cite{Paranjape:2026kix} motivates the possibility that higher-loop cycles are made of HPLs i.e. that they too are examples of well-behaved source terms.} (e.g. for $\Box\mathcal{F}_4(z)$), we should expect to find that the $\frac{1}{g^2}$ drop in $F(g,z)$ after resumming ladder attachments is a general feature. Say we have a generic ``core" diagram with $\Box \mathcal{F}_c(z) = h(z)$. If the core has $L$ loops, the differential equation at hand reads
\begin{equation}
    \Box F_{c,l}(g,z) +g^2 F_{c,l}(g,z)=g^{2L}h(z)
\end{equation}
where $\mathcal{F}_{c,l}(g,z)=-\frac{1}{g^2}F_{c,l}(g,z)$ corresponds to resumming all ladder attachments to the core diagram. If $h(z)$ is non-singular for $z>0$, then $F_{p,cl}(g,z) \sim g^{2L-2}$ at most in large $g$. The homogeneous solution should then be of the form $F_{h,cl}(g,z)=c(g) \cos(\sqrt{2}g\log(z))$, where $c(g)$ must be solved by $F_{c,l}(g,z=e^{i\pi})=g^{2L}h(z=e^{i\pi})$ and the sine term's coefficient is set to 0 by $\frac{d}{dz}F_{c,l}(z)|_{z=1}=0$. If we find a branch cut at $z=0$ again, we can repeat the procedure from above, in which case $c(g)$ picks up $\frac{1}{\cosh{(\sqrt{2}g\pi)}}$ in order to match the leading term. However, for $z>0$, this exponentially suppresses the homogeneous solution for large $g$ and kills any polynomial dependence for $g$ we have the numerator in $F_h(g,z)$. Hence, the particular solution should control the asymptotics and give $F_{c,l}(g \gg 1,z>0) \sim g^{2L-2}$ at most. 

\subsection*{Necessary Sums for \texorpdfstring{$\mathcal{I}[{\cal F}_{\rm ladder}(g,z)\mathcal{F}_3(z)]$}{I[F\_ladder(g,z)F\_3(z)]}}\label{app:sums}

For $|z|<1$, the harmonic polylogarithms that appear in our $\mathcal{F}(z)$ functions have the series expansion \cite{REMIDDI_2000}
\begin{equation}
    H_{-m_n,..,-m_1}(z)\log(z)^a = (-1)^n \sum_{j_n>...>j_1} \frac{\log(z)^a (-z)^{j_n}}{j^{m_n}_n..j_1^{m_1}}\,.
\end{equation}
So in order to compute $\mathcal{I}[{\cal F}_3(z)\mathcal{F}_{\rm ladder}(g,z)]$, we must compute integrals of the form
\begin{equation}
  (-1)^n \sum_{j_n>...>j_1}  \int^\pi_{-\pi} \frac{\cosh\sqrt{2}g\phi(i\phi)^a (-e^{i\phi j_n})}{j^{m_n}_n..j_1^{m_1}}\,.
\end{equation}
While the integration itself is easy, the nested sums that appear are non-trivial to sum into analytical functions. We are able to perform these nested sums for depth up to 2. Here is a list of sums we will find useful:
\begin{align}
    \sum_{j_1=1}^\infty \frac{1}{j_1^2(a+ j_1)}= &\frac{\zeta_2}{a}-\frac{1}{a^2}H_a\nonumber\\
    \sum_{j_2>j_1=1}^\infty \frac{1}{j_1j_2(a+ j_2)}= &\frac{1}{2a}H_a-\frac{1}{2a}H_{a,2}\nonumber\\
    \sum_{j_1=1}^\infty \frac{1}{j_1(a+ j_1)}= &\frac{1}{a}H_a\nonumber\\
    \sum_{j_2>j_1=1}^\infty \frac{1}{j_1(a+ j_2)}= &\zeta_{1,1}-\frac{1}{2}H_a^2-\frac{1}{2}H_{a,2}\nonumber\\
    \sum_{j_1=1}^\infty \frac{1}{(a+ j_1)}= &\zeta_1-H_a\nonumber\\
    \sum_{j_1=1}^\infty \frac{1}{j_1^3(a+ j_1)}= &\frac{\zeta_3}{a}-\frac{\zeta_2}{a^2}+\frac{H_b}{b^3}\frac{1}{a}H_a\nonumber\\
    \sum_{j_1=1}^\infty \frac{j_1^2}{(a+ j_1)^4}= &\zeta_2-2 a\zeta_3+a^2\zeta_4-H_{a,2}+2aH_{a,3}-a^2 H_{a,4}\nonumber\\
    \sum_{j_1=1}^\infty \frac{j_1}{(a+ j_1)^4}= &\zeta_3-a\zeta_4-H_{a,3}+a H_{a,4}\nonumber\\
    \sum_{j_1=1}^\infty \frac{1}{(a+ j_1)^4}= &\zeta_4-H_{a,4}\nonumber\\
    \sum_{j_1=1}^\infty \frac{1}{j_1(a+ j_1)^4}= &\zeta_2-2 a\zeta_3+a^2\zeta_4-H_{a,2}+2aH_{a,3}-a^2 H_{a,4}\nonumber\\
    \sum_{j_1=1}^\infty \frac{1}{j_1^2(a+ j_1)^4}= &\frac{4\zeta_2}{a^4}+\frac{2\zeta_3}{a^3}+\frac{\zeta_4}{a^2}-\frac{4 H_a}{a^5}-\frac{3 H_{a,2}}{a^4}-\frac{2 H_{a,3}}{a^3}-\frac{H_{a,4}}{a^2}\nonumber\\
    \sum_{j_1=1}^\infty \frac{1}{j_1^2(a+ j_1)^3}= &\frac{3\zeta_2}{a^3}+\frac{\zeta_3}{a^2}-\frac{3 H_a}{a^4}-\frac{2 H_{a,2}}{a^3}-\frac{H_{a,3}}{a^2}\nonumber
\end{align}
For the 1-cycle geometries we have, namely one that have the triangle or the box as a core, the most complicated nested sums have the form \begin{equation}
    \sum^{\infty}_{j_2 > j_1 > 0} \frac{j^a_2}{j_1^b(j^2_2+2g^2)^c}\,.
\end{equation} 
We can express these sums in terms of Polygamma functions and their derivatives. Let's start with $a = -1, b = 2, c = 2$ as an illustrative example. First, we can re-write the double sum in a simpler form as follows
\begin{equation}
    \sum^\infty_{j_2 > j_1 > 0} \frac{1}{j^2_1 j_2 (j^2_2+2g^2)^2} = \sum^\infty_{j_2 = 1 \ j_1 = 0} \frac{1}{j^2_1 (j_2+j_1) ((j_2+j_1)^2+2g^2)^2}\,.
\end{equation}
Doing a partial decomposition, the summand can be written as
\begin{align}
       \frac{1}{j^2_1 j_2 (j^2_2+2g^2)^2} &=\frac{1}{4 g^4 j_1^2 \left(j_1+j_2\right)}-\frac{1}{8 g^4 j_1^2 (-i\sqrt{2} g+j_1+j_2)}-\frac{1}{8 g^4 j_1^2 (i \sqrt{2} g+j_1+j_2)} \nonumber \\ 
      &+\frac{i}{8 \sqrt{2} g^3 j_1^2 (-i \sqrt{2} g+j_1+j_2)^2}-\frac{i}{8 \sqrt{2} g^3 j_1^2 (i \sqrt{2} g+j_1+j_2){}^2}\,.
\end{align}

These sums can be written in terms of polygamma functions, which have the following series representation 
\begin{equation}
    \psi^{(m)}(z) = (-1)^{m+1} m! \sum^\infty_{k=0} \frac{1}{(z+k)^{m+1}}\,.
\end{equation}
The sum over $j_2$ then becomes
\begin{align}
    &\frac{2 \psi ^{(0)}\left(i g  \sqrt{2}+j_1+1\right)+2 \psi ^{(0)}\left(-i g \sqrt{2}+j_1+1\right)-i \sqrt{2} g \psi ^{(1)}\left(ig \sqrt{2}+j_1+1\right)}{16 g^4 j_1^2} \nonumber \\ 
      &+\frac{i \sqrt{2} g \psi ^{(1)}\left(-i g \sqrt{2}+j_1+1\right)-4 \psi ^{(0)}\left(j_1+1\right)}{16 g^4 j_1^2}\,.
\end{align}
The $\frac{\psi ^{(0)}\left(-i g \sqrt{2}+j_1+1\right)}{16 g^4 j_1^2}$ sums can be evaluated using the identity $\psi^{(0)}(1+z) = \psi^{(0)}(z)+\frac{1}{z}$. More specifically, $\psi ^{(0)}(-i g \sqrt{2}+j_1+1) = \psi(1-ig\sqrt{2})+\sum^{j_1-1}_{k=0} \frac{1}{1-ig\sqrt{2}+k}$ and the sum becomes
\begin{align}
     &\sum^\infty_{j_2 > j_1 > 0} \frac{1}{j^2_1 j_2 (j^2_2+2g^2)^2}=-\frac{12 \ {}_4{F}_3^{(b_3,1)}\left(\left\{1,1,1,2-i \sqrt{2} g\right\},\left\{2,2,2-i \sqrt{2} g\right\},1\right)}{96 g^4} \nonumber \\ 
    &-\frac{6 {}_4F_3^{(a_4,1)}\left(\left\{1,1,1,2-i \sqrt{2} g\right\},\left\{2,2,2-i \sqrt{2} g\right\},1\right)+\pi ^2 \psi ^{(0)}\left(2-i \sqrt{2} g\right)}{96 g^4}\,,
\end{align}
where the notation ${}_pF_q^{(b_3,0)}$ means the generalized hypergeometric function 
\begin{align}
 {}_pF_{q}(\{a_1,...,a_p\},\{b_1,...,b_q\},z)\,,   
\end{align}
is being differentiated with respect to the index $b_3$ a single time. While these hypergeometric terms frequently appear in sums of our interest, they always do so in a sub-leading way in $g$ and fall off in the strong $g$ limit. The sums involving higher order polygamma functions can be evaluated by taking appropriate derivatives in $g$ of sums of the form above, using the identity $\frac{d^m}{dz^m} \psi^{(m)}(1-ig\sqrt{2}+j_1) = (-ig\sqrt{2})^m \psi^{(0)}(1-ig\sqrt{2}+j_1) $. 

\subsection*{Coefficients in our single cycle+ladder results}\label{app:boxCoeff}

We have the general $\mathcal{F}$ function for $n-$ladders attached to a triangle as follows
\begin{align}
    \mathcal{F}_{3,n}(z) = (-2)^n \Biggl[ &8H_{\Vec{0}_{2n},0,0,0,0}(z)+8H_{\Vec{0}_{2n},{-}1,0,0,0}(z){-}16H_{\Vec{0}_{2n},{-}1,{-}1,0,0}(z){+}8H_{\Vec{0}_{2n},0,{-}1,0,0}(z) \nonumber\\
    &
    -16\zeta_3\,H_{\Vec{0}_{2n},{-}1}(z)
    +4 \pi^2\Bigl(H_{\Vec{0}_{2n},{-}1,0}(z)-2 H_{\Vec{0}_{2n},{-}1,{-}1}(z)
    +H_{\Vec{0}_{2n},{-}2}(z)\Bigr) \nonumber\\
    &+ \frac{13 \pi^4}{45} \frac{1}{(2n)!}\log^{2n}(z) 
    +8\zeta_3 \frac{1}{(2n+1)!} \log^{2n+1}(z) \nonumber\\
    &+ \sum^{n}_{k=1}  \biggl( \frac{C_k}{[2(n-k)+1]!} 
    \log{z}^{2(n-k)+1}
    + \frac{D_k}{[2(n-k)]!} \log{z}^{2(n-k)} \biggr) \Biggr]\,.
\end{align}

Now, let's define the sequence $\eta_n$ recursively by $\eta_0 = 8$ and for $n>0$,
\begin{align}
\eta_n &\equiv - \frac{8 (-1)^n}{(2n+1)!} \Bigl[ 1 
  + (2n-1)(2n+1) 2^{2n} B_{2n} \Bigr]
  - (-1)^n\sum_{k=1}^{n-1} 
  \frac{(-1)^k \eta_k}{(2n-2k+1)!} \\
 &=  4,\ \frac{17}{15},\ \frac{457}{1890},\ \frac{3287}{75600},\ \ldots\,.
\end{align}

Then the $C_n$ coefficients of $\log z$ are given for all $n$ by
\begin{equation}
  C_n = \sum_{k=0}^n \eta_k \, \pi^{2k} \, \zeta_{2n-2k+3} \,.
\end{equation}
The Bernoulli numbers $B_{2N}$ appear because powers of $\pi^2$ are generated both by analytic continuation of $\log z$ and from Riemann zeta values $\zeta_{2k}$.

The first few $C_n$ look simpler using even zeta values instead of powers of $\pi^2$:
\begin{align}
C_0 &= 8 f_3 \,, \\
C_1 &= (8 f_5+24\zeta_2 f_3) \,, \\
C_2 &= 8 f_7 + 24\zeta_2 f_5 + 102\zeta_4 f_3
\,, \\
C_3 &=  \Bigl(8f_9+24\zeta_2 f_7 + 102\zeta_4 f_5 + \frac{457}{2}\zeta_6 f_3 \Bigr) \,.
\end{align}
Here we also introduced the $f$ alphabet arising in a motivic decomposition of MZVs~\cite{Brown:2011ik}, as encoded in the package {\sc HyperlogProcedures} by Oliver Schnetz~\cite{Schnetz:2016fhy}. For odd Riemann zeta values, the $f$-alphabet is rather trivial,
\begin{equation}
    f_{2k+1} \equiv \zeta_{2k+1} \,.
\end{equation}

The constants $D_n$ can also be written in an all-orders recursive form:
\begin{equation}
D_n = \biggl[ \rho_n \, \pi^{2n+4}
  \ -\ \sum_{j=0}^n \xi_{j+1} \, \pi^{2j} 
  \sum_{k=1}^{n-j} 
    k(2k-1) \, f_{2(n-j-k)+3,2k+1} 
     \biggr] \,,
\end{equation}
where the pure-$\pi$ coefficients $\rho_n$ are given recursively by $\rho_0 = 13/45$, $\rho_1 = 7/45$, and 
\begin{align}
    \rho_n &= (-1)^{n+1} \biggl[ \frac{8}{(2n+4)!} \Bigl( 1 - 2^{2n+3} B_{2n+4} \Bigr) 
    + \frac{13}{45}\frac{1}{(2n)!} 
    + \sum_{k=1}^{n-1} (-1)^k \frac{\rho_k}{[2(n-k)]!} \biggr] \\
    &= \frac{7}{45},\ \frac{4891}{75600},\  \frac{97987}{3742200},\ 
    \frac{866066051}{81729648000}, \ldots.
\end{align}
The seed sequence for the depth-two MZV terms, $\xi_n$, is given by $\xi_1 = 16$
\begin{align}
\xi_n &= \sum_{k=1}^{n-1} (-1)^{n+k+1} \frac{\xi_k}{[2(n-k)]!} \\
 &= 16,\ 8,\ \frac{10}{3},\ \frac{61}{45},\ \frac{277}{504},\ldots.    
\end{align}
The depth-two MZV's are encoded in the $f$-alphabet.

The first few $D_n$'s are given by:
\begin{align}
D_0 &= 26 \zeta_4 \,, \\
D_1 &= -16 f_{3,3} +147\zeta_6 \,, \\
D_2 &= -\Bigl(
16f_{5,3}+96f_{3,5}
+48\zeta_2 f_{3,3} 
- \frac{4891}{8} \zeta_8 \Bigr) \,, \\
D_3 &= -(16f_{7,3}+96f_{5,5}+240f_{3,7}
+\zeta_2(48 f_{5,3}+288 f_{3,5})
+300\zeta_4 f_{3,3} 
- \frac{97987}{40}\zeta_{10}) \,.
\end{align}
\par
Once we have the $\mathcal{F}$ for a diagram, we can compute the $\Gamma_{\rm cusp}$ contributions from the prescription in \cite{Arkani-Hamed:2021iya} given by
\begin{equation}
    g \partial_g \Gamma_{\rm cusp}(g) = -8 \mathcal{I} [{\cal F}(g,z) ]\,,
\end{equation}
where $\mathcal{I}$ integrates $\mathcal{F}$ through a contour along the unit circle, such that 
\begin{align}
    \mathcal{I}[f(z)] = -\frac{g^2}{2\pi}\int^\pi_{-\pi}f(e^{i\phi}) d\phi\,.
\end{align} 
We can use this to work out
\begin{equation}
    \mathcal{I}[z^p] = \frac{-g^2\sin{\pi p}}{\pi p}\,,
\end{equation}
and similarly
\begin{equation}
     \mathcal{I}[\log{z}^n] = \frac{-g^2\pi ^n \cos \left(\frac{\pi  n}{2}\right)}{n+1}\,.
\end{equation}
The integrated answer $-g^{2}{\cal I}_n \equiv \mathcal{I}[\mathcal{F}_{3,n}(z)]$ is then given by
\begin{align}
    {\cal I}_n &= 2^n\Big(\alpha_n + 32 \, (-1)^{n+1} \sum_{k=1}^n k \, f_{2(n-k)+3,2k+1} \nonumber\\
    &\hskip0.5cm
    + (-1)^{n+1} \sum_{j=0}^n \sum^j_{m=0} \frac{(-1)^m}{(2m+1)!}\xi_{j+1-m} \pi^{2j} \sum_{k=1}^{n-j} \, k(2k-1) \, f_{2(n-j-k)+3,2k+1}\Big) \,,
\end{align}
and the $\Gamma_{\rm cusp}$ contribution of each diagram is
\begin{align}\label{gammaF3n}
    \Gamma_{\rm cusp}(\mathcal{F}_{3,n}(z)) &= \frac{2^{2+n}g^{2(n+3)}}{(n+3)}\Big( \alpha_n + 32 \, (-1)^{n+1} \sum_{k=1}^n k \, f_{2(n-k)+3,2k+1} \nonumber\\
    &\hskip0.5cm
    + (-1)^{n+1} \sum_{j=0}^n \sum^j_{m=0} \frac{(-1)^m}{(2m+1)!}\xi_{j+1-m} \pi^{2j} \sum_{k=1}^{n-j} \, k(2k-1) \, f_{2(n-j-k)+3,2k+1} \,\Big)\,,
\end{align}
where the pure-$\pi$ terms are given by $\alpha_0 = 2/15 \times \pi^4$, 
$\alpha_1 = -176/2835 \times \pi^6$, and for $n>1$,
\begin{align}
    \alpha_n &= (-1)^{n+1} \biggl[ 
      \frac{4}{3} \frac{(2n+5)(2n^2+n+3)}{2n+1}  \zeta_{2n+4}
    + \frac{8}{3} \pi^2 \zeta_{2n+2} \biggr] 
   \nonumber\\
    &\hskip0.5cm 
    + \pi^{2n+4} \biggl[ \frac{8}{(2n+5)!} + \frac{13}{45(2n+1)!}
    +\sum_{k=1}^n (-1)^k 
   \frac{\rho_k}{[2(n-k)+1]!} \biggr] \,.
\end{align}
We could also replace the even zeta values with Bernoulli numbers in this formula.

Here is an alternate version of ${\cal I}_n$ which uses conventional MZVs, namely $\zeta_{m,n} = \sum_{j>i>0} \frac{1}{j^mi^n}$, instead of the $f$-alphabet \cite{Schnetz:2016fhy}:
\begin{align}
    {\cal I}_n &= 2^n\beta_n + (-2)^n \biggl[16\zeta_{2n+2,2}
    - 32 \sum_{k=1}^n k \, \zeta_{2k+1} \, \zeta_{2(n-k)+3}
    + \sum_{j=0}^n \sum^j_{m=0} \frac{(-1)^m}{(2m+1)!}\xi_{j+1-m} \, \pi^{2j}
    \Bigl( - \zeta_{3,2(n-j)+1}
    \nonumber\\
    & +[2(n-j)+1] \zeta_{2(n-j)+2,2}
     + (n-j+1) [2(n-j)+1] \zeta_{2(n-j)+3,1} \Bigr) \biggr] \,,
\end{align}
where the pure-$\pi$ terms are now given by:
\begin{align}
    \beta_n &= (-1)^{n+1} \biggl[ 
      - 8 \, \zeta_{2n+4}
    + \frac{8}{3} \, \pi^2 \, \zeta_{2n+2} \nonumber\\
    &\hskip2cm + \sum_{k=0}^n \sum^k_{m=0} \frac{(-1)^m}{(2m+1)!}\xi_{l+1-m} \, \pi^{2k} \, \zeta_{2(n-k)+4} \Bigl( \frac{1}{6} [2(n-k)+1] [2(n-k)^2+2(n-k)-3] \nonumber\\
    &\hskip6.5cm + \tilde\delta_{2(n-k)+1} \Bigr) \biggr] 
   \nonumber\\
    &\hskip0.5cm 
    + \pi^{2n+4} \biggl[ \frac{8}{(2n+5)!} + \frac{13}{45(2n+1)!}
    +\sum_{k=1}^n (-1)^k 
   \frac{\rho_k}{[2(n-k)+1]!} \biggr] \,.
\end{align}
and
\begin{align}
    \tilde\delta_n &= \frac{5}{4} \,, \quad n=1, \nonumber\\
    &= \frac{1}{2} \,, \quad n=3, \nonumber\\
    &= 0, \quad \hbox{otherwise}.
\end{align}
As an example, we show the $\Gamma_{\rm cusp}$ contributions of the triangle with one, two and three ladders attached on one end below. To start off, the triangle with no ladders attached ($n=0$) contributes
\begin{equation}
   \Gamma_{\rm cusp} \left( g^4 \begin{tikzpicture}
[baseline=-0.5ex,scale=1.,transform shape]
    \draw[Maroon, ultra thick] (0,-0.75)--(0,0.75);
    \draw[Maroon, ultra thick](0,-0.75)--(.75,0) node[at start,black] {\Huge$\bullet$}; 
    \draw[Maroon, ultra thick](0,0.75)--(.75,0) node[at start,black] {\Huge$\bullet$};
    \filldraw[white] (0.75,0) circle (5pt) node[]{};
    \node[black] at (.75,0) {\Large$\otimes$};
\end{tikzpicture} \right) = \frac{ 8 g^6 \pi^4}{45}\,.
\end{equation}
Next, we have the triangle with a single ladder attached, contributing
\begin{align}
    \Gamma_{cusp}\left(-g^6\begin{tikzpicture}[baseline={(0,0)cm}]
    \draw[Maroon, ultra thick] (-0.75,0.75)--(-0.75,-0.75);
    \draw[Maroon, ultra thick] (0,0)--(-0.75,0.75) node[at end, black] {\huge$\bullet$};
    \draw[Maroon, ultra thick] (0,0)--(-0.75,-0.75) node[at end, black] {\huge$\bullet$};
    \draw[Maroon, ultra thick] (0,0)--(.75,0) node[at start, black] {\huge$\bullet$};
    \node[white] at (.75,0) {\Huge$\bullet$};
    \node[black] at (.75,0) {\large$\otimes$};
\end{tikzpicture}\right) = -g^8(\frac{352 \pi ^6}{2835}-48 \zeta^2_3).
\end{align}
Adding another branch, we get
\begin{align}
 \Gamma_{\rm cusp}\left(g^8\begin{tikzpicture}[baseline={(0,0)cm},scale=0.8]
    \draw[Maroon, ultra thick] (-0.75,0.75)--(-0.75,-0.75);
    \draw[Maroon, ultra thick] (0,0)--(-0.75,0.75) node[at end, black] {\Huge$\bullet$};
    \draw[Maroon, ultra thick] (0,0)--(-0.75,-0.75) node[at end, black] {\Huge$\bullet$};
    \draw[Maroon, ultra thick] (0,0)--(1.5,0) node[at start, black] {\Huge$\bullet$} node[pos=0.5, black] {\Huge$\bullet$};
    \node[white] at (1.5,0) {\Huge$\bullet$};
    \filldraw[white] (1.5,0) circle (6pt) node[]{};
    \node[black] at (1.5,0) {\Large$\otimes$};
\end{tikzpicture}\right) = \frac{4g^{10}}{5} \left(64 \zeta_{5,3}+384 \zeta_{6,2}-\frac{32}{3}  \pi ^2 \zeta^2_3-1408 \zeta_5 \zeta_3+\frac{136 \pi ^8}{567}\right).
\end{align}
Note that we can't reduce this expression to single zetas, so we get our first true multiple-zeta value at 5-loops. We can re-write $\zeta_{5,3}$ in terms of $\zeta_{6,2}$ for a cleaner expression and get
\begin{align} 
    \Gamma_{\rm cusp}\left(g^8\begin{tikzpicture}[baseline={(0,0)cm},scale=0.8]
    \draw[Maroon, ultra thick] (-0.75,0.75)--(-0.75,-0.75);
    \draw[Maroon, ultra thick] (0,0)--(-0.75,0.75) node[at end, black] {\Huge$\bullet$};
    \draw[Maroon, ultra thick] (0,0)--(-0.75,-0.75) node[at end, black] {\Huge$\bullet$};
    \draw[Maroon, ultra thick] (0,0)--(1.5,0) node[at start, black] {\Huge$\bullet$} node[pos=0.5, black] {\Huge$\bullet$};
    \node[white] at (1.5,0) {\Huge$\bullet$};
    \filldraw[white] (1.5,0) circle (6pt) node[]{};
    \node[black] at (1.5,0) {\Large$\otimes$};
\end{tikzpicture}\right) =  \frac{4 g^{10}}{5} \left(224 \zeta_{6,2}-\frac{32}{3} \pi ^2 \zeta^2_3-1088 \zeta_5\zeta_3+\frac{2812 \pi ^8}{14175}\right).
\end{align}
Whether we choose to express everything in terms of $\zeta_{5,3}$ or $\zeta_{6,2}$, we get an answer we can't reduce to only single zetas. For the triangle + 3-ladder, we have
\begin{align}
   \Gamma_{\text{triangle},3} &= \frac{2g^{12}}{3}  \Big(\frac{128}{15} \left(5 \pi ^2 \zeta_{5,3}+\pi ^4 \zeta^2_3+30 \pi ^2 \zeta_5 \zeta_3 +555 \zeta_7 \zeta_3+270 \zeta^2_5\right)-128 \zeta_{7,3}-1024 \zeta_{8,2} \nonumber\\ &-\frac{25952 \pi ^{10}}{155925}\Big) 
 \,. 
\end{align}
This pattern continues and for any triangle with 2 or more branches attached, and we find contributions with depth-2 MZVs that can't be reduced to only single zetas.

We can do the same type of analysis for the box. The inverted Laplacian acting on the box $n$-times gives:
\begin{align}
    \mathcal{F}_{4,n}(z)=&({-}2)^{n{+}4}\bigg[3 H_{\vec{0}_{2 n{+}6}}(z)+ 2 H_{\vec{0}_{2n},{-}1,\vec{0}_5}(z) + 3 H_{\vec{0}_{2n}, {-}1, {-}1, \vec{0}_4}(z)-\frac{1}{2} H_{\vec{0}_{2 n{+}1},{-}1,\vec{0}_4}(z)\nonumber\\
    &{-}4 H_{\vec{0}_{2 n+1},{-}1,{-}1,\vec{0}_3}(z){+}3
   H_{
   \vec{0}_{2 n+2},-1,\vec{0}_3}(z){+}\frac{3}{2} H_{\vec{0}_{2 n+3},-1,0,0}(z){-}2 H_{\vec{0}_{2
   n},-1,-2,\vec{0}_3}(z)\nonumber\\
   &{+}2 H_{\vec{0}_{2 n+1},-1,-2,0,0}(z){+}\frac{1}{6} \pi ^2 H_{\vec{0}_{2
   n+4}}(z){+}\frac{7}{6} \pi ^2 H_{\vec{0}_{2 n},-1,\vec{0}_3}(z){+}\frac{1}{2} \pi ^2 H_{\vec{0}_{2
   n},-1,-1,0,0}(z)\nonumber\\
   &-5 H_{\vec{0}_{2 n},-1,-3,0,0}(z)-\frac{5}{12} \pi ^2 H_{\vec{0}_{2
   n+1},-1,0,0}(z)-\frac{2}{3} \pi ^2 H_{\vec{0}_{2 n+1},-1,-1,0}(z)\nonumber\\
   &+\frac{1}{2} \pi ^2 H_{\vec{0}_{2
   n+2},-1,0}(z)+\frac{3}{4} \pi ^2 H_{\vec{0}_{2 n+3},-1}(z)+2 \zeta_3 H_{\vec{0}_{2 n+3}}(z)+4 \zeta_3 H_{\vec{0}_{2 n},-1,0,0}(z)\nonumber\\
   &-8 \zeta_3 H_{\vec{0}_{2 n+1},-1,0}(z)-\frac{1}{3} \pi ^2 H_{\vec{0}_{2
   n},-1,-2,0}(z)+\pi ^2 H_{\vec{0}_{2 n+1},-1,-2}(z)+\frac{5}{72} \pi ^4 H_{\vec{0}_{2
   n+2}}(z)\nonumber\\
   &{+}\frac{11}{45} \pi ^4 H_{\vec{0}_{2 n},-1,0}(z){+}\frac{1}{8} \pi ^4 H_{\vec{0}_{2
   n},-1,-1}(z){-}\frac{5}{2} \pi ^2 H_{\vec{0}_{2 n},-1,-3}(z){-}\frac{43}{144} \pi ^4 H_{\vec{0}_{2
   n+1},-1}(z)\nonumber\\
   &+\frac{31}{2} \zeta_5 H_{\vec{0}_{2 n+1}}(z)-19 \zeta_5 H_{\vec{0}_{2 n},-1}(z)+\frac{3}{2}
   \pi ^2 \zeta_3 H_{\vec{0}_{2 n+
   1}}(z)-\frac{1}{3} \pi ^2 \zeta_3 H_{\vec{0}_{2 n},-1}(z)\nonumber\\
   &{+}11 \zeta_3^2
   H_{\vec{0}_{2 n}}(z){+}\frac{505 \pi ^6 H_{\vec{0}_{2 n}}(z)}{9072}{+}\sum_{k{=}1}^n \hat{C}_{k} \frac{\log^{2(n{-}k){+}1}(z)}{(2(n{-}k){+}1)!} {+}\sum_{k{=}1}^n \hat{D}_{k} \frac{\log^{2(n{-}k)}(z)}{(2(n{-}k))!}\bigg]\,.
\end{align}
The coefficients $\hat{C}_n$ and $\hat{D}_n$ are determined recursively using the boundary condition $\mathcal{F}_{4,n}(-1)=0$. The imaginary and real parts of the boundary condition fix $\hat C_n$ and $\hat D_n$ respectively. Let us start with 
\begin{align}
    \hat C_n = 2 \sum_{a=0}^{n-1} \frac{1}{(2a)!} \left|2^{2a-1}-1\right| \pi^{2a} \left|B_{2a}\right|\hat f_{n-a} \,,
\end{align}
where $B_n$ is the $n^\text{th}$ Bernoulli number and $\hat f_n$ is the imaginary part of the other functions in $I_{4,n}$ evaluated at -1. The first few coefficients $\hat C_n$ are then
\begin{align}
    \hat C_1=&\hat f_1\,,\nonumber\\
    \hat C_2=&\frac16 \pi^2 \hat f_1 + \hat f_2\,,\nonumber\\
    \hat C_3=&\frac7{360} \pi^4 \hat f_1 + \frac16 \pi^2 \hat f_2 + \hat f_3\,,\nonumber\\
    \hat C_4=&\frac{31}{15120}\pi^6 \hat f_1 + \frac{7}{360} \pi^4 \hat f_2 + \frac16 \pi^2 \hat f_3 + \hat f_4\,.
\end{align}
The function $\hat f_n$ can be be expressed purely in terms of zeta and multiple zeta values by using the shuffle identities satisfied by polylogarithms. Simplified, this gives
\begin{align}
\label{eq:f_box}
    \hat f_n =&{-}\frac{({-}1)^n \left(6 n^2{+}15 n{+}7\right) \pi ^{2 n{+}2}\zeta _3 }{(2 n{+}3)!}{-}\frac{31 ({-}1)^n \pi ^{2 n}\zeta _5 }{2 (2 n{+}1)!}{+}\left(10 n^2{+}27
   n{+}17\right) \zeta _{2 n{+}3,2}\nonumber\\
   &{+}\left(4 n^3{+}20 n^2{+}31 n{+}15\right) \zeta _{2 n{+}4,1}{-}6 \zeta _{2 n{+}1,4}{+}2 (2 n{+}5) \zeta _{2 n{+}2,3}{+}\pi ^2
   \left(\frac{5 n^2}{3}{+}\frac{19 n}{6}{+}\frac{3}{2}\right) \zeta _{2 n{+}3}\nonumber\\
   &{+}\left(\frac{4 n^4}{3}{+}\frac{22 n^3}{3}{+}\frac{62
   n^2}{3}{+}\frac{92 n}{3}{+}16\right) \zeta _{2 n{+}5}{-}4 \zeta _3 (2 n{+}3) \zeta _{2 n{+}2}{+}\frac{1}{15} \pi ^4 \zeta _{2 n{+}1}\,.
\end{align}

Putting this together, gives the first few $\hat C_i$:
\begin{align}
    \hat C_1 = &\frac{13 \pi ^4 \zeta _3}{30}+\frac{107 \pi ^2 \zeta _5}{12}+38 \zeta _7\,,\nonumber\\
    \hat C_2 = &\frac{989 \pi ^6 \zeta _3}{15120}+\frac{1961 \pi ^4 \zeta _5}{720}+\frac{125 \pi ^2 \zeta _7}{6}+\frac{141 \zeta _9}{2}\,,\nonumber\\
\hat C_3=&\frac{167 \pi ^8 \zeta _3}{100800}+\frac{1997 \pi ^6 \zeta _5}{3360}+\frac{283 \pi ^4 \zeta _7}{45}+\frac{151 \pi ^2 \zeta _9}{4}+113
   \zeta _{11}\,.
\end{align}
Note that the $\hat{C}_i$ only contain single zeta values because every multiple zeta value in \eqref{eq:f_box} is of the form $\zeta_{\text{odd},\text{even}}$ or $\zeta_{\text{even},\text{odd}}$.

The coefficients $\hat D_n$ can similarly be written now using Euler numbers $E_n$,
\begin{align}
    \hat D_n = \sum_{a=0}^{n-1} \frac{1}{(2a)!}\pi^{2a}(-1)^{a+1} E_{2a}\, \hat g_{n-a}\,.
\end{align}
As before, the function $\hat g_n$ is the value of the other polylogarithms in $I_{4,n}(z)$ evaluated at $z=-1$. The first few coefficients are
\begin{align}
    \hat D_1=&-\hat g_1\nonumber\\
    \hat D_2=&-\frac12 \pi^2 \hat g_1 - \hat g_2\nonumber\\
    \hat D_3=& -\frac{5}{24}\pi^4 \hat g_1 - \frac12 \pi^2 \hat g_2 - \hat g_3\nonumber\\
    \hat D_4=& -\frac{61}{720} \pi^6 \hat g_1 -\frac{5}{24} \pi^4 \hat g_2 - \frac12 \pi^2 \hat g_3 - \hat g_4\,.
\end{align}
Using the shuffle identities to write these in terms of zeta and multiple zeta values, we can simplify the result to
\begin{align}
    \hat g_n = &\frac{({-}1)^n \left(4040 n^6{+}42420 n^5{+}175490 n^4{+}359835 n^3{+}373331 n^2{+}172998 n{+}19368\right) \pi ^{2 n{+}6}}{1134 (2 n{+}6)!}\nonumber\\
    &{+}\frac{11 \zeta
   _3^2 ({-}1)^n \pi ^{2 n}}{(2 n)!}{-}\pi ^2 \left(2 n^2{+}\frac{17 n}{3}{+}\frac{11}{3}\right) \zeta _{2 n{+}3,1}{+}\left(4 n^2{+}22 n{+}18\right)
   \zeta _{2 n{+}3,3}\nonumber\\
   &{+}\left(\frac{20 n^3}{3}{+}32 n^2{+}\frac{145 n}{3}{+}23\right) \zeta _{2 n{+}4,2}{+}\left(2 n^4{+}\frac{46 n^3}{3}{+}\frac{83
   n^2}{2}{+}\frac{283 n}{6}{+}19\right) \zeta _{2 n{+}5,1}\nonumber\\
   &{-}\frac{7}{3} \pi ^2 \zeta _{2 n{+}1,3}{-}19 \zeta _{2 n{+}1,5}{-}\pi ^2 \left(\frac{10
   n}{3}{+}3\right) \zeta _{2 n{+}2,2}{+}4 (1{-}3 n) \zeta _{2 n{+}2,4}\nonumber\\
   &{-}\zeta _3 \left(8 n^2{+}28 n{+}20\right) \zeta _{2 n{+}3}{-}\pi ^2 \left({-}\frac{2
   n^3}{9}{-}n^2{+}\frac{5 n}{9}{+}\frac{7}{3}\right) \zeta _{2 n{+}4}{-}\pi
   ^4 \left(\frac{23 n}{45}{+}\frac{13}{90}\right) \zeta _{2 n{+}2}\nonumber\\
   &{+}\left(\frac{8 n^5}{15}{+}\frac{13 n^4}{3}{+}\frac{53 n^3}{3}{+}\frac{503
   n^2}{12}{+}\frac{981 n}{20}{+}\frac{39}{2}\right) \zeta _{2 n{+}6}{+}\frac{7}{3} \pi ^2 \zeta _3 \zeta _{2 n{+}1}{+}19 \zeta _5 \zeta _{2 n{+}1}\,.
\end{align}
Putting this together gives the first few $D_i$:
\begin{align}
\hat D_1 = &-19 \zeta _{5,3}+5 \pi ^2 \zeta _3^2-39 \zeta _5 \zeta _3+\frac{157517 \pi ^8}{5443200}\,,\nonumber\\
\hat D_2 = &-\frac{331}{30} \pi ^2 \zeta _{5,3}-12 \zeta _{7,3}+\frac{49}{24} \pi ^4 \zeta _3^2-\frac{51}{2} \pi ^2 \zeta _5 \zeta _3-60 \zeta _7
   \zeta _3-\frac{313 \zeta _5^2}{2}+\frac{2398813 \pi ^{10}}{199584000}\,,\nonumber\\
   \hat D_3 = &-\frac{189}{40} \pi ^4 \zeta _{5,3}-\frac{155}{21} \pi ^2 \zeta _{7,3}-6 \zeta _{9,3}+\frac{149}{180} \pi ^6 \zeta _3^2-\frac{89}{8}
   \pi ^4 \zeta _5 \zeta _3-45 \pi ^2 \zeta _7 \zeta _3+14 \zeta _9 \zeta _3\nonumber\\
   &-\frac{6949}{84} \pi ^2 \zeta _5^2-641 \zeta _5 \zeta
   _7+\frac{29357017 \pi ^{12}}{6604416000}\,.
\end{align}
We see that all coefficients are products of zeta values times integers over powers of 2 with a two exceptions:
\begin{enumerate}
    \item The coefficient of $H_{\vec{0}_{2n}}$.
    \item The coefficient of $\zeta_{6+2n}$ in $\hat{D}_n$.
\end{enumerate}
Both of these are related to the constant term in the box integral which cannot be written as a zeta value times an integer over a power of 2.

Now that we have the Wilson loop expectation value $\mathcal{F}(z)$ for this set of graphs, we can also calculate their contribution to $\Gamma_{\rm cusp}$.
\begin{align}
    &\frac{({-}1)^n \left(4040 n^6{+}54540 n^5{+}296690 n^4{+}826965 n^3{+}1233131 n^2{+}917964 n{+}256536\right) \pi ^{2 n{+}6}}{1134 (2 n{+}7)!}\nonumber\\
    &{+}\frac{11
   \zeta _3^2 ({-}1)^n \pi ^{2 n}}{(2 n{+}1)!}{-}\left(10 n^2{+}37 n{+}33\right) \zeta _{2 n{+}4,2}{-}2 \left(2 n^3{+}13 n^2{+}27 n{+}18\right) \zeta
   _{2 n{+}5,1}{+}6 \zeta _{2 n{+}2,4}\nonumber\\
   &{-}4 (n{+}3) \zeta _{2 n{+}3,3}{-}\frac{1}{6} (2 n{+}3) \left(4 n^3{+}24 n^2{+}65 n{+}75\right) \zeta _{2 n{+}6}{+}8 \zeta _3
   (n{+}2) \zeta _{2 n{+}3}{-}\frac{1}{15} \pi ^4 \zeta _{2 n{+}2}\nonumber\\
   &{-}\frac{1}{6} \pi ^2 (2 n{+}3) (5 n{+}7) \zeta _{2 n{+}4}{{+}}\sum _{a=1}^n \frac{({-}1)^{n{-}k}\hat D_{k}\pi^{2(n{-}k)}}{(2 (n{-}k){+}1)!}\,,
\end{align}
where $\hat D_i$ are as defined above and one can check that the expression contains irreducible multiple zeta values.

\subsection*{\texorpdfstring{$\Gamma_{\rm cusp}$}{Gamma\_cusp} Contribution of \texorpdfstring{$\mathcal{F}_{3}(z){\cal F}_{\rm tree}(g,z)$}{F\_3(z)F\_tree(g,z)}}\label{app:tretriangle}

We can apply the same summation techniques for the $\mathcal{I}[{\cal F}_{\rm tree}(z)\mathcal{F}_{3}(z)]$, which now gives us triple sums instead of double sums. First, we use
\begin{equation}
   {\cal F}_{\rm tree}(z) = \frac{A^2z^A}{g^2(1{+}z^A)^2} = \frac{A}{g^2}\sum_{m} (mA)(-1)^m z^{mA}\,.
\end{equation}
Then the integrals of interest are of the form
\begin{equation}
    \int^\pi_{-\pi}H_{-m_2,-m_1}(e^{i\phi})(i\phi)^a (e^{i\phi mA}) d\phi\,.
\end{equation}
Using the methods outlined in Appendix \eqref{app:sums}, we can analytically perform the sums over the HPL terms, but we still have leftover summation terms from writing ${\cal F}_{\rm tree}(e^{i\phi})$ as an infinite sum. More specifically, we get terms of the form
\begin{equation}
    \sum_m \frac{8 (-1)^m\cos (\pi  A m) \psi ^{(i)}(A m+1)}{A^j m^j}\,,
\end{equation}
where the integers $i,j$ run between 1 and 4. We can check that each of these sums are convergent and we can also perform the sum for the term that gives the leading order contribution analytically, which gives
\begin{equation}
    \mathcal{I}[g^4{\cal F}_{\rm tree}\mathcal{F}_{\rm 3}(z)]_{g>>1} \sim \frac{22 \pi ^3}{45}g^5+\mathcal{O}(g^4\log(g))\,.
\end{equation}
Interestingly, while the leading $g$ term is free of logarithms, like before, the sub-leading log terms don't cancel out. While we can't sum the sub-leading terms analytically, we can perform the summations after taking the asymptotic expansions of polylogarithms, i.e. $\psi^{(k)}(z>>1) \sim \frac{1}{z^k}$ or by approximating $1+Am \sim 1+m$ for large $g$. Doing this approximation gives us sub-leading log terms. This is confirmed by evaluating the integral numerically for different values of g. The full result for $g>>1$ is
\begin{align}
    \mathcal{I}[g^4{\cal F}_{\rm tree}(g,z)\mathcal{F}_{3}(z)] \sim &\frac{22 \pi ^3}{45}g^5+5.13 g^4{-}16. g^4 \log (g){-}5.5 g^3{-}13.8 g^3 \log (g){-}8.5 g^2 \log (g)\nonumber\\
    &-4.8 g^2-3.2 g-4.5 g \log (g)-2.1 \log (g)\,+\mathcal{O}\Big(\frac{1}{g}\Big)\,.
\end{align}
\section{Usage of ancillary files}
\label{app:anc}
The symbol and function-level results of all negative geometries in section \ref{sec:integrated_neg_geom} are documented in the {\tt Mathematica} notebook {\tt demo\_HPL.nb}, which loads a source file {\tt all\_HPL\_data.m}

We use the frozen node to specify the inequivalent negative geometries. For example, there are three triangle diagrams, according to $AB$, $CD$, or $EF$ being frozen in the following diagram. 
\begin{align} \label{eq:D3,1}
   \begin{tikzpicture}[baseline=-0.6ex,scale=1.25,transform shape]
    \draw[Maroon, ultra thick] (0.5,-0.5)--(0.5,0.5);
    \draw[Maroon, ultra thick](0.5,-0.5)--(1,0) node[at start,black] {\huge$\bullet$}; 
    \draw[Maroon, ultra thick](0.5,0.5)--(1,0) node[at start,black] {\huge$\bullet$};
    \draw[Maroon, ultra thick](1,0)--(1.5,0) node[at end,black] {\huge$\bullet$} node[at end, right]{};
    \node[black] at (1.0,0) {\huge$\bullet$};
    \node[black] at (2.,0) {\scriptsize$AB$};
    \node[black] at (1.2,0.35) {\scriptsize$CD$};
    \node[black] at (0.3,0.85) {\scriptsize$EF$};
    \end{tikzpicture}
\end{align}

The object {\tt triangleAB} in the {\tt Mathematica} notebook gives the HPL result of the triangle diagram with the AB-node being frozen, similarly for CD and EF. Its symbol {\tt SBtriangleAB} is directly given in the notebook, e.g.
\begin{align}
 {\rm SBtriangleAB} = 16\, (&{\rm SB}[z, z, z, z, z, z] + {\rm SB}[z, z, z, 1 + z, z, z] \nonumber\\
 &+ {\rm SB}[z, z, 1 + z, z, z, z] - 2 {\rm SB}[z, z, 1 + z, 1 + z, z, z]);
\end{align}
Similarly, one can get the HPL expressions of all negative geometries by entering
\begin{align}
    \{&{\tt tree, triangleAB, triangleCD, triangleEF, boxOneCycleAB,}\nonumber\\
&{\tt boxTwoCycleAB, boxTwoCycleCD, boxThreeCycleAB}\}, 
\end{align}
which sums to the three-loop Wilson loop observable with Lagrangian insertion, {\tt F3} in the {\tt Mathematica} notebook.

\bibliographystyle{JHEP}
\bibliography{refs}
\end{document}